\documentclass[a4paper,11pt]{article}
\usepackage{jinstpub} 
\usepackage{lineno}

\title{A Unified Approach for Coupled Beam Optics in Accelerators}

\author[a,b]{Onur Gilanliogullari}
\author[a]{Brahim Mustapha}
\author[b]{Pavel Snopok}
\affiliation[a]{Argonne National Laboratory}
\affiliation[b]{Illinois Institute of Technology}

\emailAdd{ogilanliogullari@anl.gov}

\abstract{	
Coupled beam optics can be geometrically described in terms of invariant eigenmode planes of a stable symplectic ``one-turn'' map $\mathcal M\in Sp(4)$. We show that the non-uniqueness of symplectically normalized bases within each eigenmode plane constitutes an in-plane gauge freedom $Sp(2)\times Sp(2)$, and that many coupled-optics parametrizations differ primarily by gauge choice. Building on this fact, we identify basis-independent descriptors of lattice and beam optics and introduce bounded, gauge-invariant coupling parameters or fractions $u_{k,\mathrm{inv}}$ computed from orthogonal projectors onto the eigenmode planes. To obtain smooth $s$-dependent optics functions and consistent mode labeling, we present a unifying and practical approach based on an $SO(2)$ continuity gauge (Procrustes alignment), together with diagnostics for stability and invariance. We further relate Edwards--Teng, Mais--Ripken, Lebedev--Bogacz, Wolski, and Sagan--Rubin parametrizations as gauge-equivalent representations within the respective $Sp(2)\times Sp(2)$ gauge freedom. Numerical examples of coupled lattices and beam optics illustrate the proposed invariants and show how representation-dependent scalar coupling parameters (e.g.\ in the Lebedev--Bogacz gauge) can leave their nominal bounds while $u_{k,\mathrm{inv}}$, defined here, remain bounded and physically interpretable.
}


\begin{document}
\maketitle
\flushbottom

\section{Introduction}\label{sec:intro}

Much of accelerator beam optics is built around the uncoupled Courant--Snyder theory, in which the transverse motion separates into independent horizontal and vertical degrees of freedom described by the Twiss parameters~\cite{courant1958theory}. However, many realistic lattices exhibit coupling---from solenoids, skew quadrupoles, magnet roll errors, and dispersion/correction schemes---making it necessary to extend the Courant--Snyder framework to coupled dynamics. In the uncoupled setting, Twiss parameters and normalized coordinates provide a compact description of amplitudes, projected envelopes, and stability. They form the standard language for lattice design and correction. An analogous, convention-robust description is desirable in the coupled case. Background on uncoupled linear optics and its extensions can be found in standard accelerator physics textbooks~\cite{lee2018accelerator,wiedemann2015particle}. More generally, because the dynamics are Hamiltonian, linear transfer maps are symplectic and preserve phase-space volume in canonical coordinates, consistent with Liouville's theorem. Mathematically, these maps are in the symplectic group $Sp(2N)$, where $N$ is the degree of freedom.

Several formalisms exist for describing linear coupled transverse dynamics. Widely used approaches include Edwards--Teng~\cite{edwards1973parametrization}, Mais--Ripken~\cite{willeke1989methods}, Lebedev--Bogacz~\cite{lebedev2010betatron}, Wolski~\cite{Wolskicoupled}, and Sagan--Rubin~\cite{sagan1999linear}. In transverse linear optics, the one-turn (or cell) transfer map $\mathcal M$ is symplectic and therefore lies in $Sp(4)$, including the effects of solenoids and skew quadrupoles. In the stable non-degenerate case, $\mathcal M$ lends itself to a decomposition of the transverse phase space into two $\mathcal M$-invariant two-dimensional symplectic subspaces (eigenmode planes) $\mathcal P_1$ and $\mathcal P_2$, generalizing the horizontal and vertical planes of uncoupled motion. While the eigenmode planes are uniquely defined in this case, the choice of a symplectically normalized basis within each plane is not unique. However, different choices can be related by independent in-plane $Sp(2)$ transformations, forming a natural $Sp(2)\times Sp(2)$ gauge freedom. 

As a consequence, physically meaningful coupled-optics descriptors should be constructed from gauge-invariant quantities that depend only on the planes and on the dynamics restricted to them (Sec.~\ref{subsec:geometricsubsection}). By contrast, Twiss-like coupled lattice functions and certain phase conventions are representation-dependent and vary across parametrizations. Different coupled-optics formalisms can thus be viewed as different gauge fixings of the same underlying invariant structure. Accordingly, all physical observables (e.g.\ projected beam sizes/second moments and eigenmode tunes) must agree across parametrizations even when the reported Twiss-like parameters differ. 

We call a quantity \emph{gauge invariant} if it is unchanged under the full in-plane $Sp(2)$ freedom, i.e.\ under symplectic reparametrizations of a chosen basis spanning a given eigenmode plane. After adopting a particular gauge, we may additionally define \emph{derived} (convention-dependent) quantities—e.g.\ Twiss-like functions and actions—that are invariant under the remaining residual gauge (often $SO(2)$ phase rotations) but depend on the chosen convention. 

Having identified this gauge structure, we introduce a basis-independent (gauge-invariant) coupling measure that quantifies the overlap of each eigenmode plane with the horizontal/vertical coordinate subspaces (Sec.~\ref{subsec:geometricsubsection}). We also relate our parametrization to historical coupled-optics formalisms (Appendix~\ref{app:formalismrelations} and Sec~\ref{subsec:comparisonwithothers}), showing that they correspond to different symplectic basis reparametrizations (gauge choices) built on the same invariant eigenmode-plane structure. In this work, we adopt the convenient eigenvector parametrization of Lebedev--Bogacz~\cite{lebedev2010betatron}, and we emphasize that the corresponding scalar coupling parameter is gauge dependent. Meaning that if the intended in-plane gauge convention is not enforced (or becomes ill-conditioned near degeneracy), this parameter can leave its nominal $[0,1]$ range even though the underlying eigenmode plane is unchanged.

This paper is organized as follows: we give a brief review of uncoupled beam optics and the Courant--Snyder parametrization in Sec.~\ref{sec:background}. Section ~\ref{sec:coupledparam} introduces coupled beam dynamics, with subsections that show gauge-invariant quantities and the definition of gauge-invariant coupling measures. Section~\ref{sec:Results} presents our formalism's applications and comparisons with other formalisms using optics codes such as MADX~\cite{MADXManual} and OptiMX~\cite {OptiMX}. 

\section{Brief Review of Uncoupled Beam Optics}\label{sec:background}

\subsection{Courant--Snyder Formalism}\label{subsec:CSformalism}
The Courant-Snyder parametrization employs Floquet theory to solve Hill's equation~\cite{courant1958theory,lee2018accelerator}. Hill's equation~\cite{hill1878researches} provides parametric solutions to uncoupled systems that are represented in terms of transfer matrices. The transfer matrices, for stable oscillations, are symplectic and their eigenvalues lie on the complex unit circle. Symplecticity results in conserved area of the phase space and constrained eigenvalues result in sinusoidal solution with intial conditions. An uncoupled system has the transfer matrix format as $\mathcal{M}=\mathrm{diag}(M,N)$, where $M$ is transfer matrix for horizontal plane $(x,x')$ and $N$ is transfer matrix for vertical plane $(y,y')$. Symplectic unit matrices are defined as 
\begin{equation}
    S_4 = \begin{pmatrix}
        S_2 & 0 \\
        0 & S_2
    \end{pmatrix}\,, \qquad S_2 = \begin{pmatrix}
        0 & 1 \\
        -1 & 0
    \end{pmatrix}.
    \label{eq:symplecticunitmatrices}
\end{equation}
Transfer matrix symplecticity condition is written as $\mathcal{M}^TS_4\mathcal{M}=S_4$ in 4D, or $M^TS_2M=S_2$ in 2D representation. Symplecticity results in $\mathrm{det}(\mathcal{M})=1$ for 4D representation and $\mathrm{det}(M)=\mathrm{det}(N)=1$ for 2D representation. For stability, the eigenvalues of the transfer map lie on the complex unit circle and relate to tunes of the system as
\begin{equation}
    \begin{split}
        \mathcal{M}\vec{v}_i = \lambda_i\vec{v}_i, \qquad \lambda_i = e^{\pm i\mu_i} = e^{\pm i2\pi Q_i}. 
    \end{split}
    \label{eq:eigenvectorrelationshipunc}
\end{equation}
Here, the eigenvalues are complex conjugated pairs $\lambda_x = (\lambda_x^*)^{-1}$ and $Q_{x,y}$ are called the tunes of the system. Based on the sympleciticity and stability criteria, Courant and Snyder parametrize the phase space vector as
\begin{equation}
    \begin{split}
        x&= \sqrt{2J_x\beta_x}\cos\psi_x, \quad y=\sqrt{2J_y\beta_y}\cos\psi_y, \\
        x'&= -\sqrt{\frac{2J_x}{\beta_x}}\left(\alpha_x\cos\psi_x + \sin\psi_x \right), \quad y'=-\sqrt{\frac{2J_y}{\beta_y}}\left(\alpha_y\cos\psi_y + \sin\psi_y\right).
    \end{split}
\end{equation}
Here, $J_{x,y}$ are the particle amplitudes (actions), $\beta_{x,y}$ the betatron amplitude functions, $\alpha_{x,y}=-\frac{1}{2}\beta'_{x,y}$ the alpha functions, and $\psi_{x,y}$ the betatron phases. The particle's amplitude, $J$ is expressed as the invariant of phase space:
\begin{equation}
    2J_x = \left( \frac{1+\alpha_x^2}{\beta_x} \right)x^2 + 2\alpha_xxx' + \beta_xx'^2,
    \label{eq:CSinvariant}
\end{equation}
for $x$ dimension, and similar for $y$ dimension. Furthermore, the optics functions $(\beta,\alpha,\gamma)$ are called Twiss parameters, where $\gamma=\tfrac{1+\alpha^2}{\beta}$. The existence of conserved quantity, allows one to write a transformation law for the optics functions as
\begin{equation}
    \begin{pmatrix}
        \beta \\
        \alpha\\
        \gamma
    \end{pmatrix}_f = \begin{pmatrix}
        M_{11}^2 & -2M_{11}M_{12} & M_{12}^2 \\
        -M_{11}M_{21} & (M_{11}M_{22} + M_{12}M_{21}) & -M_{12}M_{22} \\
        M_{21}^2 & -2M_{21}M_{22} & M_{22}^2
    \end{pmatrix}\cdot \begin{pmatrix}
        \beta \\
        \alpha \\
        \gamma
    \end{pmatrix}_i.
    \label{eq:Twissfunctionstransport}
\end{equation}
Here, $f,i$ subscripts refer to initial and final locations. Equation~\eqref{eq:Twissfunctionstransport} is found in almost every accelerator textbook~\cite{wiedemann2015particle,lee2018accelerator} and if the fundamental descriptor for matching and transporting the uncoupled Twiss functions that characterizes the particle motion.  

\subsection{Geometric Description of Uncoupled Dynamics}\label{subsec:geomuncrel}

To prepare for the coupled case, we recast the uncoupled Courant--Snyder formalism in the same geometric language used below.  
In canonical coordinates $\vec z=(x,p_x,y,p_y)^T\in\mathbb R^4$, an uncoupled lattice has a block-diagonal one-turn map
$\mathcal M=\mathrm{diag}(M,N)$.
In the stable (non-degenerate) case, $\mathcal M$ has two complex-conjugate eigenpairs, one supported entirely in the horizontal subspace and one entirely in the vertical subspace.  
The associated real invariant eigenmode planes are
\begin{equation}
\mathcal P_x \equiv \mathrm{span}\{\Re \vec v_x,\Im \vec v_x\}=\mathcal X,
\qquad
\mathcal P_y \equiv \mathrm{span}\{\Re \vec v_y,\Im \vec v_y\}=\mathcal Y,
\end{equation}
where $\mathcal X=\{(x,p_x,0,0)^T\}$ and $\mathcal Y=\{(0,0,y,p_y)^T\}$.
Therefore the phase space decomposes as the direct sum
\begin{equation}
\mathbb R^4=\mathcal P_x\oplus\mathcal P_y.
\end{equation}
Let $\vec v_x$ and $\vec v_y$ denote the normalized complex eigenvectors associated with the two eigenpairs
$\lambda_x=e^{\pm i\mu_x}$ and $\lambda_y=e^{\pm i\mu_y}$.
A Courant--Snyder representative choice is
\begin{equation}
\vec v_x=
\begin{pmatrix}
\sqrt{\beta_x}\\[2pt]
-\dfrac{\alpha_x+i}{\sqrt{\beta_x}}\\[2pt]
0\\[2pt]
0
\end{pmatrix},
\qquad
\vec v_y=
\begin{pmatrix}
0\\[2pt]
0\\[2pt]
\sqrt{\beta_y}\\[2pt]
-\dfrac{\alpha_y+i}{\sqrt{\beta_y}}
\end{pmatrix},
\label{eq:uncoupled_evecs_CS}
\end{equation}
with symplectic normalization $\vec v_{x,y}^\dagger S_4 \vec v_{x,y}=-2i$.
From each complex eigenvector we construct a real symplectic generator pair spanning the corresponding plane,
\begin{equation}
W_x \equiv [\vec z_1\ \vec z_2],\qquad \vec z_1=\Re\,\vec v_x,\ \ \vec z_2=-\Im\,\vec v_x,
\qquad
W_y \equiv [\vec z_3\ \vec z_4],\qquad \vec z_3=\Re\,\vec v_y,\ \ \vec z_4=-\Im\,\vec v_y,
\label{eq:WxWy_from_evecs}
\end{equation}
so that
\begin{equation}
W_x^T S_4 W_x=S_2,\qquad W_y^T S_4 W_y=S_2,\qquad W_x^T S_4 W_y=0.
\label{eq:uncoupled_sympl_orth}
\end{equation}
The last condition expresses the symplectic orthogonality of the planes $\mathcal P_x$ and $\mathcal P_y$.

The usual Twiss functions can be recovered directly from the generators.
Writing $\vec z_1=(x_1,p_{x1},0,0)^T$ and $\vec z_2=(x_2,p_{x2},0,0)^T$, one obtains
\begin{equation}
\beta_x = x_1^2+x_2^2,\qquad
\alpha_x = -(x_1 p_{x1}+x_2 p_{x2}),\qquad
\gamma_x = p_{x1}^2+p_{x2}^2,
\label{eq:twiss_from_generators_unc_x}
\end{equation}
and similarly for the vertical plane from $\vec z_3,\vec z_4$.

Finally, the phase advance can be defined in a form that will generalize to the coupled case.
Because $\mathcal P_x$ is invariant under $\mathcal M$, there exists a unique $2\times2$ reduced map $R_x$ such that
\begin{equation}
\mathcal M W_x = W_x R_x,
\qquad
R_x = W_x^{+}\mathcal M W_x,
\qquad
W_x^{+}\equiv -S_2 W_x^T S_4,
\label{eq:reduced_map_unc_x}
\end{equation}
and likewise for $R_y$.
The reduced maps satisfy $R_{x,y}\in Sp(2)$ and have eigenvalues $e^{\pm i\mu_{x,y}}$,
so $\mu_{x,y}$ coincide with the usual Courant--Snyder phase advances (and $Q_{x,y}=\mu_{x,y}/2\pi$ for one turn).
This construction will be used verbatim in the coupled case, with $W_x,W_y$ replaced by the eigenmode-plane bases $W_{1,2}$.

The eigenvectors are defined only up to a complex phase $\vec v_{x,y}\mapsto e^{i\theta}\vec v_{x,y}$, which corresponds to a rotation of the real generator pair $W_{x,y}\mapsto W_{x,y}R(\theta)$; all Twiss functions and the reduced-map eigenvalues are invariant under this choice.

\section{Coupled Beam Optics Parametrization}\label{sec:coupledparam}

\subsection{Overview and relation to standard parametrizations}
There are five widely used coupled-optics parametrizations: Edwards--Teng~\cite{edwards1973parametrization}, Mais--Ripken~\cite{willeke1989methods}, Lebedev--Bogacz~\cite{lebedev2010betatron}, Wolski~\cite{Wolskicoupled}, and Sagan--Rubin~\cite{sagan1999linear}. A summary of coupled parametrizations is given in~\cite{vanwelde2023parametrizations}. Edwards--Teng aims to decouple the one-turn map $\mathcal{M}$ into normal modes. Mais--Ripken and Lebedev--Bogacz provide Twiss-like parametrizations that extend the Courant--Snyder description to coupled motion. Wolski's approach starts from the covariance (beam) matrix $\Sigma$ and constructs coupled lattice functions directly related to second moments.

We begin from geometric properties common to all approaches. For linear Hamiltonian transport the one-turn map $\mathcal{M}\in \mathrm{Sp}(4)$ is symplectic, $\mathcal{M}^TS_4\mathcal{M}=S_4$ (with $S_4$ defined in Eq.~\eqref{eq:symplecticunitmatrices}). For stable motion its spectrum consists of two complex-conjugate eigenpairs on the unit circle. All these approaches ultimately rely on the same geometric fact: a stable symplectic one-turn map admits two invariant symplectic planes (``eigenmode planes''), and the dynamics restricted to each plane is $2$D symplectic. 

We now formalize this geometric backbone and emphasize the gauge freedom within the defined geometry. We also define the reduced dynamics $R_k$ on each eigenmode plane.

\subsection{Geometric basis: eigenmode planes and reduced dynamics}\label{subsec:geometricsubsection}

\begin{figure}
    \centering
    \includegraphics[width=\linewidth]{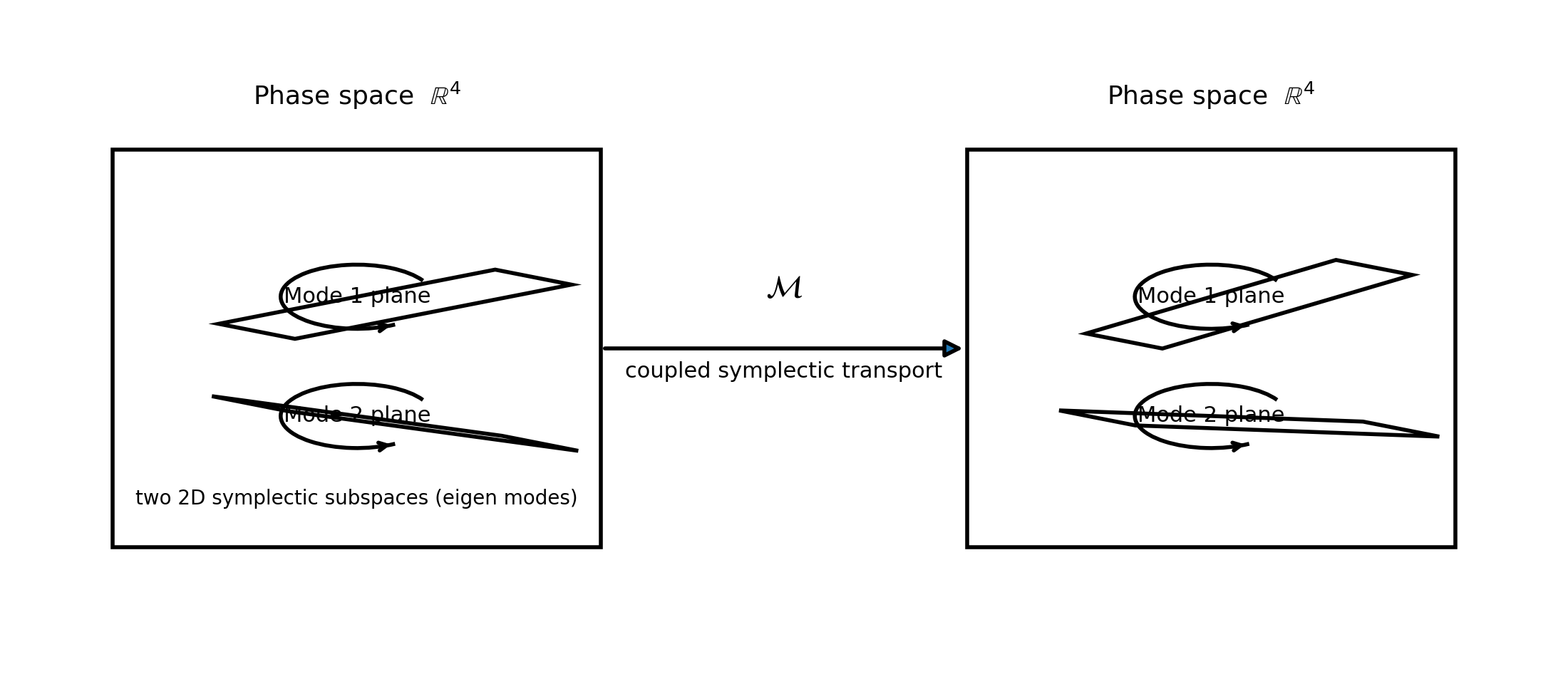}
    \caption{Schematic representation of a coupled linear symplectic transfer map $\mathcal{M}\in Sp(4)$ acting on the transverse phase space $\mathbb{R}^4=(x,p_x,y,p_y)$. For a stable map, the phase space admits a decomposition into two $\mathcal{M}$-invariant two-dimensional symplectic subspaces (normal-mode or eigenmode planes) $\mathcal{P}_1$ and $\mathcal{P}_2$, such that $\mathbb{R}^4=\mathcal{P}_1\oplus\mathcal{P}_2$ and $\mathcal{M}\mathcal{P}_k=\mathcal{P}_k$. Motion corresponds to an elliptic linear map (conjugate to a rotation) on each plane with phase advances $\mu_1$ and $\mu_2$.}
    \label{fig:mapill}
\end{figure}

The one-turn coupled map is
\begin{equation}
    \mathcal{M} = \begin{pmatrix}
        M & m \\
        n & N
    \end{pmatrix},
\end{equation}
where $M,N,m,n$ are $2\times 2$ blocks. We assume $\mathcal{M}\in \mathrm{Sp}(4)$ and that the motion is stable, so its spectrum consists of two complex-conjugate eigenvalue pairs on the unit circle. Let $\mathcal{M}\vec{v}_k=\lambda_k\vec{v}_k$ ($k=1,2$) with $\lambda_k=e^{\pm i\mu_k}=e^{\pm i2\pi Q_k}$, where $Q_k$ is the eigenmode tune and $\mu_k$ is the eigenmode phase advance, $Q_k=\mu_k/2\pi$. In the non-degenerate (simple-spectrum) case, each complex pair defines a unique real two-dimensional invariant subspace (eigenmode plane)
\begin{equation}
\mathcal{P}_k=\mathrm{span}\{\Re\vec v_k,\Im\vec v_k\}.
\end{equation}
In this case the transverse phase space decomposes as $\mathbb{R}^4=\mathcal{P}_1\oplus\mathcal{P}_2$. The illustration of the map and the eigenplanes in 4D phase space is given in Fig.~\ref{fig:mapill}.

Although the invariant planes $\mathcal{P}_{1,2}$ are uniquely defined in the stable simple-spectrum case, the choice of symplectically normalized bases within each plane is not unique. If $W_k\in\mathbb{R}^{4\times2}$ spans $\mathcal{P}_k$ and satisfies the symplectic normalization $W_k^TS_4W_k=S_2$, then for any $G_k\in Sp(2)$ the transformed basis $W_k\mapsto W_kG_k$ spans the same plane and preserves the normalization. This within-plane freedom forms a gauge group $Sp(2)\times Sp(2)$: it changes the coordinate representation inside each eigenmode (e.g.\ phase origins and Twiss-like parametrizations) but leaves invariant all physical, basis-independent quantities such as tunes, invariant subspaces/projectors, and second moments (the beam matrix $\Sigma$). 

With the within-plane gauge freedom $W_k\mapsto W_kG_k$ ($G_k\in Sp(2)$) understood as a pure reparametrization within the invariant eigenmode planes, we call a quantity \emph{gauge invariant} if it is unchanged under this transformation. Physically meaningful coupled-optics descriptors must therefore be constructed from such invariants, which depend only on the planes $\mathcal P_k$ and on the dynamics restricted to them. Representative examples include:
\begin{itemize}
    \item \textbf{Dynamics invariants (lattice):} the eigenmode tunes $Q_{1,2}$ (from the eigenvalues of the one-turn map), and the invariant eigenmode subspaces $\mathcal P_{1,2}$ in the simple-spectrum stable case. In the degenerate/near-degenerate case the individual planes may not be uniquely defined, but the invariant subspace associated with the degenerate pair is. Equivalently, one may use basis-independent projectors onto these subspaces, e.g.\ the Euclidean projectors $\Pi_k=W_k(W_k^TW_k)^{-1}W_k^T$, or the symplectic (oblique) projectors $\Pi_k^{(S)}=W_kW_k^{+}$ once the complementary splitting $\mathbb R^4=\mathcal P_1\oplus\mathcal P_2$ is fixed (see Appendix~\ref{app:InvariantProjectors}). 
    \item \textbf{Beam invariants (distribution):} the eigen-emittances $\varepsilon_{1,2}$, i.e.\ the Williamson invariants\cite{Williamson1936} of the covariance matrix $\Sigma$. Equivalently, $\pm i\varepsilon_{1,2}$ are the eigenvalues of $S_4\Sigma$ (or $\Sigma S_4$).
    \item \textbf{Measurable second moments:} physical-coordinate second moments such as $\langle x^2\rangle,\langle y^2\rangle,\langle xx'\rangle,$ etc.\ (entries of $\Sigma$ in the chosen laboratory coordinates) and derived projected quantities such as $\sigma_x=\sqrt{\langle x^2\rangle}$, $\sigma_y=\sqrt{\langle y^2\rangle}$, and projected emittances.
    \item \textbf{Coupling invariants (plane content):} for each eigenmode plane $\mathcal P_k$ there exists a \emph{basis-independent coupling fraction} $u_{k,\mathrm{inv}}\in[0,1]$ that quantifies how strongly the plane overlaps the $(y,p_y)$ coordinate subspace (equivalently, $1-u_{k,\mathrm{inv}}$ measures overlap with $(x,p_x)$). This scalar depends only on $\mathcal P_k$ (or equivalently its projector) and is therefore gauge invariant.
    \item \textbf{Dynamical invariants along the lattice:} for any reference point $s_0$, the accumulated eigenmode phase advances $\mu_k(s\leftarrow s_0)$ obtained from the eigenvalues of the accumulated reduced maps $R_k(s\leftarrow s_0)$ are gauge invariant and satisfy $\mu_k(s_0+L\leftarrow s_0)=2\pi Q_k \; (\mathrm{mod}\;2\pi)$.
    \item \textbf{Scalar invariants of the beam matrix:} besides the eigen-emittances, any symmetric polynomial of the spectrum of $S_4\Sigma$ (e.g.\ $\det\Sigma$) is invariant and encodes equivalent information about $\varepsilon_{1,2}$.
\end{itemize}
The list above gives examples of quantities that are invariant under the full within-plane gauge freedom $W_k\mapsto W_kG_k$ with $G_k\in Sp(2)$ (i.e.\ under $Sp(2)\times Sp(2)$ acting independently in each eigenmode plane), and therefore represent genuine basis-independent content of the dynamics and of the beam distribution.

To work with the eigenmode planes we choose, for each $\mathcal{P}_k$, a real $4\times2$ basis matrix $W_k=[\vec z_{2k-1}\ \vec z_{2k}]$ such that $\mathrm{span}(W_k)=\mathcal{P}_k$ and
\begin{equation}
W_k^T S_4 W_k = S_2,\qquad k=1,2,
\label{eq:Wk_norm}
\end{equation}
where $S_{2n}$ denotes the standard symplectic form. A convenient representative is obtained from a normalized complex eigenvector $\vec v_k$ of $\mathcal M$ associated with the stable eigenvalue $e^{i\mu_k}$ by taking
\begin{equation}
\vec z_{2k-1}=\Re\,\vec v_k,\qquad \vec z_{2k}=-\Im\,\vec v_k,
\qquad\text{so that}\qquad
W_k=[\Re\,\vec v_k\ \ -\Im\,\vec v_k],
\label{eq:Wk_from_vk}
\end{equation}
which satisfies \eqref{eq:Wk_norm} provided $\vec v_k^\dagger S_4 \vec v_k=-2i$. The remaining freedom $\vec v_k\to e^{i\theta_k}\vec v_k$ induces $W_k\to W_k R(\theta_k)$, i.e. the residual $SO(2)$ phase gauge in the normalized representation. The eigenvector normalization also implies $\vec{z}_{2k-1}^TS_4\vec{z}_{2k}=1$ for $k=1,2$ by construction. Since $W_k$ spans the invariant (eigenmode) plane $\mathcal{P}_k$, the image of this basis under $\mathcal{M}$ remains inside the same plane, and therefore there exists a unique $2\times 2$ matrix $R_k$ such that
\begin{equation}
    \mathcal{M}W_k = W_kR_k.
    \label{eq:uniqueRmat}
\end{equation}
When $W_k$ is symplectically normalized (Eq.~\eqref{eq:Wk_norm}), we can introduce its symplectic adjoint
\begin{equation}
    W_k^{+} \equiv -S_2W_k^TS_4,
\end{equation}
which satisfies the left-inverse identity
\begin{equation}
    W_k^{+}W_k=I_2.
\end{equation}
Consequently the induced (restricted) map on the plane is obtained by projection,
\begin{equation}
    R_k =W^{+}_k\mathcal{M}W_k.
\end{equation}
Because $\mathcal{M}\in\mathrm{Sp}(4)$ and $W_k$ is symplectically normalized, one finds $R_k\in\mathrm{Sp}(2)$. Moreover, under any within-plane gauge transformation $W_k\mapsto W_kG_k$ with $G_k\in Sp(2)$, the reduced map transforms by similarity,
\begin{equation}
R_k \mapsto G_k^{-1}R_kG_k ,
\end{equation}
so its eigenvalues are gauge invariant. For stable motion these eigenvalues lie on the unit circle and can be written as $e^{\pm i\mu_k}$; thus the eigenmode phase advance $\mu_k$ (and tune $Q_k=\mu_k/2\pi$ for a one-turn map) is a gauge-invariant quantity defined directly from the dynamics restricted to $\mathcal{P}_k$. When computing an $s$-dependent ``phase advance'', it is important to distinguish a local notion from an accumulated one. A short segment map need not represent a rotation: for example, a drift is a symplectic shear (with unit eigenvalues) rather than an elliptic rotation, so an ``instantaneous'' phase advance is not uniquely defined at every location. A robust gauge-invariant definition is instead obtained from the accumulated transfer map from a reference point $s_0$ to $s$. Let $\mathcal M(s\leftarrow s_0)$ denote the transfer matrix from $s_0$ to $s$. With a symplectically normalized basis $W_k(s_0)$ spanning $\mathcal P_k$ at $s_0$, we define the accumulated reduced map
\begin{equation}
R_k(s\leftarrow s_0)\equiv W_k^{+}(s_0)\,\mathcal M(s\leftarrow s_0)\,W_k(s_0).
\end{equation}
The accumulated phase advance $\mu_k(s)$ is then defined as the argument of the stable eigenvalues of $R_k(s\leftarrow s_0)$,
$e^{\pm i\mu_k(s)}$. By construction, $\mu_k(s)$ reproduces the one-turn phase advance when $s=s_0+L$ and is invariant under the within-plane gauge choice at $s_0$. This accumulated definition is the natural object to compare against tunes extracted directly from the eigenvalues of the full one-turn map.

To obtain a gauge-invariant measure of coupling associated with the plane itself, we use the Euclidean (orthogonal) projector onto the subspace $\mathcal P_k$,
\begin{equation}
\Pi_k = W_k\,(W_k^T W_k)^{-1}W_k^T,
\end{equation}
which depends only on the subspace $\mathcal P_k=\mathrm{span}(W_k)$ and is invariant under any within-plane change of basis $W_k\mapsto W_kG_k$ with $G_k\in GL(2)$ (in particular $G_k\in Sp(2)$), as shown in Appendix~\ref{app:InvariantProjectors}. By construction, $\Pi_k^T=\Pi_k$ and $\Pi_k^2=\Pi_k$, so $\Pi_k$ is the orthogonal projector with respect to the standard Euclidean inner product. We first introduce coordinate projectors as
\begin{equation}
    P_X = \begin{pmatrix}
        I_2 & 0 \\
        0 & 0
    \end{pmatrix}, \quad P_Y=\begin{pmatrix}
        0 & 0 \\
        0 & I_2
    \end{pmatrix}, \quad P_X+P_Y=I_4.
    \label{eq:projectionsontophasespace}
\end{equation}
We then define the invariant coupling fractions
\begin{equation}
u_{k,\mathrm{inv}}\equiv \frac{1}{2}\mathrm{tr}(P_Y\Pi_k),\qquad
1-u_{k,\mathrm{inv}}=\frac{1}{2}\mathrm{tr}(P_X\Pi_k),
\label{eq:invariantcouplingdef}
\end{equation}
so that $u_{k,\mathrm{inv}}\in[0,1]$ (since $0\le \mathrm{tr}(P_Y\Pi_k)\le 2$) and $(1-u_{k,\mathrm{inv}})+u_{k,\mathrm{inv}}=1$. The scalar $u_{k,\mathrm{inv}}$ is a basis-independent coupling fraction that quantifies the Euclidean overlap (in the chosen laboratory coordinates) of the eigenmode plane $\mathcal P_k$ with the $(y,p_y)$ subspace: it equals the average squared projection of an orthonormal basis of $\mathcal P_k$ onto $(y,p_y)$. In the uncoupled limit, one has $\mathcal P_1=\mathcal X$ and $\mathcal P_2=\mathcal Y$ (up to mode labeling), and therefore $u_{1,\mathrm{inv}}=0$, $u_{2,\mathrm{inv}}=1$ (or vice versa after swapping mode labels). More generally:
\begin{itemize}
\item $u_{k,\mathrm{inv}}\approx 0$: mode $k$ is predominantly horizontal (the plane is close to $\mathcal X$);
\item $u_{k,\mathrm{inv}}\approx 1$: mode $k$ is predominantly vertical (the plane is close to $\mathcal Y$);
\item $u_{k,\mathrm{inv}}\approx 1/2$: the plane has comparable content in $\mathcal X$ and $\mathcal Y$.
\end{itemize}

With these constructions, the coupled linear optics is characterized by (i) the invariant eigenmode planes $\mathcal P_k$ and their basis-independent coupling content $u_{k,\mathrm{inv}}$, and (ii) the reduced symplectic dynamics $R_k$ on each plane, whose eigenvalues define the gauge-invariant phase advances $\mu_k$ (and tunes $Q_k$). In the next subsection we adopt a convenient gauge fixing to parameterize eigenvectors/generators in a Twiss-like form (Lebedev--Bogacz style), introducing derived quantities such as projected $(\beta,\alpha,\gamma)$ functions and relative coupling phases that are invariant only under the residual $SO(2)$ phase gauge.

\subsection{Gauge fixing and Twiss-like eigenvector parametrization}\label{subsec:twisslikeparam}

The constructions in Sec.~\ref{subsec:geometricsubsection} provide gauge-invariant geometric and dynamical objects (the planes $\mathcal P_k$, the reduced maps $R_k$, and the invariant coupling fractions $u_{k,\mathrm{inv}}$). To obtain a practical Twiss-like description of the coupled motion—analogous to the uncoupled Courant--Snyder formalism—we now adopt a convenient \emph{gauge fixing} within each eigenmode plane. Specifically, we choose complex eigenvectors $\vec v_k$ (and real generators $W_k$) such that the remaining freedom is reduced to an $SO(2)$ phase rotation. This allows us to define derived projected functions $(\beta_{kx},\alpha_{kx},\gamma_{kx})$, $(\beta_{ky},\alpha_{ky},\gamma_{ky})$ and relative coupling phases, which are invariant under the residual phase gauge but not under general $Sp(2)$ reparametrizations. 

We choose Lebedev--Bogacz parametrization~\cite{lebedev2010betatron} of eigenvectors as a convenient gauge choice~/ representative form. The eigenvectors from Lebedev--Bogacz are parametrized as
\begin{equation}
    \vec{v}_1 = \begin{pmatrix}
        \sqrt{\beta_{1x}} \\
        -\dfrac{\alpha_{1x}+ iA_{x}(W_1)}{\sqrt{\beta_{1x}}} \\
        \sqrt{\beta_{1y}}\,e^{i\nu_1} \\
        -\dfrac{\alpha_{1y}+ iA_{y}(W_1)}{\sqrt{\beta_{1y}}}\,e^{i\nu_1}
    \end{pmatrix}, \qquad
    \vec{v}_2=\begin{pmatrix}
        \sqrt{\beta_{2x}}\,e^{i\nu_2} \\
        -\dfrac{\alpha_{2x}+ iA_{x}(W_2)}{\sqrt{\beta_{2x}}}\,e^{i\nu_2} \\
        \sqrt{\beta_{2y}} \\
        -\dfrac{\alpha_{2y}+ iA_{y}(W_2)}{\sqrt{\beta_{2y}}}
    \end{pmatrix}.
\label{eq:vk_general_form}
\end{equation}
The parameters $(\beta_{kx},\alpha_{kx},\beta_{ky},\alpha_{ky})$ describe the projections of the eigenmode plane onto the coordinate subspaces, while $\nu_k$ is a relative phase between the $(x,p_x)$ and $(y,p_y)$ components. The quantities $A_{x}(W_k),A_{y}(W_k)$ encode the within-plane canonical normalization and are constrained by $\vec v_k^\dagger S_4 \vec v_k=-2i$; they should therefore be viewed as gauge-dependent partitions rather than independent physical invariants. The quantities $A_{x}(W_k)$ and $A_y(W_k)$ represent signed symplectic areas and are defined as
\begin{equation}
A_x(W_k)\equiv \vec z_{2k-1}^{\,T}S_4P_X\,\vec z_{2k},\qquad
A_y(W_k)\equiv \vec z_{2k-1}^{\,T}S_4P_Y\,\vec z_{2k},
\label{eq:AxAy_signed}
\end{equation}
which satisfy $A_x(W_k)+A_y(W_k)=1$ by construction. The parametrization of eigenvectors in Eq.~\eqref{eq:vk_general_form} is a gauge choice that resembles Courant--Snyder eigenvectors, and it is not a unique definition. The eigenmode planes $\mathcal{P}_{1,2}$ are unique, but the choice of basis inside the planes are not. Since the basis $W_k$ is constructed from the eigenvectors as defined in Eq.~\eqref{eq:Wk_from_vk},  we define projected (coupled) Twiss functions by quadratic forms that are invariant under the within-plane phase gauge $W_k \mapsto W_kR(\theta_k)$ ($R(\theta_k) \in SO(2)$) as:
\begin{equation}
    \begin{split}
        \beta_{kx} &\equiv \|E_xW_k \|_F^2, \quad \gamma_{kx}\equiv \|E_{p_x}W_k \|_F^2, \quad \alpha_{kx}\equiv -\langle E_xW_k,E_{p_x}W_k \rangle_F, \\
        \beta_{ky} &\equiv \|E_yW_k \|_F^2, \quad \gamma_{ky}\equiv \|E_{p_y}W_k \|_F^2, \quad \alpha_{ky}\equiv -\langle E_yW_k,E_{p_y}W_k \rangle_F.
    \end{split}
    \label{eq:coupledTwissfunctions}
\end{equation}
Here, $E_x=\mathrm{diag}(1,0,0,0)$, $E_{p_x}=\mathrm{diag}(0,1,0,0)$, $E_y=\mathrm{diag}(0,0,1,0)$, $E_{p_y}=\mathrm{diag}(0,0,0,1)$, $\|A\|_{F}\equiv \sqrt{\mathrm{tr}(A^{T}A)}$ is the Frobenius norm, and $\langle A,B\rangle_{F}\equiv \mathrm{tr}(A^{T}B)$ is the Frobenius inner product. 

The coupling phases $\nu_k$ can be extracted from the complex ratio of the $(y,p_y)$ and $(x,p_x)$ components of $\vec v_k$ or, equivalently, from the relative orientation between $P_XW_k$ and $P_YW_k$. To make the relative phase between the $(x,p_x)$ and $(y,p_y)$ projections explicit, we split the complex eigenvector into its two canonical blocks,
\begin{equation}
    \vec{v}_k =\begin{pmatrix}
        v_{kX} \\
        v_{kY}
    \end{pmatrix}, \qquad v_{kX},v_{kY}\in\mathbb{C}^2,
\end{equation}
where $v_{kX}$ contains the $(x,p_x)$ components and $v_{kY}$ contains the $(y,p_y)$ components. Since the eigenvector is defined only up to an overall complex phase $\vec{v}_k\mapsto e^{i\theta_k}\vec{v}_k$, any meaningful “coupling phase” must be invariant under this transformation. A convenient choice is to define $\nu_k$ from the relative complex orientation of the two blocks,
\begin{equation}
    \nu_k \equiv \arg\!\left(v_{kY}^\dagger v_{kX}\right),
\end{equation}
(or equivalently using the normalized inner product $\arg\!\left(v_{kY}^\dagger v_{kX}/(\|v_{kY}\|\|v_{kX}\|)\right)$).
This definition is unchanged by the eigenvector phase gauge $\vec v_k\mapsto e^{i\theta_k}\vec v_k$ and reduces to the familiar $\nu_k$ in the Lebedev--Bogacz representative form, where the $(y,p_y)$ components are written with an $e^{i\nu_k}$ factor relative to $(x,p_x)$ for mode~1.
However, $\nu_k$ becomes ill-conditioned when either projection is vanishingly small (uncoupled limit), and it can vary under equally valid choices of eigenvector basis when the modes are nearly degenerate (e.g. when $\mu_1\simeq \mu_2$, allowing small rotations/mixings within the associated invariant subspace without changing the underlying dynamics).
Therefore $\nu_k$ is best interpreted as a meaningful relative phase only when both projections are non-negligible and the eigenmodes are well separated.

To maintain mode-labeling continuously, we also introduce a continuity gauge. In Lebedev-Bogacz formalism, the eigenmode~1 is defined as the plane with larger $x$ contribution and the eigenmode~2 plane is defined as the plane with larger $y$ contribution to the dynamics. This notion can result in mode flips/swaps when the coupling is strong and close to eigenvalue degeneracy. Mode flips in the dynamics can result in discontinuities that can result in tedious bookkeeping of mode swaps if they happen. To resolve this, we introduce a continuity gauge that keeps the mode labels and invariant planes the same from initial definitions. In practice we fix the gauge so that the residual freedom is $SO(2)$, which corresponds to a phase convention inside $\mathcal P_k$ and leaves all quadratic invariants unchanged. Consider a discrete set of longitudinal locations $\{s_i\}$ with transfer maps $\mathcal M_{i\to i+1}\equiv \mathcal M(s_{i+1}\leftarrow s_i)$. Starting from a symplectically normalized basis $W_k(s_i)$, we propagate it by
\begin{equation}
\widetilde W_k(s_{i+1}) = \mathcal M_{i\to i+1}\,W_k(s_i),
\label{eq:propagate_W}
\end{equation}
and then choose a rotation $R(\theta)\in SO(2)$ such that the updated basis
\begin{equation}
W_k(s_{i+1}) = \widetilde W_k(s_{i+1})\,R(\theta_k^\star)
\label{eq:continuity_update}
\end{equation}
is as close as possible (in a Euclidean sense) to the previous basis $W_k(s_i)$. Concretely, $\theta_k^\star$ is defined by the Procrustes criterion~\cite{schoenemann1966procrustes,gower1975generalizedprocrustes}
\begin{equation}
\theta_k^\star \;=\;\arg\min_{\theta\in(-\pi,\pi]}
\big\|\,W_k(s_i)-\widetilde W_k(s_{i+1})R(\theta)\,\big\|_F,
\label{eq:procrustes}
\end{equation}
where $\|\cdot\|_F$ is the Frobenius norm. This is equivalent to maximizing $\mathrm{tr}\!\left(W_k(s_i)^T\widetilde W_k(s_{i+1})R(\theta)\right)$ and yields a unique $\theta_k^\star$ except in pathological cases where the propagated columns are nearly orthogonal to the previous ones. The continuity gauge therefore selects a smooth representative $W_k(s)$ of each invariant plane and removes arbitrary sign flips and phase jumps caused by eigenvector phase choices.

In addition to the continuous within-plane gauge, there is a discrete ambiguity:
interchanging the two planes $(\mathcal P_1,\mathcal P_2)$ produces an equally valid
decomposition of $\mathbb R^4$. We fix the labels using a gauge-invariant plane
attribute at a reference location $s_0$, for example the invariant coupling fraction
$u_{k,\mathrm{inv}}$ defined in \eqref{eq:invariantcouplingdef}. One convenient convention is
\begin{equation}
u_{1,\mathrm{inv}}(s_0)\le u_{2,\mathrm{inv}}(s_0),
\label{eq:label_rule_uinv}
\end{equation}
so that ``mode~1'' is the plane with smaller overlap with the $(y,p_y)$ subspace at $s_0$.
The labels are then transported by continuity: at each step we choose the assignment
$(W_1,W_2)$ at $s_{i+1}$ that maximizes the total overlap with $(W_1,W_2)$ at $s_i$,
\begin{equation}
\max\left\{
\|W_1(s_i)^T\widetilde W_1(s_{i+1})\|_F^2+\|W_2(s_i)^T\widetilde W_2(s_{i+1})\|_F^2,\;
\|W_1(s_i)^T\widetilde W_2(s_{i+1})\|_F^2+\|W_2(s_i)^T\widetilde W_1(s_{i+1})\|_F^2
\right\},
\label{eq:label_swap_test}
\end{equation}
and then apply the within-plane rotations \eqref{eq:procrustes} to enforce the phase
continuity. This prevents spurious mode swaps driven purely by numerical eigensystem
choices, while still allowing the modes to become ambiguous in the physically singular
case $Q_1\simeq Q_2$, where the invariant planes themselves cease to be uniquely defined.

It is important to emphasize that all gauge-invariant observables—such as $u_{k,\mathrm{inv}}(s)$, the restricted-map eigenvalues $e^{\pm i\mu_k}$, and beam second moments reconstructed from $W_k(s)$—are independent of the particular $SO(2)$ phase choices made by \eqref{eq:procrustes}. The continuity gauge instead serves to make the remaining gauge-dependent quantities (such as the signed partitions $A_{x,y}(W_k)$, coupled Twiss functions, and relative phases) well behaved and continuous whenever the eigenmodes remain well separated and both plane projections are non-negligible.

\subsection{Mode coordinates, actions, and reconstruction of phase space}
\label{subsec:mode_coordinates}

To parametrize the mode coordinates and actions, let $W=[W_1\ W_2]\in\mathbb R^{4\times 4}$ be a symplectic frame adapted to the invariant eigenmode planes, with $W_k\in\mathbb R^{4\times 2}$ spanning $\mathcal P_k$ and
\begin{equation}
W^T S_4 W = S_4,
\qquad\text{equivalently}\qquad
W_i^T S_4 W_j = \delta_{ij}S_2.
\label{eq:W_full_norm}
\end{equation}
The second condition expresses \emph{the symplectic orthogonality} of the planes: it is automatically satisfied when $W$ is constructed from a stable symplectic eigenbasis
of the one-turn map and then is symplectically orthonormalized. Under \eqref{eq:W_full_norm}, $W$ is invertible with the explicit symplectic inverse
\begin{equation}
W^{-1} = -S_4W^T S_4.
\label{eq:W_inverse}
\end{equation}
Any phase-space vector $\vec z\in\mathbb R^4$ admits a unique decomposition
\begin{equation}
\vec z = W_1 \vec a_1 + W_2 \vec a_2,
\qquad \vec a_k\in\mathbb R^2,
\label{eq:z_decomposition}
\end{equation}
where the mode coordinates are obtained by the symplectic adjoints
\begin{equation}
\vec a_k = W_k^{+}\vec z,
\qquad
W_k^{+}\equiv -S_2W_k^T S_4,
\label{eq:mode_coordinates}
\end{equation}
so that $W_k^{+}W_k=I_2$ and $W_k^{+}W_j=0$ for $j\neq k$ (from \eqref{eq:W_full_norm}).
Thus $\vec a_k$ are the canonical coordinates of $\vec z$ in the eigenmode plane $\mathcal P_k$ (i.e. the amplitudes of the decomposition $\vec z=W_1\vec a_1+W_2\vec a_2$). If we define $\vec a\equiv(\vec a_1,\vec a_2)\in\mathbb R^4$, then $\vec a=W^{-1}\vec z$ and the symplectic form is preserved, $d\vec z^T S_4 d\vec z = d\vec a^T S_4 d\vec a$.

In the mode coordinates, the natural quadratic invariants are
\begin{equation}
J_k \equiv \tfrac12\,\vec a_k^T\vec a_k
=\tfrac12\,\vec z^T\,(W_k^{+})^T W_k^{+}\,\vec z,
\label{eq:actions_def_expanded}
\end{equation}
which generalize the Courant--Snyder invariants.
Because $(W_k^{+})^T W_k^{+}$ is unchanged under within-plane rotations
$W_k\mapsto W_kR(\theta)$, the action $J_k$ is invariant under the $SO(2)$ phase gauge.
In the uncoupled limit, $\mathcal P_1=\mathcal X$ and $\mathcal P_2=\mathcal Y$, one recovers
$J_1=J_x$ and $J_2=J_y$.

We parameterize each $\vec a_k$ by polar variables
\begin{equation}
\vec a_k = \sqrt{2J_k}\,(\cos\phi_k,\,-\sin\phi_k)^T,
\label{eq:ak_polar}
\end{equation}
so that the phase-space vector can be reconstructed as
\begin{equation}
\vec z(s)=\sqrt{2J_1}\Big(\vec z_1(s)\cos\phi_1(s)-\vec z_2(s)\sin\phi_1(s)\Big)
+\sqrt{2J_2}\Big(\vec z_3(s)\cos\phi_2(s)-\vec z_4(s)\sin\phi_2(s)\Big),
\label{eq:z_floquet_like}
\end{equation}
where $W_1=[\vec z_1\ \vec z_2]$ and $W_2=[\vec z_3\ \vec z_4]$.
Equation \eqref{eq:z_floquet_like} makes explicit that \emph{the geometry} (the transported generators)
and \emph{the invariants} (the actions $J_k$) together determine particle motion, similar to Mais-Ripken formalism~\cite{willeke1989methods}.

For a beam with mode-coordinate covariance $\Sigma_a=\langle \vec a\vec a^T\rangle$, the physical covariance is
\begin{equation}
\Sigma_z \equiv \langle \vec z\vec z^T\rangle = W\,\Sigma_a\,W^T.
\label{eq:sigma_transport}
\end{equation}
In particular, if the two modes are uncorrelated in the chosen gauge, $\Sigma_a=\mathrm{diag}(\varepsilon_1 I_2,\varepsilon_2 I_2)$ with $\varepsilon_k=\langle 2J_k\rangle$, then
\begin{equation}
\Sigma_z = \varepsilon_1\,W_1W_1^T + \varepsilon_2\,W_2W_2^T,
\label{eq:sigma_uncorrelated_modes}
\end{equation}
and the projected rms sizes follow directly: $\sigma_x^2 = e_x^T\Sigma_z e_x$, $\sigma_y^2=e_y^T\Sigma_z e_y$, etc. Thus the quantities $\beta_{kx}(s),\beta_{ky}(s)$ extracted from $W_k(s)$ determine how each eigen-emittance contributes to the observable projected beam envelopes.

\subsection{Comparison with other standard parametrizations}\label{subsec:comparisonwithothers}

The formalism developed above is based on the geometric fact that, for a stable one-turn map $\mathcal M\in Sp(4)$ with simple spectrum, the transverse phase space admits a unique decomposition into two $\mathcal M$-invariant symplectic planes $\mathcal P_1\oplus\mathcal P_2$. While the planes $\mathcal P_k$ are uniquely defined (up to mode labeling), the choice of a symplectically normalized basis within each plane is not: if $W_k$ spans $\mathcal P_k$ and satisfies $W_k^TS_4W_k=S_2$, then $W_k\mapsto W_kG_k$ with any $G_k\in Sp(2)$ spans the same plane and preserves the normalization. This within-plane freedom constitutes a gauge group $Sp(2)\times Sp(2)$ acting independently on the two eigenmode planes. Accordingly, physically meaningful coupled-optics descriptors must be built from gauge-invariant objects (Sec.~\ref{subsec:geometricsubsection}).

Once a particular gauge fixing is adopted---e.g.\ the common normalized (rotation) gauge in which the reduced maps take a rotation form---the remaining freedom reduces to a residual phase rotation $SO(2)$ in each plane. In such a gauge one may introduce derived Twiss-like functions and mode actions that are invariant under the residual $SO(2)$ phase gauge but not under general within-plane $Sp(2)$ transformations.

Physical observables---such as the eigenmode tunes (from the spectrum of $\mathcal M$) and the beam second moments encoded in the covariance matrix $\Sigma$---must agree across all parametrizations, even though the associated Twiss-like parameters and other representation-dependent quantities may differ. Accordingly, the standard coupled-optics parametrizations can be viewed as different gauge fixings (i.e.\ different symplectic basis conventions within the invariant planes) built on the same underlying gauge-invariant structure. 

\subsubsection{Relationship with Edwards--Teng parametrization}\label{subsubsec:ETrel}

For instance, Edwards--Teng parametrization states: there exists a matrix $T$ that diagonalizes the one-turn map $\mathcal{M}$ and parametrizes the diagonal map similar to Courant--Snyder form. The matrix $T$ exists, however the form that Edwards--Teng chooses to parametrize the $T$ matrix is a gauge-fixing. According to Edwards--Teng:
\begin{equation}
    \mathcal{M}_{\mathrm{diag}} = T^{-1}\cdot\mathcal{M}\cdot T.
\end{equation}
Here, the form of $T$ is chosen as:
\begin{equation}
    T = \begin{pmatrix}
        \cos\phi\, I_{2\times2} & -\sin\phi\, D^{-1} \\
        \sin\phi \, D & \cos\phi I_{2\times2}
    \end{pmatrix},
\end{equation}
where $D$ is a $2\times 2$ symplectic matrix, $I_{2\times 2}$ is a unit matrix, $\phi$ is the decoupling angle. As we know, the decoupling matrix is not unique and any gauge transformation of $T$ results in a diagonal matrix with preserved eigenvalues (tunes). In the formalism mentioned in Sec~\ref{subsec:geometricsubsection}, the $W=[W_1\;W_2]$ matrix is a symplectic basis matrix, $W^TS_4W = S_4$. From Eq.\eqref{eq:uniqueRmat}, we can write the diagonal reduced map, $R =\mathrm{diag}(R_1,R_2)$, as:
\begin{equation}
    R = W^{-1}\cdot\mathcal{M}\cdot W.
\end{equation}
Therefore, the basis $W$ also decouples the one-turn map $\mathcal{M}$. The $W$ and $T$ matrices are related with a gauge transformation
\begin{equation}
    T = W\cdot G, \quad \text{where} \quad G=\mathrm{diag}(G_1,G_2)\in Sp(2)\times Sp(2). 
\end{equation}
For the stable non-degenerate eigenvalues of the one-turn map $\mathcal{M}$, there exists a $G$ that will re-parametrize $W$ matrix to be in the form of Edwards--Teng decoupling matrix $T$. For numerical example, see Appendix~\ref{app:formalismrelations}. 

\subsubsection{Relationship with Wolski Parametrization}\label{subsubsec:wolskirel}

Wolski's parametrization starts from the beam covariance matrix $\Sigma$ and its invariants which we call eigenmode emittances or formally Williamson invariants. The beam matrix can be written as:
\begin{equation}
    \Sigma = N \begin{pmatrix}
        \epsilon_1 I_2 &0 \\
        0 & \epsilon_2 I_2
    \end{pmatrix}N^T.
\end{equation}
Here, $\epsilon_{1,2}$ are invariants of the beam matrix and called eigenmode emittances. Wolski then defines mode-contribution matrices as
\begin{equation}
    B_1 \equiv N\begin{pmatrix}
        I_2 & 0 \\
        0 & 0
    \end{pmatrix}N^T, \qquad B_2 \equiv N \begin{pmatrix}
        0 & 0 \\
        0 & I_2
    \end{pmatrix}N^T,
\end{equation}
so that
\begin{equation}
    \Sigma = \epsilon_1B_1 + \epsilon_2B_2.
\end{equation}
We constructed the beam matrix $\Sigma$ from mode coordinates in Eq.~\eqref{eq:sigma_transport} with $\vec{a}=W^{-1}\vec{z}$ and take an uncorrelated mode covariance in that gauge, $\Sigma_a=\mathrm{diag}(\epsilon_1I_2,\epsilon_2I_2)$ that yields:
\begin{equation}
    \Sigma_z = W\Sigma_aW^T = \epsilon_1W_1W_1^T + \epsilon_2W_2W_2^T.
\end{equation}
The comparison yields: $N=W$ and $B_k=W_kW_k^T$. However, more generally $N$ and $W$ are related by within-plane gauge choices. The eigenmode emittances $\epsilon_{1,2}$ and $\Sigma$ is gauge invariant, however the symplectic normalizer $N$ is not unique.

Wolski’s parametrization is a beam-matrix-first formulation of the same eigenmode geometry. Given a symplectic normalizing matrix $N$ such that $\Sigma = N\,\mathrm{diag}(\epsilon_1 I_2,\epsilon_2 I_2)\,N^T$, the mode contribution matrices $B_k=N E_k N^T$ satisfy $\Sigma=\epsilon_1 B_1+\epsilon_2 B_2$. In our formalism, choosing mode coordinates $\vec a=W^{-1}\vec z$ with uncorrelated mode covariance $\Sigma_a=\mathrm{diag}(\epsilon_1 I_2,\epsilon_2 I_2)$ yields $\Sigma = \epsilon_1 W_1W_1^T+\epsilon_2 W_2W_2^T$, so $B_k$ correspond to $W_kW_k^T$ up to within-plane gauge choices. Thus Wolski’s “lattice functions” encode the same plane content and projected envelopes as our generator-based quantities, but packaged directly at the level of the covariance matrix.

\subsubsection{Relationship with Sagan--Rubin formalism}

Sagan and Rubin~\cite{sagan1999linear} construct a symplectic normal-mode frame along the lattice that decouples the linear map into two $2\times2$ normal-mode maps, with an explicit convention for mode labeling and with special handling of ``mode flips'' near strong coupling. In our language, their construction is a particular \emph{gauge fixing} of a symplectic basis spanning the invariant eigenmode planes. 

We construct the invariant planes by $W=[W_1\; W_2]$ bases that relate to eigenvectors of the one-turn map $\mathcal{M}$. Similarly, Sagan and Rubin produces another symplectic frame denoted by $V=[V_1\;V_2] \in Sp(4)$, with the defining property that the transfer map between two locations
becomes block diagonal in this basis,
\begin{equation}
V(s)^{-1}\,\mathcal M(s\leftarrow s_0)\,V(s_0)
=\begin{pmatrix}
E(s\leftarrow s_0) & 0\\[2pt]
0 & F(s\leftarrow s_0)
\end{pmatrix},
\qquad E,F\in Sp(2),
\label{eq:SR_block_diag}
\end{equation}
up to the discrete possibility of a \emph{mode flip} (swap of the two $2\times2$ blocks).
Equation \eqref{eq:SR_block_diag} implies that $\mathrm{span}(V_k(s))$ are invariant/transported mode planes: the image of each plane under $\mathcal M(s\leftarrow s_0)$ remains inside the corresponding plane at $s$.
Therefore, away from degeneracy and away from a deliberate mode swap, $V_k(s)$ spans the \emph{same} plane as $W_k(s)$:
\begin{equation}
\mathrm{span}(V_k(s))=\mathcal P_k(s)=\mathrm{span}(W_k(s)),\qquad k=1,2.
\label{eq:SR_same_planes}
\end{equation}
Because $V_k(s)$ and $W_k(s)$ are both full-rank $4\times2$ bases of the same plane $\mathcal P_k(s)$, there exists an invertible $2\times2$ matrix $G_k(s)\in Sp(2)$ such that
\begin{equation}
V_k(s)=W_k(s)\,G_k(s),\qquad k=1,2.
\label{eq:SR_Gk_GL2}
\end{equation}
Equivalently, the corresponding mode coordinates are related by
\begin{equation}
\vec a_k^{\,(\mathrm{SR})}(s)=V_k(s)^{+}\,\vec z(s)
=G_k(s)^{-1}\,W_k(s)^{+}\,\vec z(s)
=G_k(s)^{-1}\,\vec a_k^{\,(\mathrm{ours})}(s),
\label{eq:SR_mode_coords_relation}
\end{equation}
and the reduced maps transform by within-plane similarity,
\begin{equation}
E(s\leftarrow s_0)=G_1(s)^{-1}\,R_1(s\leftarrow s_0)\,G_1(s_0),\qquad
F(s\leftarrow s_0)=G_2(s)^{-1}\,R_2(s\leftarrow s_0)\,G_2(s_0),
\label{eq:SR_similarity_reduced}
\end{equation}
where $R_k(s\leftarrow s_0)=W_k(s_0)^{+}\mathcal M(s\leftarrow s_0)W_k(s_0)$ are our accumulated reduced maps.
In particular, the eigenvalues of $E$ and $F$ coincide with those of $R_1$ and $R_2$; hence the tunes/phase advances are identical in the two formalisms.

To summarize this subsection, it is evident that if a symplectically normalized basis spans the invariant eigenmode plane, then it is a valid choice of basis. The choice of symplectically normalized basis is not uniquely defined and forms a gauge freedom in $Sp(2)\times Sp(2)$ group. This gauge freedom is a coordinate re-paratmetrization that leaves the eigensystem unchanged, and representation dependent parameters--- e.g. \  Twiss functions and phase advance conventions---differ among different formalisms by a gauge transformation.

\subsection{Practical computation, periodic matching, and diagnostics}
\label{subsec:methodology}
This section summarizes the computational procedure used throughout the paper to (i) compute periodic coupled optics, (ii) enforce a continuous mode labeling (continuity gauge), and (iii) compute diagnostics that verify invariance, stability, and consistency with the one-turn tunes.

We assume a lattice is given as a sequence of linear elements with known $4\times4$ symplectic transfer matrices. Let $\mathcal{M}(s\leftarrow s_0)$ denote the accumulated transfer map from a reference location $s_0$ to $s$, and let the full one-turn map be $\mathcal{M}_\mathrm{cell}\equiv \mathcal{M}(s_0+L\leftarrow s_0)$, where $L$ is the cell length. All computations below are performed in canonical coordinates $\vec z=[x,p_x,y,p_y]^T$, so that $S_4=\mathrm{diag}(S_2,S_2)$.

\textbf{Step 1: Stable eigenmode planes at the reference location.}
We compute the eigensystem of $\mathcal{M}_\mathrm{cell}$ and select the two stable complex conjugate pairs $\lambda_k=e^{\pm i\mu_k}$ ($k=1,2$), which exist when motion is stable and non-degenerate. From each selected complex eigenvector $\vec v_k$ we form a real symplectic generator pair
\begin{equation}
\vec z_{2k-1}(s_0)=\Re\,\vec v_k,\qquad \vec z_{2k}(s_0)=-\Im\,\vec v_k,
\qquad W_k(s_0)=[\vec z_{2k-1}(s_0)\ \vec z_{2k}(s_0)].
\end{equation}
We then symplectically normalize the basis so that $W_k(s_0)^TS_4W_k(s_0)=S_2$ and impose the symplectic orthogonality of the two planes,
$W_1(s_0)^TS_4W_2(s_0)=0$. The resulting $W(s_0)=[W_1(s_0)\ W_2(s_0)]$ defines a symplectic frame adapted to the eigenmode planes.

\textbf{Step 2: Transport of eigenmode bases along the lattice.}
Given $W_k(s_0)$, a natural transported basis at location $s$ is obtained by pushing forward the generators,
\begin{equation}
\widetilde W_k(s)\equiv \mathcal{M}(s\leftarrow s_0)\,W_k(s_0).
\end{equation}
In exact arithmetic $\widetilde W_k(s)$ spans the correct transported plane, but due to numerical roundoff and the fact that the within-plane gauge is not fixed, $\widetilde W_k(s)$ may drift in normalization. We therefore re-normalize each $\widetilde W_k(s)$ to enforce
$W_k(s)^TS_4W_k(s)=S_2$ and (if needed) re-enforce $W_1^TS_4W_2=0$ by symplectic Gram–Schmidt. This yields a well-defined symplectically normalized basis $W_k(s)$ spanning the propagated plane.

\textbf{Step 3: Continuity gauge and mode labeling.}
The eigenmode planes are unique in the simple-spectrum case, but the basis inside each plane is not. To obtain smooth $s$-dependent optics functions and to prevent spurious mode swaps in near-degenerate regions, we fix the within-plane gauge by a continuity criterion. Concretely, at each step we choose a $2\times2$ rotation $R_k(s)\in SO(2)$ that best aligns the new basis to the previous one in the Euclidean sense,
\begin{equation}
R(\theta_k^\star)(s) \in \arg\min_{R\in SO(2)} \|W_k(s-\Delta s)-W_k(s)R\|_F,
\qquad
W_k(s)\leftarrow W_k(s)R(\theta_k^\star)(s).
\end{equation}
and then update $W_k(s)\leftarrow W_k(s)R_k(s)$.
This is an orthogonal Procrustes alignment performed independently in each mode plane and serves as a practical gauge-fixing that produces continuous generators, continuous projected Twiss functions, and a consistent labeling of mode~1 versus mode~2.

\textbf{Step 4: Coupled Twiss functions and coupling diagnostics.}
At each location $s$, the projected Twiss-like functions for mode $k$ are computed from the transported generators via
\begin{equation}
\beta_{kx}(s)=\|E_xW_k(s)\|_F^2,\qquad
\beta_{ky}(s)=\|E_yW_k(s)\|_F^2,
\end{equation}
with $E_x=\mathrm{diag}(1,0,0,0)$ and $E_y=\mathrm{diag}(0,0,1,0)$.
The corresponding $\alpha$ and $\gamma$ functions are obtained from the generator components,
$\alpha_{kx}=-(x_{2k-1}p_{x,{2k-1}}+x_{2k}p_{x,2k})$ and $\gamma_{kx}=p_{x,2k-1}^2+p_{x,2k}^2$ (and similarly for $y$).
To quantify coupling in a basis-independent way we form the Euclidean projector onto the plane,
$\Pi_k=W_k(W_k^TW_k)^{-1}W_k^T$, and compute the invariant overlap fractions
\begin{equation}
u_{k,\mathrm{inv}}(s)=\tfrac12\,\mathrm{tr}(P_Y\Pi_k),\qquad
1-u_{k,\mathrm{inv}}(s)=\tfrac12\,\mathrm{tr}(P_X\Pi_k),
\end{equation}
which satisfy $u_{k,\mathrm{inv}}\in[0,1]$ by construction and remain invariant under within-plane gauge transformations.

A complementary diagnostic is the \emph{leakage} between the two transported mode planes. In a perfectly invariant decomposition the planes are symplectically orthogonal and do not mix; numerically we monitor
\begin{equation}
\ell_{12}(s)\equiv \|W_1(s)^+W_2(s)\|_F,\qquad
\ell_{21}(s)\equiv \|W_2(s)^+W_1(s)\|_F,
\end{equation}
where $W_k^+=-S_2W_k^TS_4$.
Small leakage throughout the lattice provides a stringent check that the computed planes are consistent and that numerical re-normalization has not introduced artificial mixing.

\textbf{Coupling phase extraction.}
In addition to the gauge-invariant coupling fractions, it is sometimes useful to track a relative
phase between the $(x,p_x)$ and $(y,p_y)$ projections of a mode. For each stable eigenmode we work with a complex eigenvector $\vec v_k(s)$ spanning $\mathcal P_k(s)$ (equivalently, a complex representation of the transported basis $W_k(s)$). We split
\begin{equation}
\vec v_k(s)=\begin{pmatrix} v_{kX}(s) \\ v_{kY}(s) \end{pmatrix},\qquad
v_{kX}(s),v_{kY}(s)\in\mathbb C^2,
\end{equation}
where $v_{kX}$ collects the $(x,p_x)$ components and $v_{kY}$ the $(y,p_y)$ components.
Because $\vec v_k$ is defined only up to an overall phase $\vec v_k\mapsto e^{i\theta_k}\vec v_k$,
the coupling phase must be invariant under this gauge.
A convenient choice is
\begin{equation}
\nu_k(s)\equiv \arg\!\left(v_{kY}(s)^\dagger\,v_{kX}(s)\right),
\label{eq:nu_def_innerprod}
\end{equation}
or, equivalently, using the normalized inner product
$\arg\!\left(v_{kY}^\dagger v_{kX}/(\|v_{kY}\|\|v_{kX}\|)\right)$.
This definition is unchanged by the overall eigenvector phase gauge and reduces to the familiar
Lebedev--Bogacz form in which the $(y,p_y)$ block carries an explicit factor $e^{i\nu_k}$ relative
to $(x,p_x)$.

In practice $\nu_k(s)$ is ill-conditioned when either projection is small (near the uncoupled limit),
and it can become noisy in nearly-degenerate regions where eigenvectors are not uniquely determined.
To avoid spurious jumps we compute $\nu_k(s)$ only when both $\|v_{kX}(s)\|$ and $\|v_{kY}(s)\|$
exceed a chosen tolerance; otherwise we either omit $\nu_k$ or hold it by continuity.
When a continuous $s$-dependent coupling phase is desired, we additionally apply a phase unwrapping
procedure so that $\nu_k(s)$ varies smoothly along the lattice.

\textbf{Step 5: Tunes and phase advance consistency checks.}
The eigenmode tunes are obtained directly from the eigenvalues of the full one-turn map, $\mathcal{M}_\mathrm{cell}$, as $Q_k=\mu_k/2\pi$. To cross-check that the reduced-plane representation is consistent, we compute the reduced one-turn map
\begin{equation}
R_k^\mathrm{cell}\equiv W_k(s_0)^+\,\mathcal{M}_\mathrm{cell}\,W_k(s_0)\in Sp(2),
\end{equation}
whose eigenvalues must match $e^{\pm i\mu_k}$. Agreement between these two tune extractions validates both the plane construction and the symplectic normalization of $W_k(s_0)$.

For an $s$-dependent accumulated phase advance we use the accumulated reduced map
\begin{equation}
R_k(s\leftarrow s_0)\equiv W_k(s_0)^+\,\mathcal{M}(s\leftarrow s_0)\,W_k(s_0),
\end{equation}
and define $\mu_k(s)$ from the argument of its stable eigenvalues. This definition is gauge invariant (it depends only on $W_k(s_0)$ up to similarity) and reproduces the one-turn phase advance when $s=s_0+L$.

\textbf{Step 6: Periodic matching.}
Periodic optics correspond to finding a symplectic frame $W(s_0)$ such that the transported basis after one cell returns to itself up to a within-plane symplectic rotation,
\begin{equation}
\mathcal{M}_\mathrm{cell}W_k(s_0)=W_k(s_0)\,R_k^\mathrm{cell},\qquad k=1,2,
\end{equation}
with $R_k^\mathrm{cell}\in Sp(2)$ stable. In practice, this periodicity is achieved by selecting $W_k(s_0)$ from the stable eigenvectors of $\mathcal{M}_\mathrm{cell}$ and enforcing the symplectic normalization and orthogonality constraints described above. The resulting transported $W_k(s)$ then generates periodic projected Twiss functions $\beta_{kx}(s+L)=\beta_{kx}(s)$ and $\beta_{ky}(s+L)=\beta_{ky}(s)$ by construction.

\textbf{Step 7: Projected phase--plane ellipses (visual diagnostic)}
In addition to scalar coupling measures and leakage norms, we visualize the coupled motion by
projecting the $4\times4$ second-moment matrix onto selected $2$D coordinate planes.
Let the canonical phase-space vector be ordered as $\vec z=[x,p_x,y,p_y]^T$ (equivalently $[x,x',y,y']^T$).
Given the physical covariance matrix
\begin{equation}
\Sigma_z(s)\equiv \langle \vec z(s)\vec z(s)^T\rangle,
\label{eq:Sigma_z_def}
\end{equation}
we form the $2\times2$ covariance associated with a coordinate pair $(i,j)$ by a linear selection map.
Define the selector
\begin{equation}
S_{ij}\equiv
\begin{pmatrix}
\vec e_i^T\\
\vec e_j^T
\end{pmatrix},
\qquad
\Sigma_{ij}(s)\equiv S_{ij}\,\Sigma_z(s)\,S_{ij}^T,
\label{eq:Sigma_ij_def}
\end{equation}
where $\{\vec e_1,\vec e_2,\vec e_3,\vec e_4\}$ is the standard basis of $\mathbb R^4$ so that
$i,j\in\{1,2,3,4\}$ correspond to $x,x',y,y'$ respectively.

The projected $n\sigma$ ellipse in the $(z_i,z_j)$-plane is then defined by
\begin{equation}
\mathcal E_{ij}(s;n)\equiv
\left\{\,
\vec u\in\mathbb R^2:\;
\vec u^T\,\Sigma_{ij}(s)^{-1}\,\vec u = n^2
\,\right\},
\label{eq:ellipse_def}
\end{equation}
and can be parameterized by an eigen-decomposition of $\Sigma_{ij}(s)$.
If $\Sigma_{ij}(s)=R\,\mathrm{diag}(\lambda_1,\lambda_2)\,R^T$ with $R\in SO(2)$ and $\lambda_{1,2}>0$, then
\begin{equation}
\vec u(\theta)= n\,R
\begin{pmatrix}
\sqrt{\lambda_1}\cos\theta\\
\sqrt{\lambda_2}\sin\theta
\end{pmatrix},
\qquad \theta\in[0,2\pi),
\label{eq:ellipse_param}
\end{equation}
yields the ellipse boundary.

When the two eigenmodes are uncorrelated in the chosen gauge, a convenient model for the physical covariance is
\begin{equation}
\Sigma_z(s)=\varepsilon_1\,W_1(s)W_1(s)^T+\varepsilon_2\,W_2(s)W_2(s)^T,
\label{eq:Sigma_from_modes}
\end{equation}
where $\varepsilon_k=\langle 2J_k\rangle$ are the eigen-emittances and $W_k(s)$ are the transported
symplectic generator pairs spanning the eigenmode planes.
This decomposition also enables \emph{mode footprints} in physical coordinates:
\begin{equation}
\Sigma^{(k)}_z(s)\equiv \varepsilon_k\,W_k(s)W_k(s)^T,
\qquad
\Sigma^{(k)}_{ij}(s)=S_{ij}\,\Sigma^{(k)}_z(s)\,S_{ij}^T,
\label{eq:mode_footprints}
\end{equation}
so that one may plot both the total projected ellipse \eqref{eq:ellipse_def} and the individual mode contributions on the same physical plane.

In this work we evaluate these projections at a set of lattice locations $s$ and plot the ellipses in the
planes $(x,x')$, $(y,y')$, $(x,y)$, $(x,y')$, $(y,x')$, and $(x',y')$ as a compact visual summary of
decoupling quality, mode content, and coupling evolution along the lattice.

\section{Applications and Comparison to Other Formalisms}\label{sec:Results}

In this section, we follow the steps described in Section~\ref{subsec:methodology} for computing the periodic solutions for coupled lattices. 

\subsection{Comparison of formalism with Lebedev--Bogacz and Mais--Ripken Parametrizations}\label{subsec:comparisonresults}

We compute the coupled optics functions with our formalism and compare it with Lebedev--Bogacz from OptiMX~\cite{OptiMX} optics code and Mais--Ripken formalism with MADX~\cite{MADXManual} optics code. The comparison here also gives tremendous insight to the gauge freedom that exists in the formalism.

\subsubsection{Derbenev's Adapter}

\begin{figure}[tbp]
    \centering
    \includegraphics[width=0.49\linewidth]{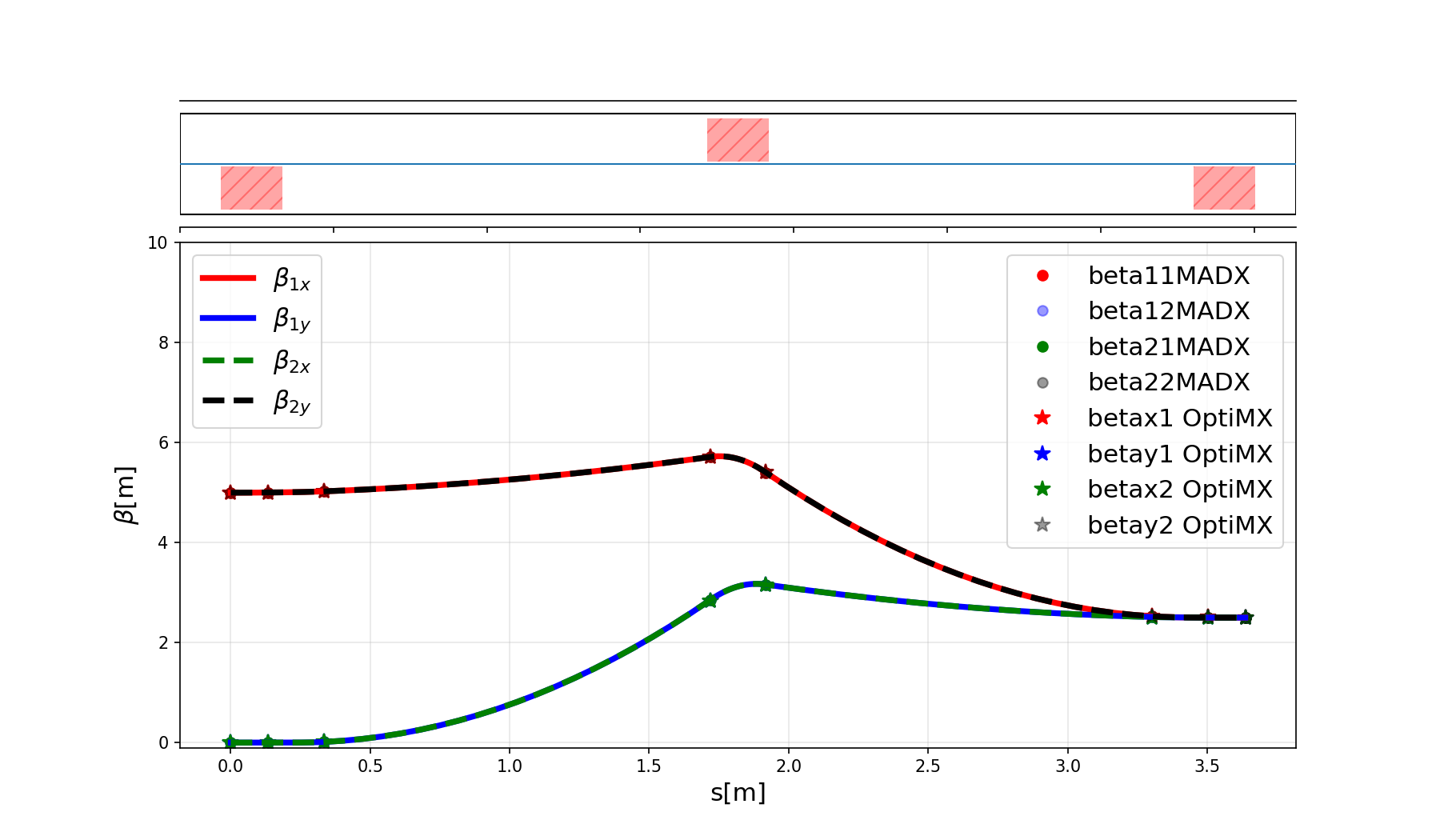}
    \includegraphics[width=0.49\linewidth]{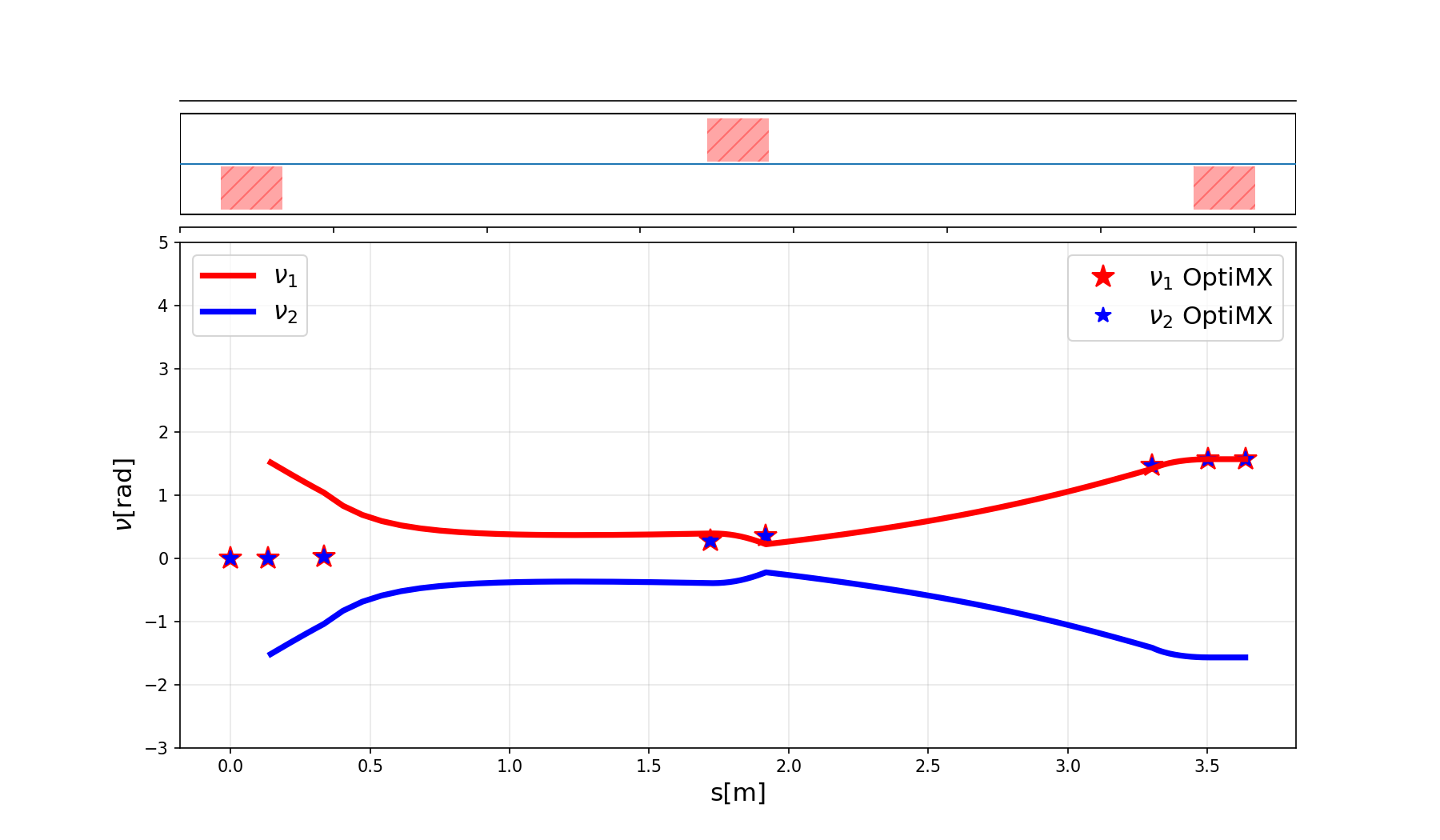}
    \includegraphics[width=0.49\linewidth]{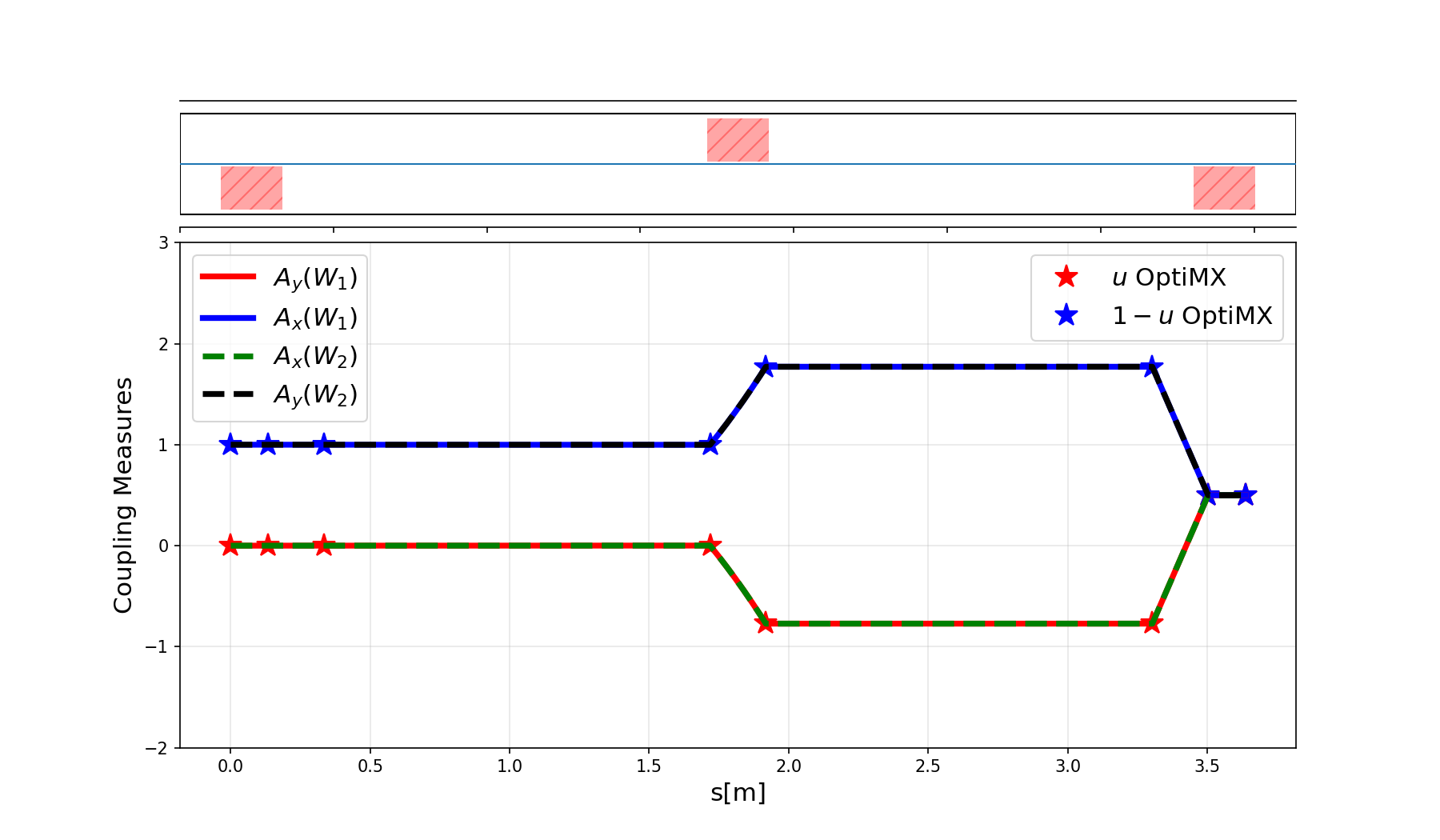}
    \includegraphics[width=0.49\linewidth]{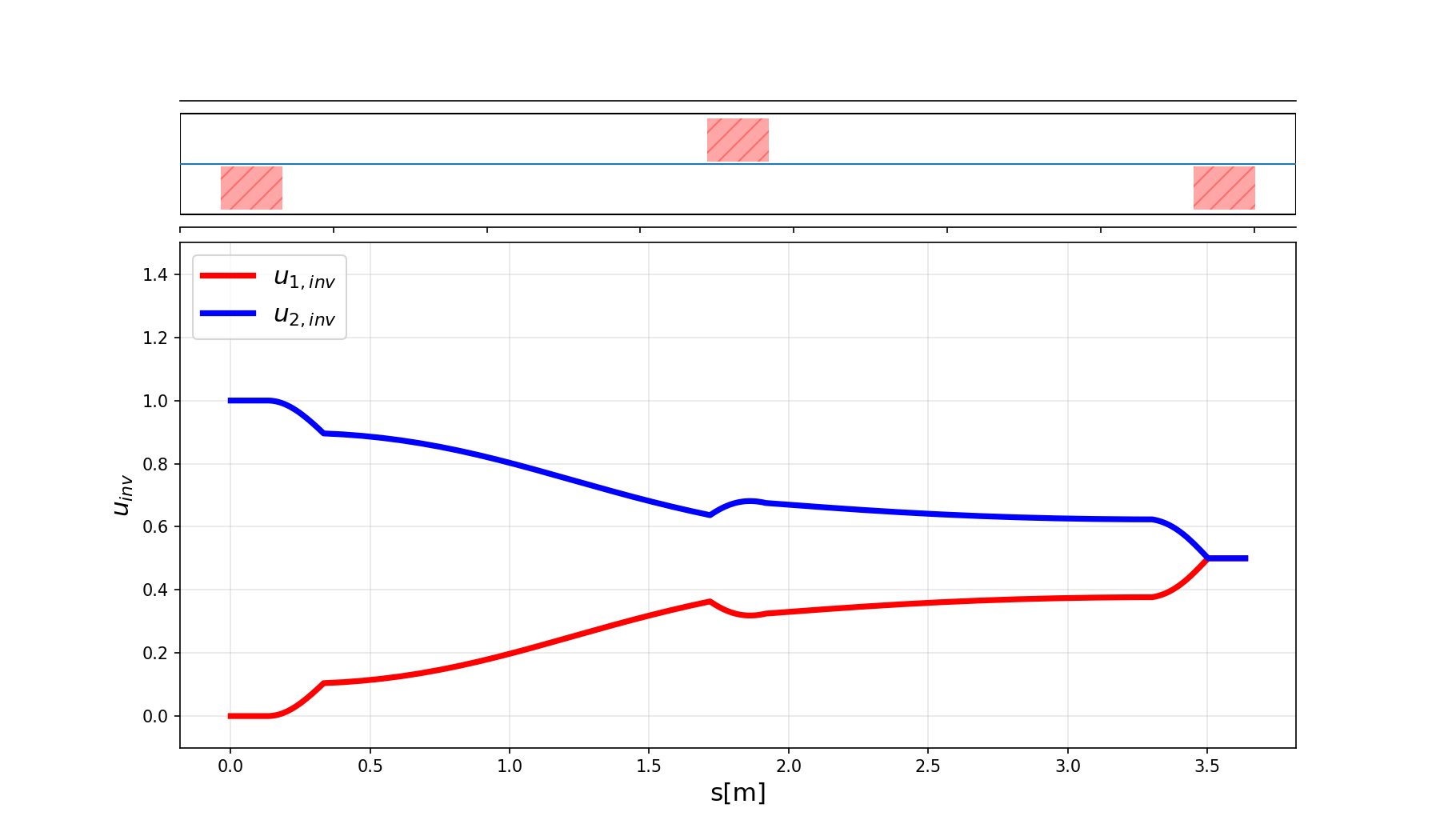}
    \caption{Derbenev's Adapter optics computation for creation of circular mode and comparison with different optics codes. Top left plot shows the coupled $\beta$ functions computed from MADX PTC, OptiMX, and our formalism. Top right plot shows the coupling phases $\nu_{1,2}$ computed from our formalism and OptiMX. Bottom left plot shows the signed symplectic area measures $A_{x,y}(W_{1,2})$ and Lebedev-Bogacz coupling strength parameter $u$. Bottom right plot shows the gauge invariant coupling measures for both eigenmode planes, $u_{1,2,inv}$.}
    \label{fig:derbenevadapteroptics}
\end{figure}

First, we will test matching and propagation of optics through Derbenev's Adapter~\cite{derbenev1998adapting}, which consists of three skew quadrupoles. Derbenev's adapter is also known as flat-to-round beam converter that can be used for creating a circular-mode beams~\cite{burov2002circular,gilanliogullaricircular}. Initially uncoupled optics state, with $(\beta_x,\beta_y=5.0)$\,m and $(\alpha_x,\alpha_y=0)$, is propagated through the Derbenev's Adapter. The coupling is created via Derbenev's Adapter and all the coupling $\beta$ functions have equal values at the end, $\beta_{il}=2.5$\,m for $i\in\{1,2\}$ and $l\in\{x,y\}$. Figure.~\ref{fig:derbenevadapteroptics} shows the propagation of optics functions computed from formalism discussed in this manuscript and compared with different optics codes such as MADX and OptiMX. The $\beta$ function plots, top left plot in Fig.~\ref{fig:derbenevadapteroptics}, shows great agreement for coupling creation. The coupling phases, $\nu_{1,2}$ the top right plot, is compared with our formalism and Lebedev-Bogacz formalism. The signed symplectic area measures $A_{x,y}(W_{1,2})$ and the Lebedev-Bogacz coupling strength is given in the bottom left plot. The invariant coupling measures are given in the bottom right plot of Fig.~\ref{fig:derbenevadapteroptics}.

\begin{figure}[tbp]
    \centering
    \includegraphics[width=\linewidth]{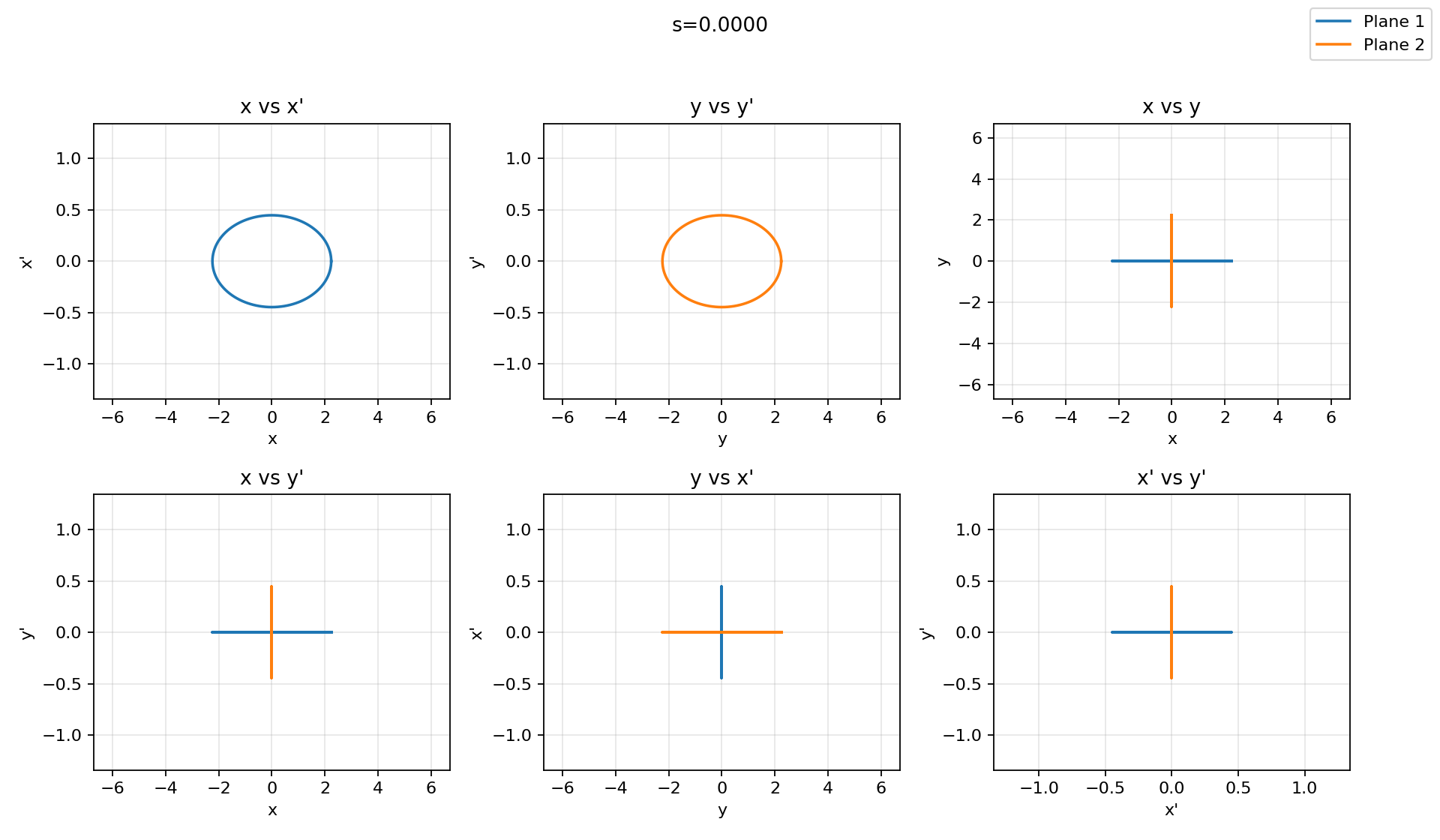}
    \includegraphics[width=\linewidth]{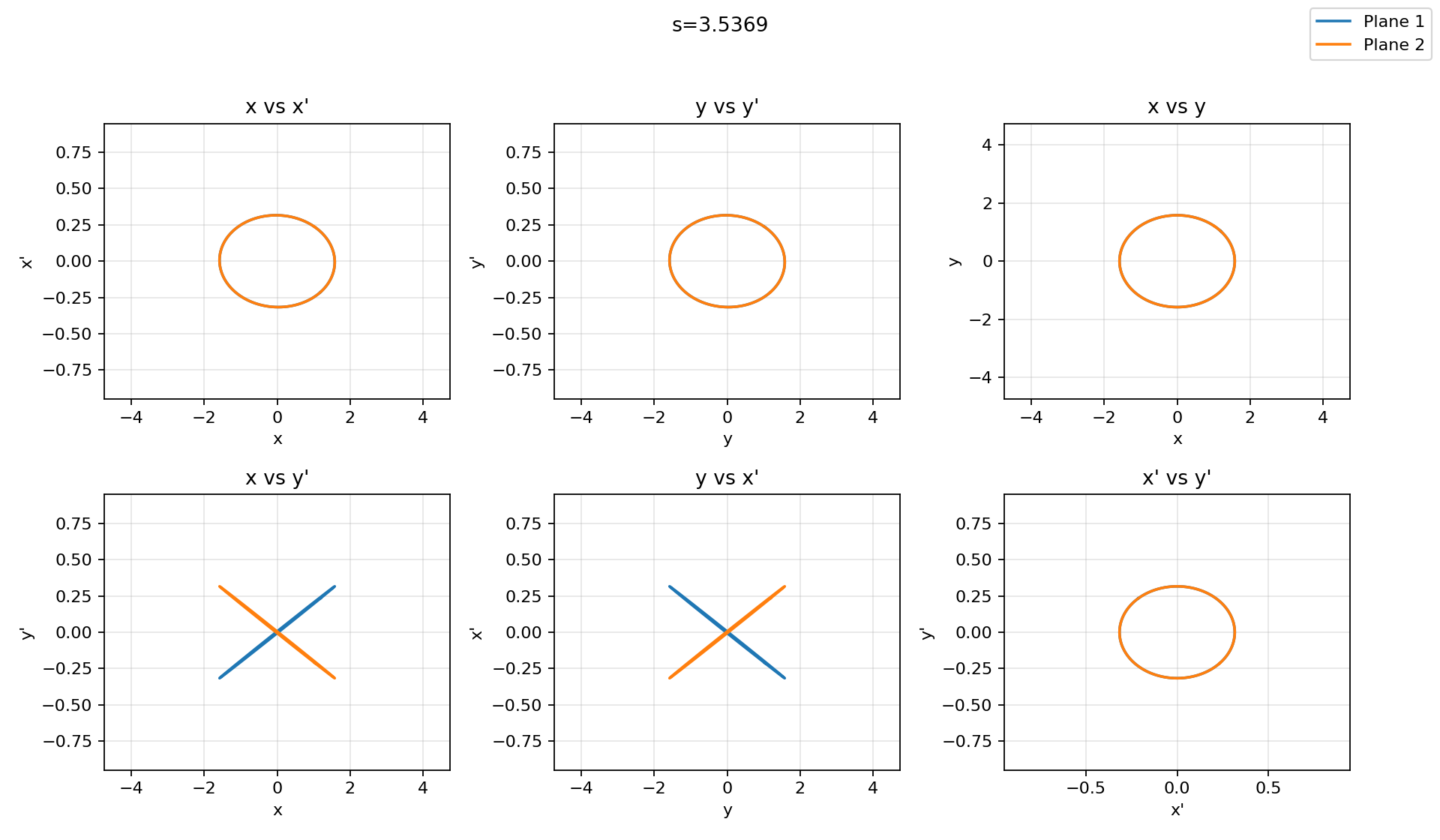}
    \caption{Ellipse Projections from the invariant planes $\mathcal{P}_1$ and $\mathcal{P}_2$ onto phase spaces for Derbenev's Adapter. Top plot is at initial location $s=0.0$ and bottom plot is at final location.}
    \label{fig:ellipseprojectionsDerbenevprop}
\end{figure}

The computation of the $\beta$ function agrees among all the formalisms, namely Mais-Ripken from the MADX-PTC module, and Lebedev-Bogacz functions from the OptiMX optics code. The first discrepancy is observed in the coupling phases computation, $\nu_{1,2}$. The coupling phase computation initially disagrees with each other, this is because we do not set a numerical value for coupling phase when the state of the optics is uncoupled. At the end of the cell, the values agree with each other at a value of $\nu_{1,2,\mathrm{LB}}=\tfrac{\pi}{2}$ and $\nu_1=\tfrac{\pi}{2}$. In this formalism, $\nu_2=-\tfrac{\pi}{2}$ because the same definition of $\nu$ is used for both planes (Eq.~\eqref{eq:nu_def_innerprod}, which yields a minus sign for the second mode. Predictably, the signed symplectic areas and the Lebedev-Bogacz coupling strength value agree with each other as shown in the bottom left plot of Fig.~\ref{fig:derbenevadapteroptics}, $A_y(W_1)=A_x(W_2)=u$. However, the signed symplectic areas are gauge dependent values and as we can see the values can go beyond $1.0$ and beloew $0.0$ as long as the sum is maintained at unity, $u+(1-u)=1$ and $A_x(W_{1,2}) + A_{y}(W_{1,2})=1$. There is no mathematical or physical reason for the need for individual area measures to be confined between $0$ and $1$, in addition to their sum. The plot on the bottom right shows the gauge-invariant coupling measures $u_{k,inv}$, computed from Eq.~\eqref{eq:invariantcouplingdef} and are always confined between $[0,1]$ and are independent of the choice of coordinates. The ellipse projections from invariant planes $\mathcal{P}_1$ and $\mathcal{P}_2$ is shown in Fig.~\ref{fig:ellipseprojectionsDerbenevprop} at intial location and final location. The initial location shows axes aligned with the $x$ and $y$ planes indicating no coupling, whereas the final location shows the creation of circular modes with circular projections in addition to the $(x,y')$ and $(y,x')$ planes. The coupling creation with Derbenev's Adapter is used for creating circular-mode beams, where the optics are not extracted from the eigenmode planes of the transfer map.

\begin{figure}[tbp]
    \centering
    \includegraphics[width=0.49\linewidth]{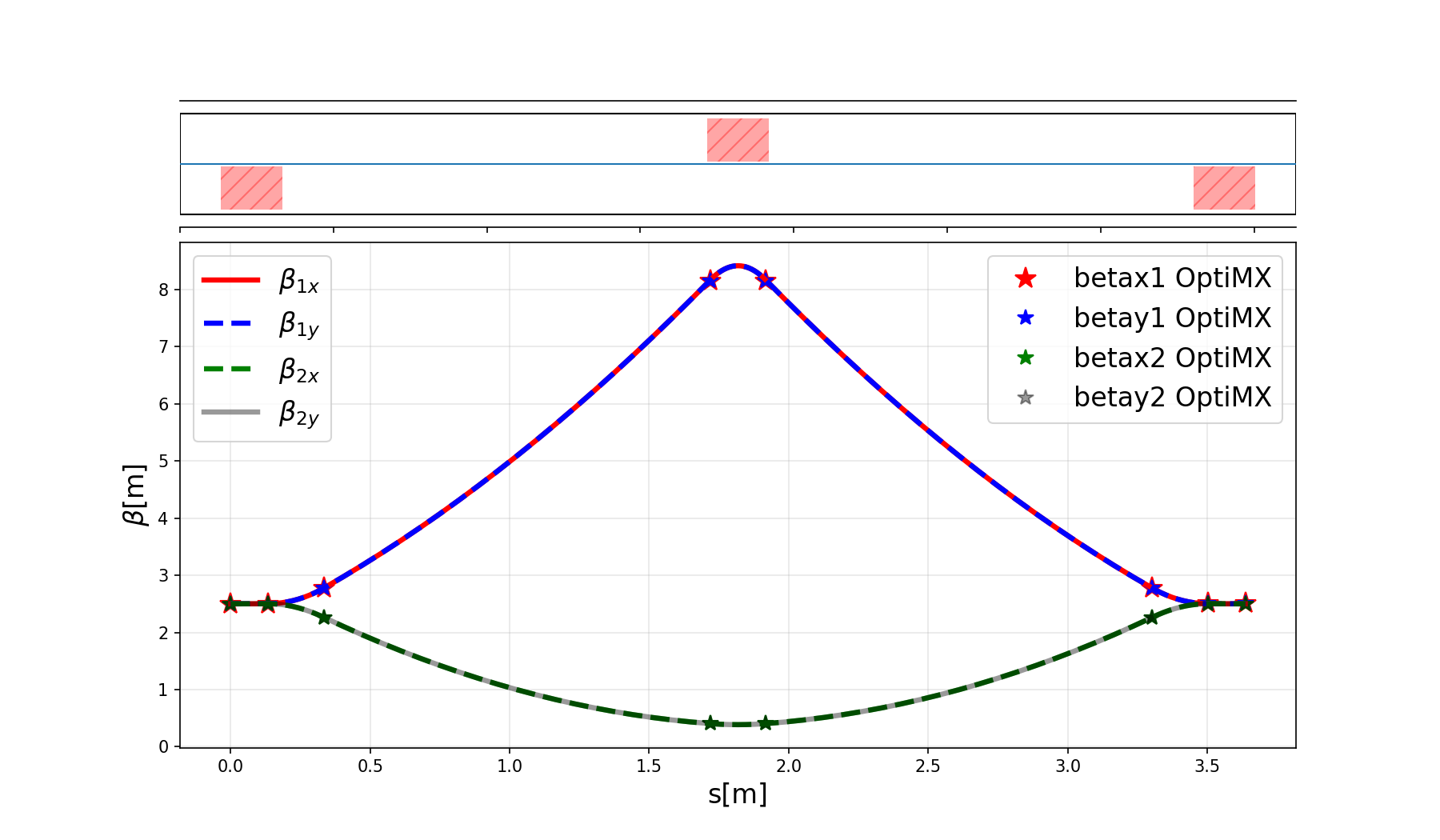}
    \includegraphics[width=0.49\linewidth]{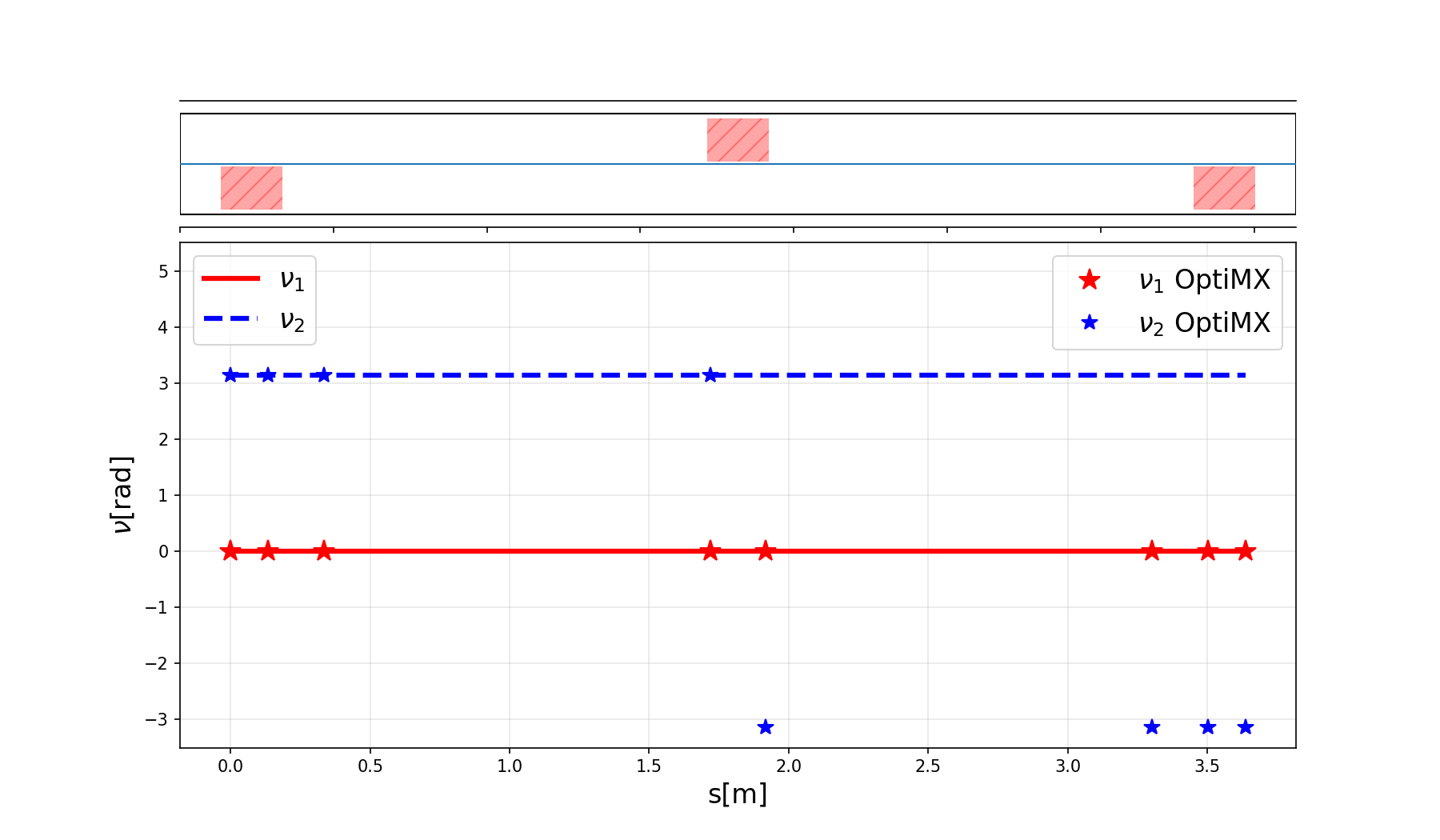}
    \includegraphics[width=0.49\linewidth]{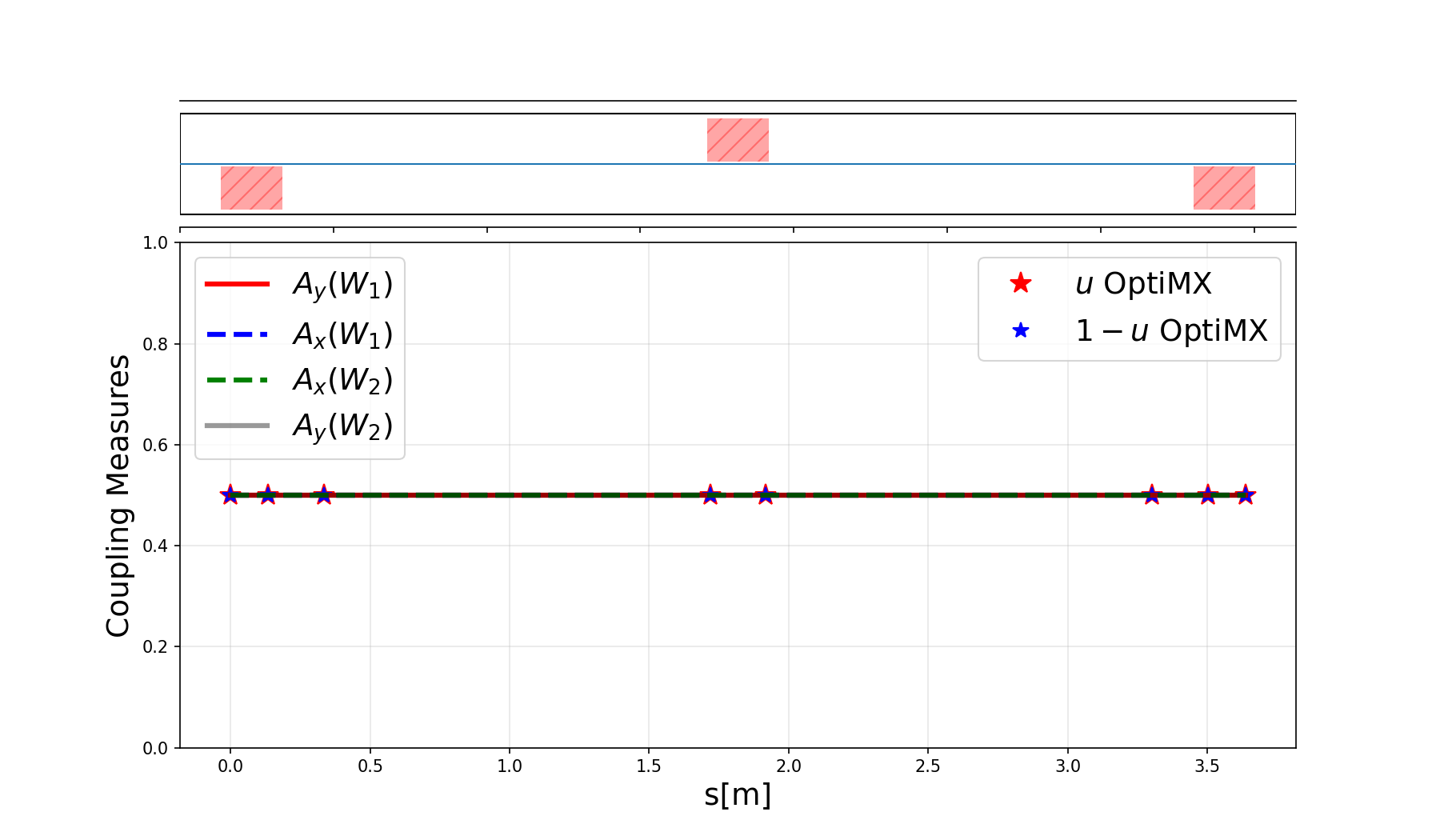}
    \includegraphics[width=0.49\linewidth]{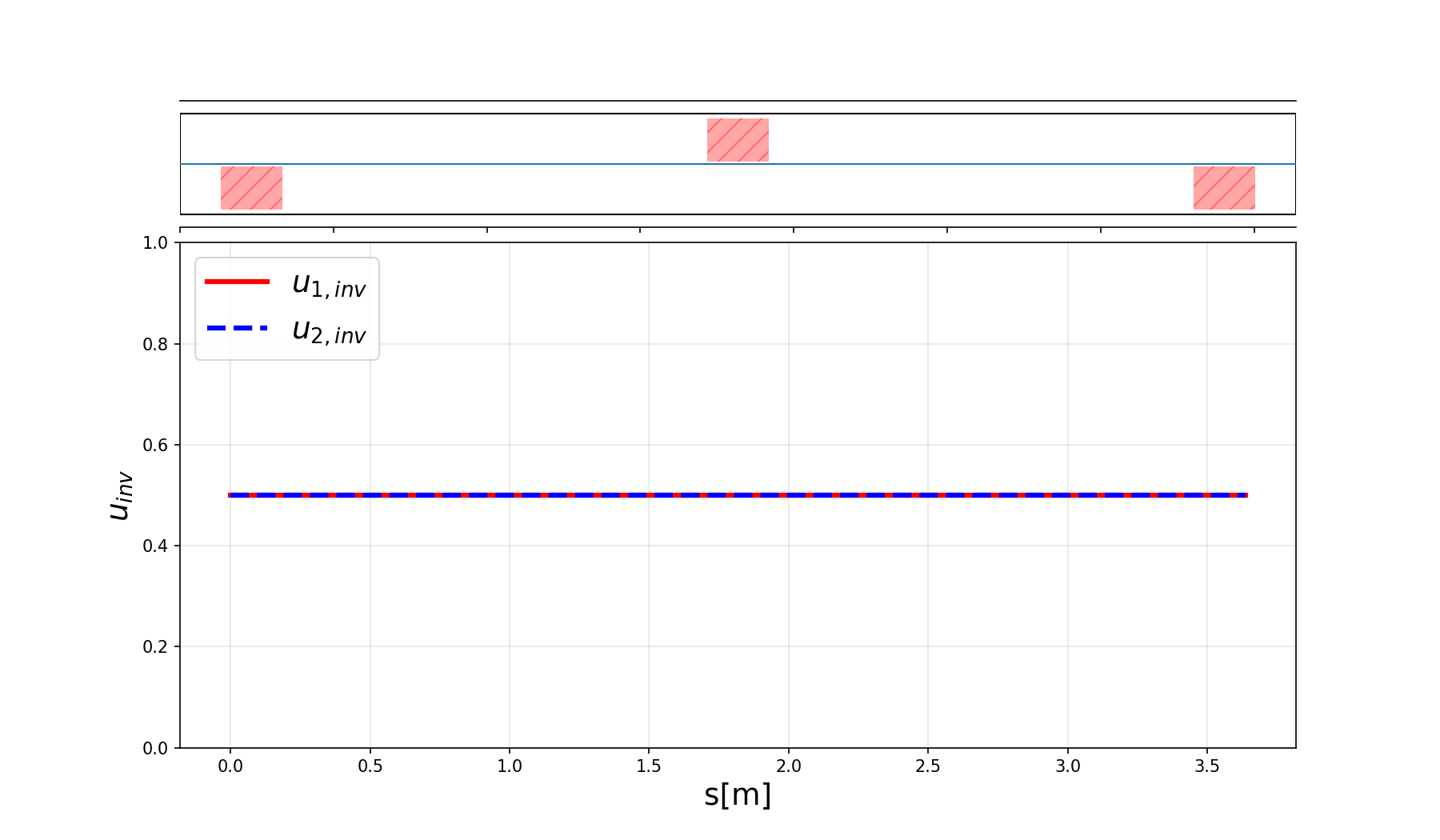}
    \caption{Skew quadrupole triplet eigenmode optics computation and comparison with Lebedev-Bogacz formalism. Top left plot shows the coupled $\beta$ functions, top right plot shows the coupling phases, bottom left shows coupling measures $A_{x,y}(W_{1,2})$ and $u,1-u$ parameters, bottom right plot shows the invariant coupling fractions $u_{1,2,\mathrm{inv}}$. }
    \label{fig:threeskeweigenmodecomp}
\end{figure}

In addition to creation of circular modes, we can compute the eigenmodes of the Derbenev's Adapter that results in periodic motion. In this case it will be skew triplet rather than Derbenev's adapter, since Derbenev's adapter is used for converting a flat beam to a round circular-mode beam. The optics is given in Fig.~\ref{fig:threeskeweigenmodecomp}, where every optics functions computed agree with Lebedev-Bogacz except the coupling phase of $\nu_2$ which jumps a branch at the middle skew quadrupole. Our formalism uses continuity gauge and the value stays the same. The coupling strength value from Lebedev-Bogacz, $u$, is identical to invariant measures and the signed symplectic areas, which shows the gauge choice lines with the invariant eigenmode planes $\mathcal{P}_1$ and $\mathcal{P}_2$.

\subsubsection{Solenoid and Skew Doublets}\label{subsubsec:solandskews}
\begin{figure}[tbp]
    \centering
    \includegraphics[width=0.49\linewidth]{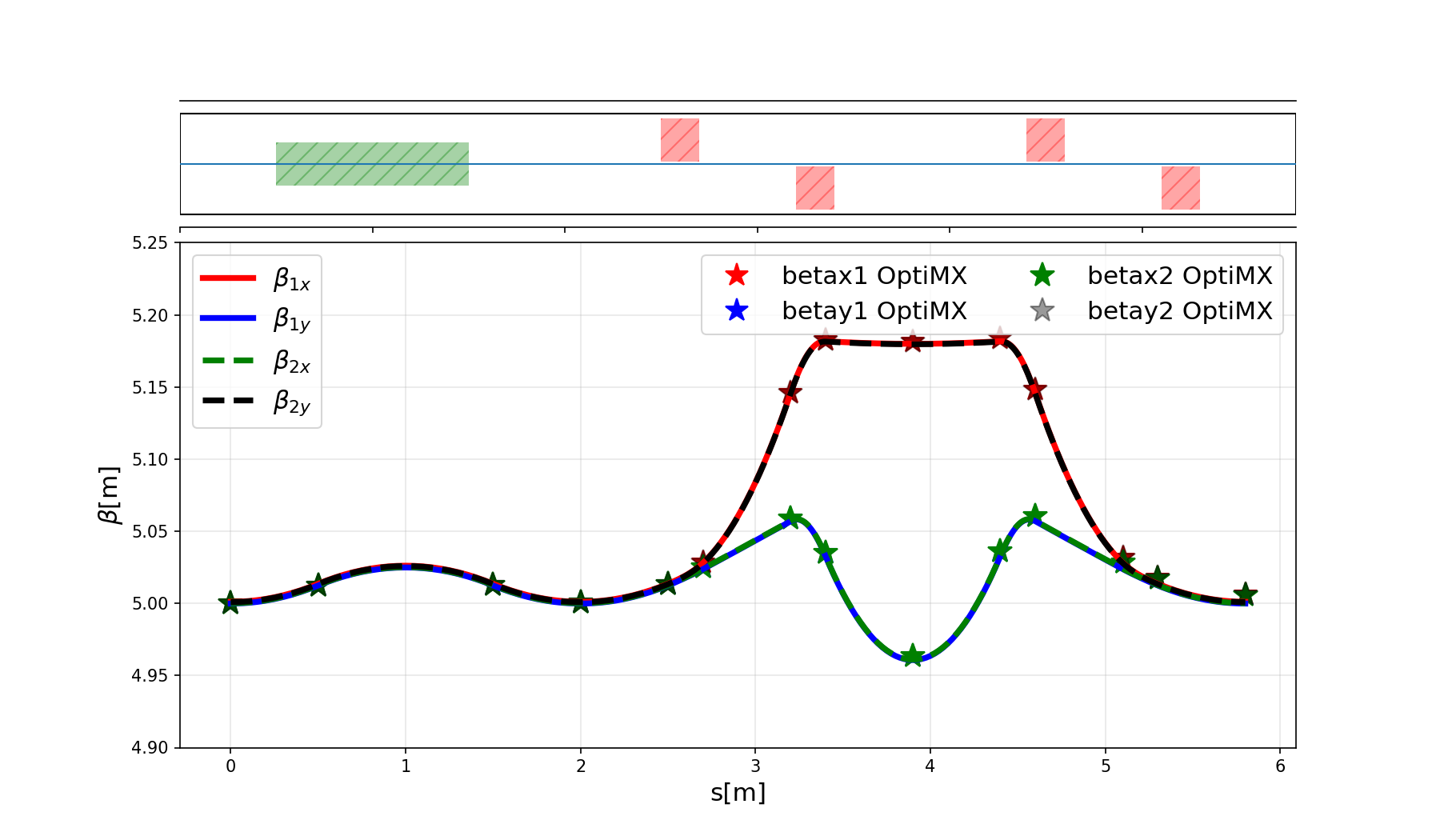}
    \includegraphics[width=0.49\linewidth]{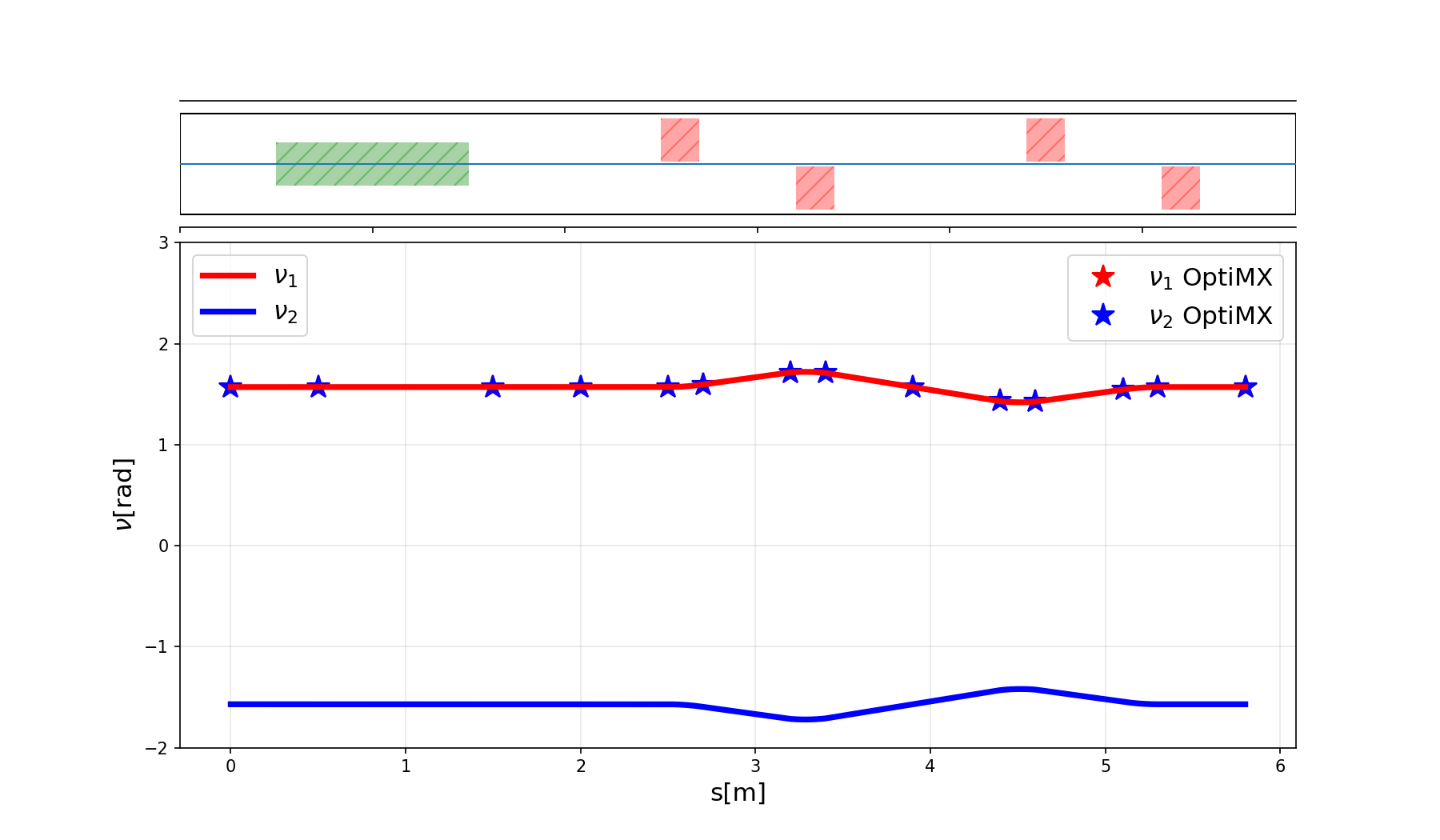}
    \includegraphics[width=0.49\linewidth]{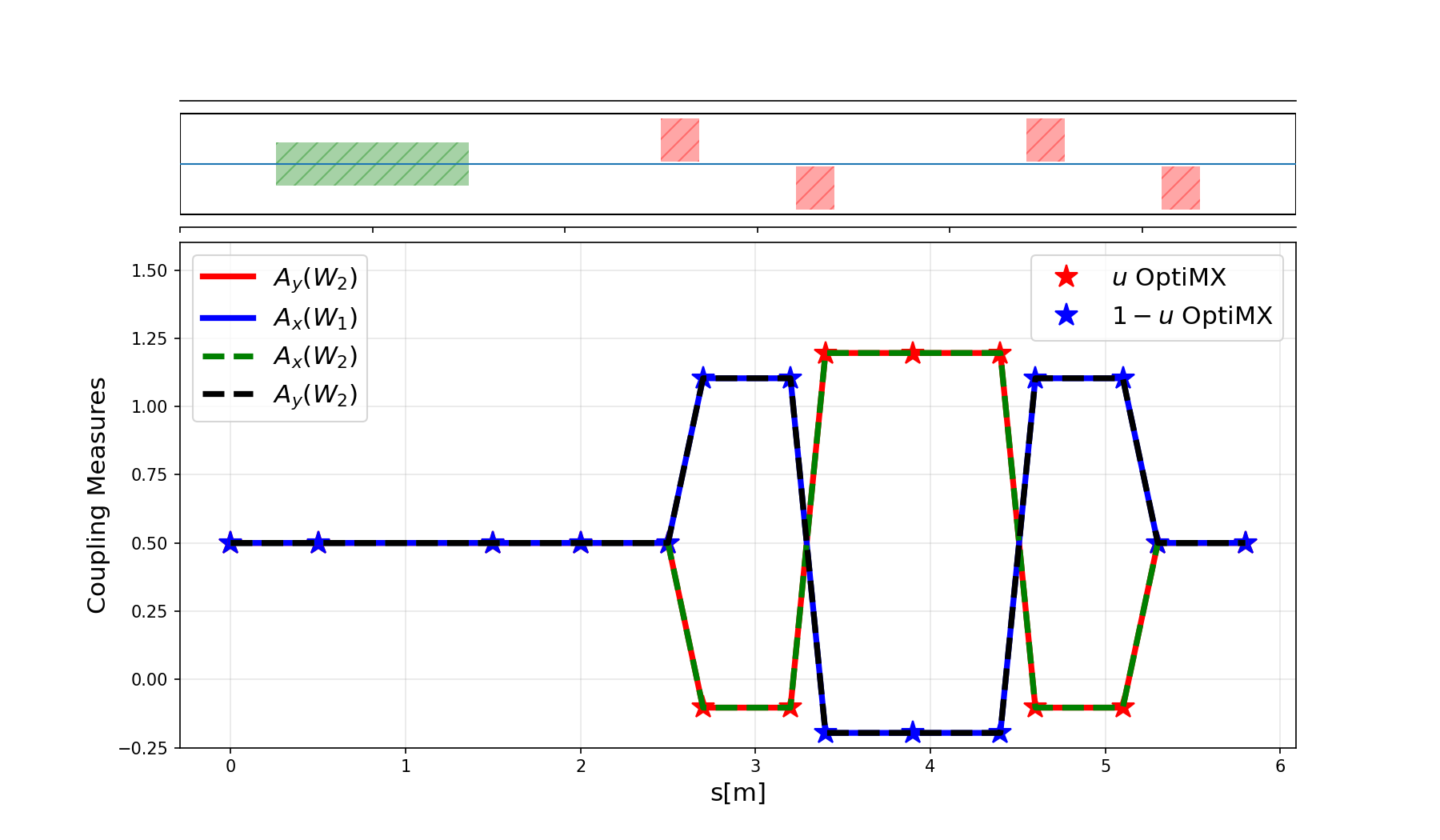}
    \includegraphics[width=0.49\linewidth]{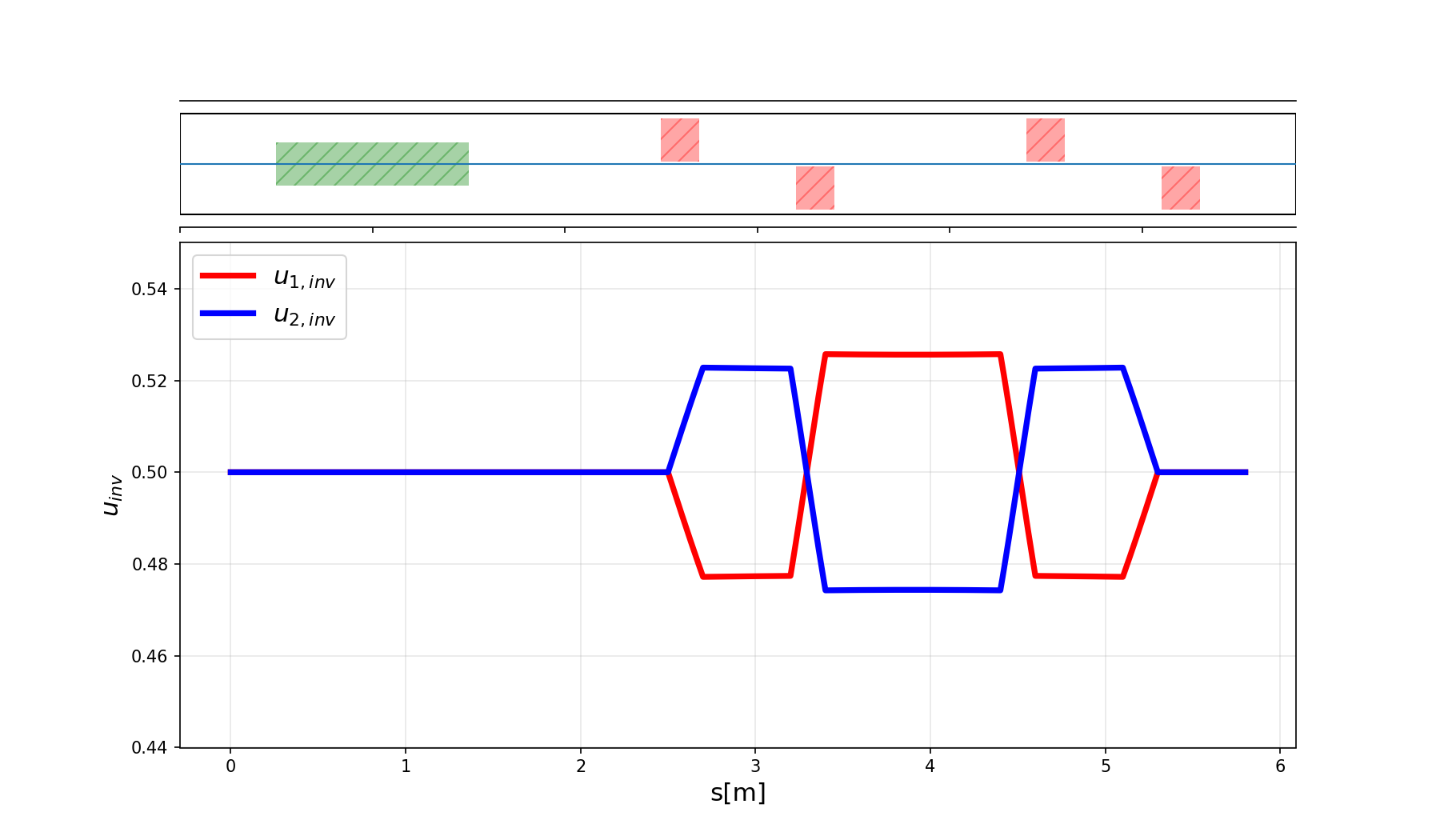}
    \caption{Solenoid and skew doublet cell. Top left plot shows the coupled $\beta$ functions, top right plot shows the coupling phases $\nu_{1,2}$, bottom left shows coupling emasures $A_{x,y}(W_{1,2})$ and $u$ from Lebedev--Bogacz parametrization, bottom right plot shows the gauge invariant coupling fractions $u_{1,2,\mathrm{inv}}$.}
    \label{fig:solandskewdoublet}
\end{figure}

For our second comparison case, we use a combination of a solenoid and skew-quadrupole doublet cell. The cell is matched periodically and the results are given in Fig.~\ref{fig:solandskewdoublet}. The solenoid is depicted in green color where the skew-quadrupoles are depicted in red. As we can see from the Fig.~\ref{fig:solandskewdoublet}, the Lebedev--Bogacz formalism and the formalism depicted here agrees with each other in terms of $\beta$ functions, coupling phases $\nu_{1,2}$ and the coupling measures $A_{x,y}(W_{1,2})$ and Lebedev--Bogacz coupling strength parameter $u$. We see that the scalar function $u$ starts from a value of 0.5 and as the coupling content of the basis changes it changes to values greater than one and smaller than zero. In the paper by Lebedev--Bogacz~\cite{lebedev2010betatron}, it is stated that the coupling strength parameter $u$ is a scalar function lower than equal to one, which implies when $u$ goes above one the other choice has to be selected as $u\to 1-u$ to maintain value of $u$ less than one. This bookkeeping creates mode swaps and discontuinities in the dynamics. On the other hand, our gauge-invariant defined coupling measures are always bounded between $[0,1]$ and do not depend on the choice of the basis. Because of this, the measure of coupling content should be separated from the choice of basis and gauge-invariant quantities must be reported separately.

\subsection{Application of Formalism to Coupled Rings}\label{subsec:appofformalismtoRings}

\subsubsection{Skew Quadrupole Triplet Ring}\label{subsubsec:Skewtripletring}

\begin{figure}[tbp]
    \centering
    \includegraphics[width=0.49\linewidth]{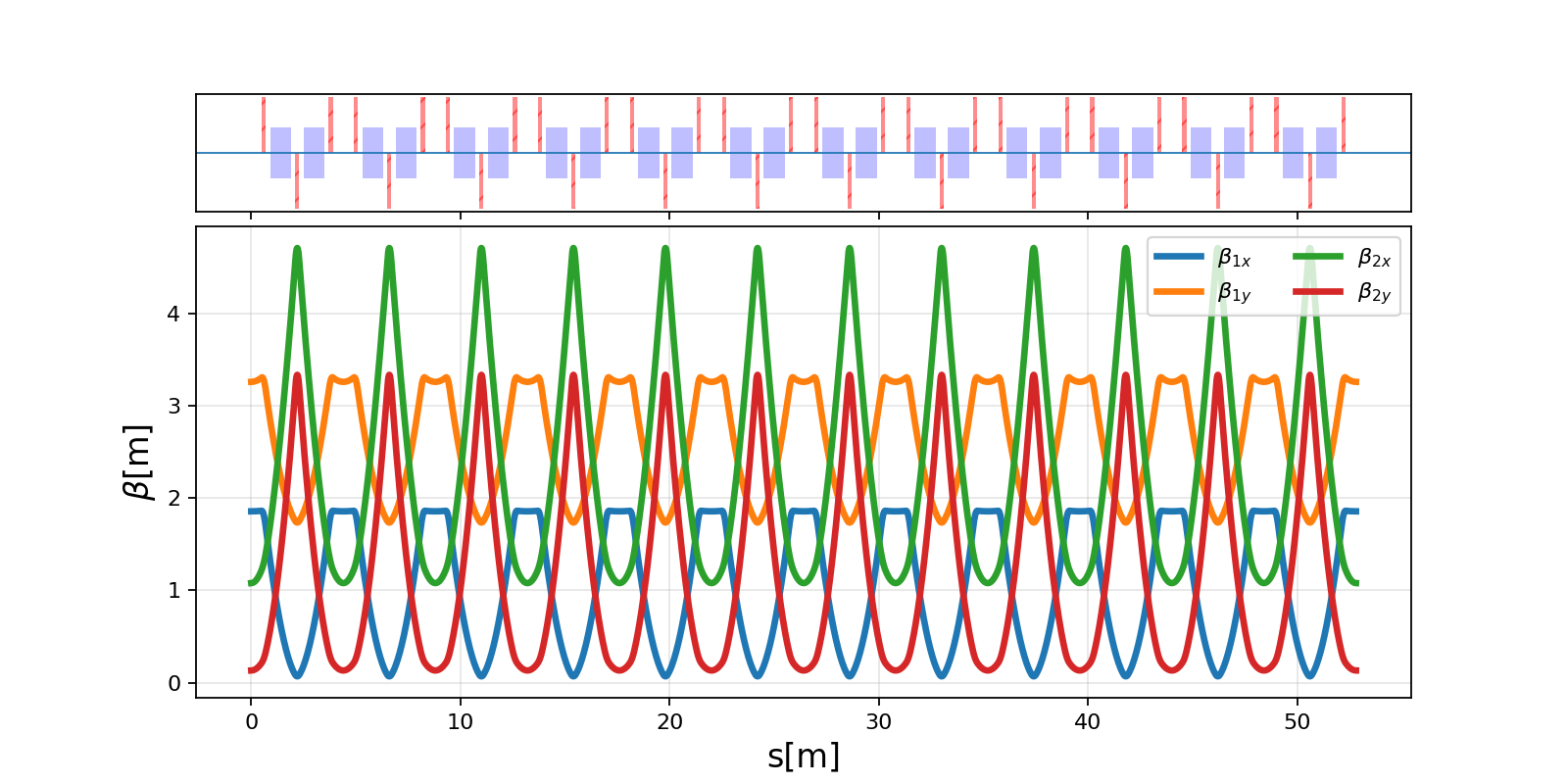}
    \includegraphics[width=0.49\linewidth]{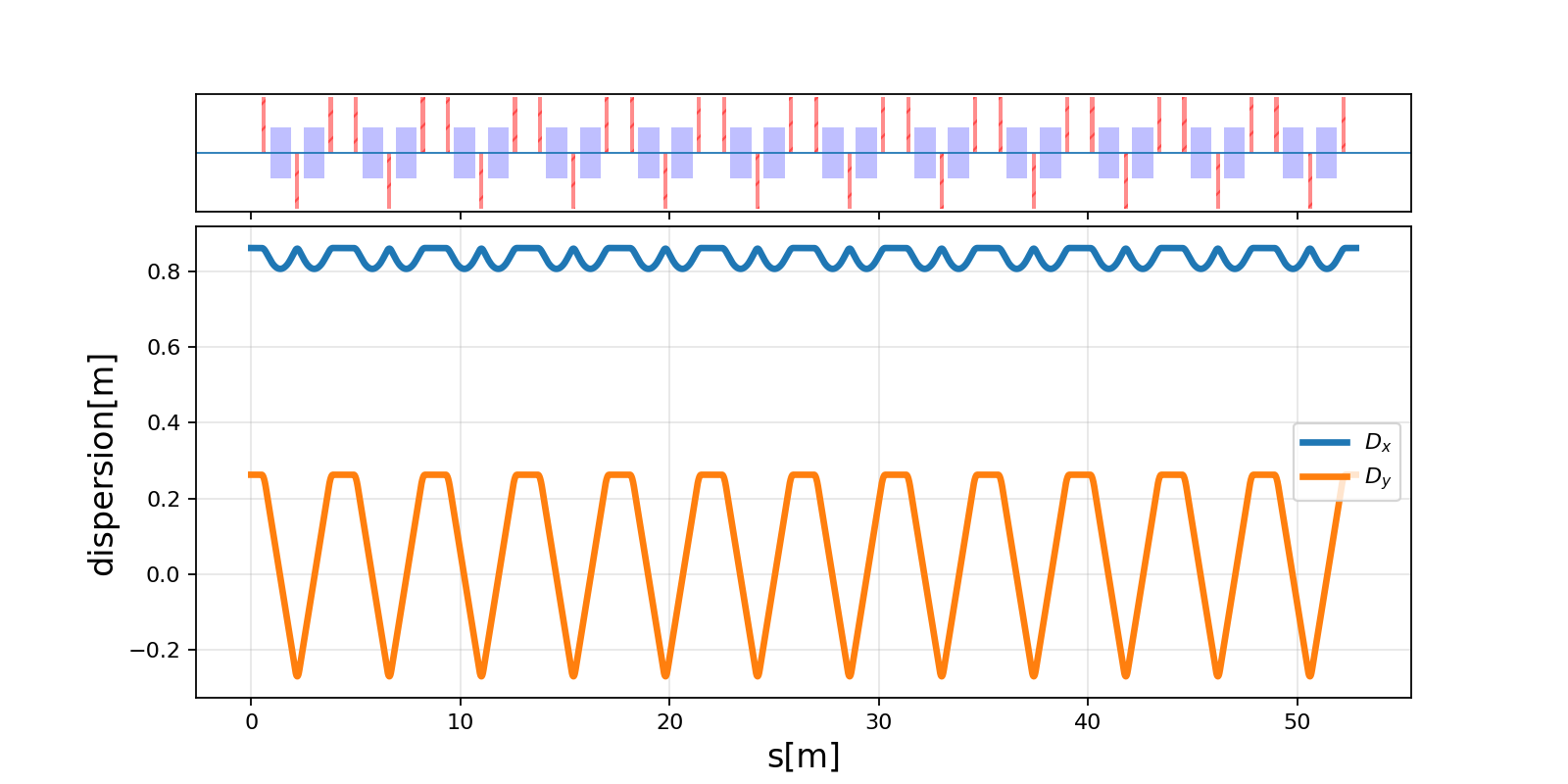}
    \includegraphics[width=0.49\linewidth]{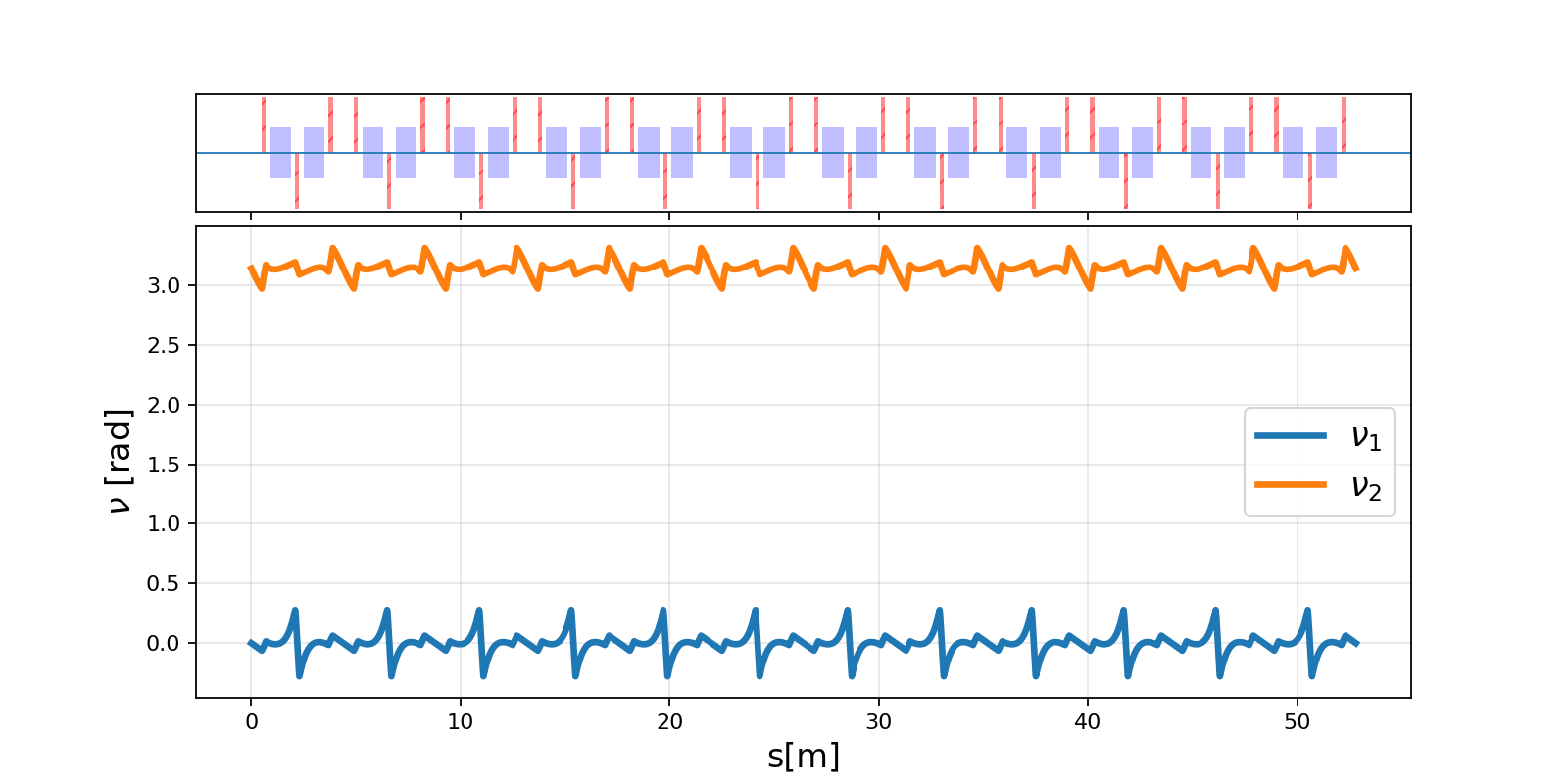}
    \includegraphics[width=0.49\linewidth]{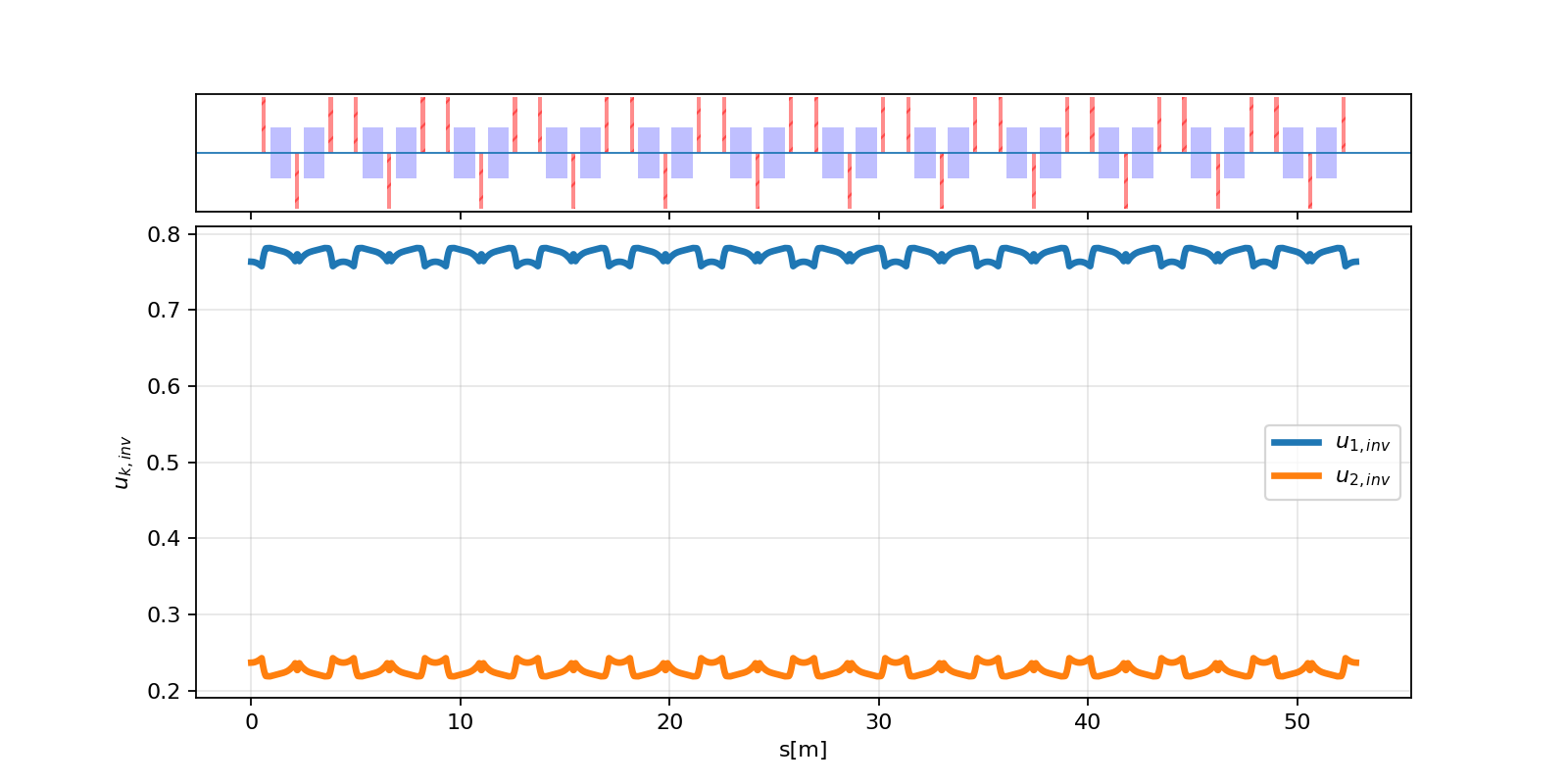}
    \caption{Coupled ring design with skew-quadrupole triplet. Top left plot shows the coupled $\beta$ functions, top right plot shows dispersion functions, bottom left plot shows the coupling phases $\nu_{1,2}$, and bottom right plot shows the invariant coupling strengths $u_{k,\mathrm{inv}}$.}
    \label{fig:skewtripletring}
\end{figure}

\begin{figure}[tbp]
    \centering
    \includegraphics[width=\linewidth]{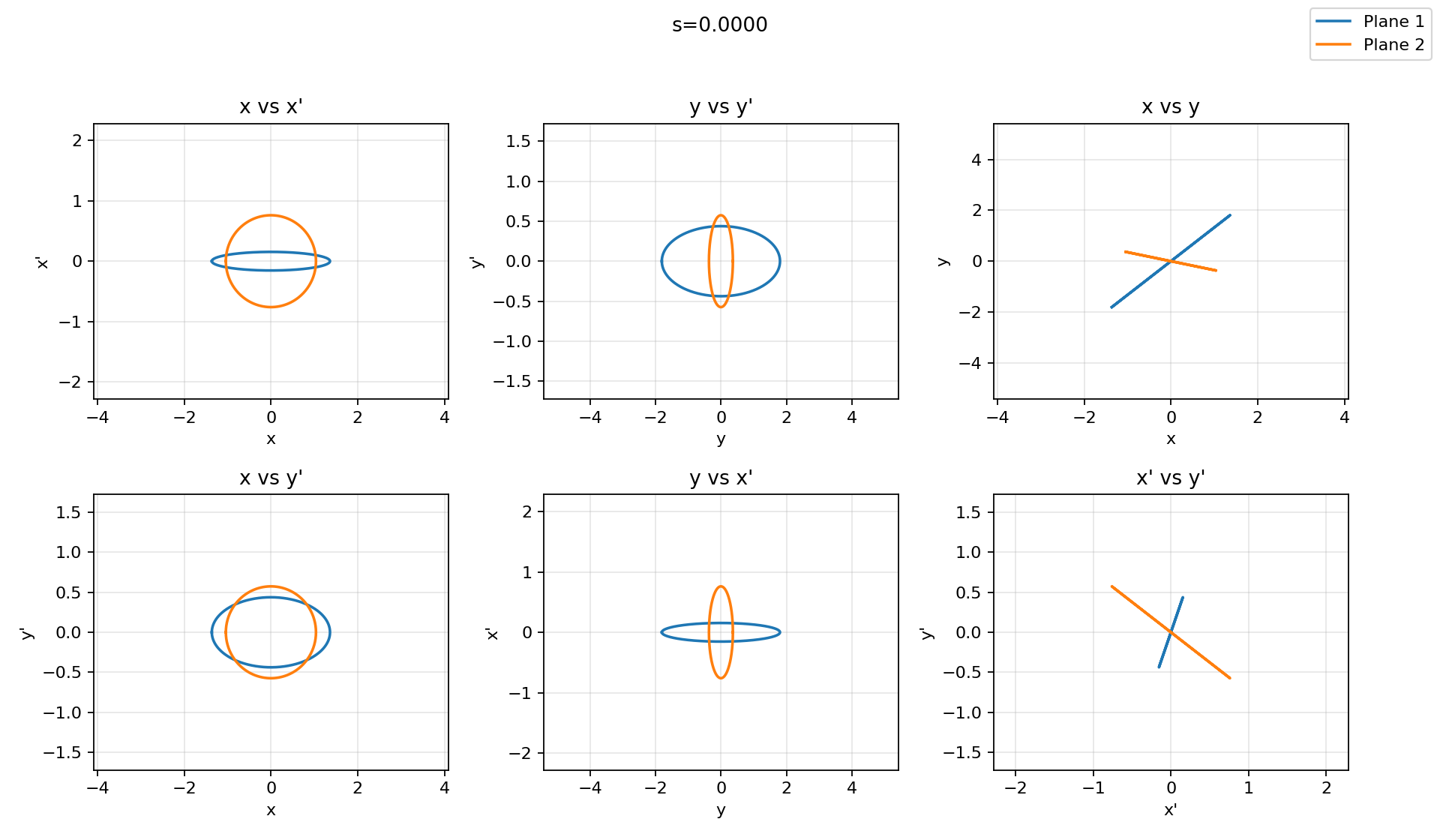}
    \includegraphics[width=\linewidth]{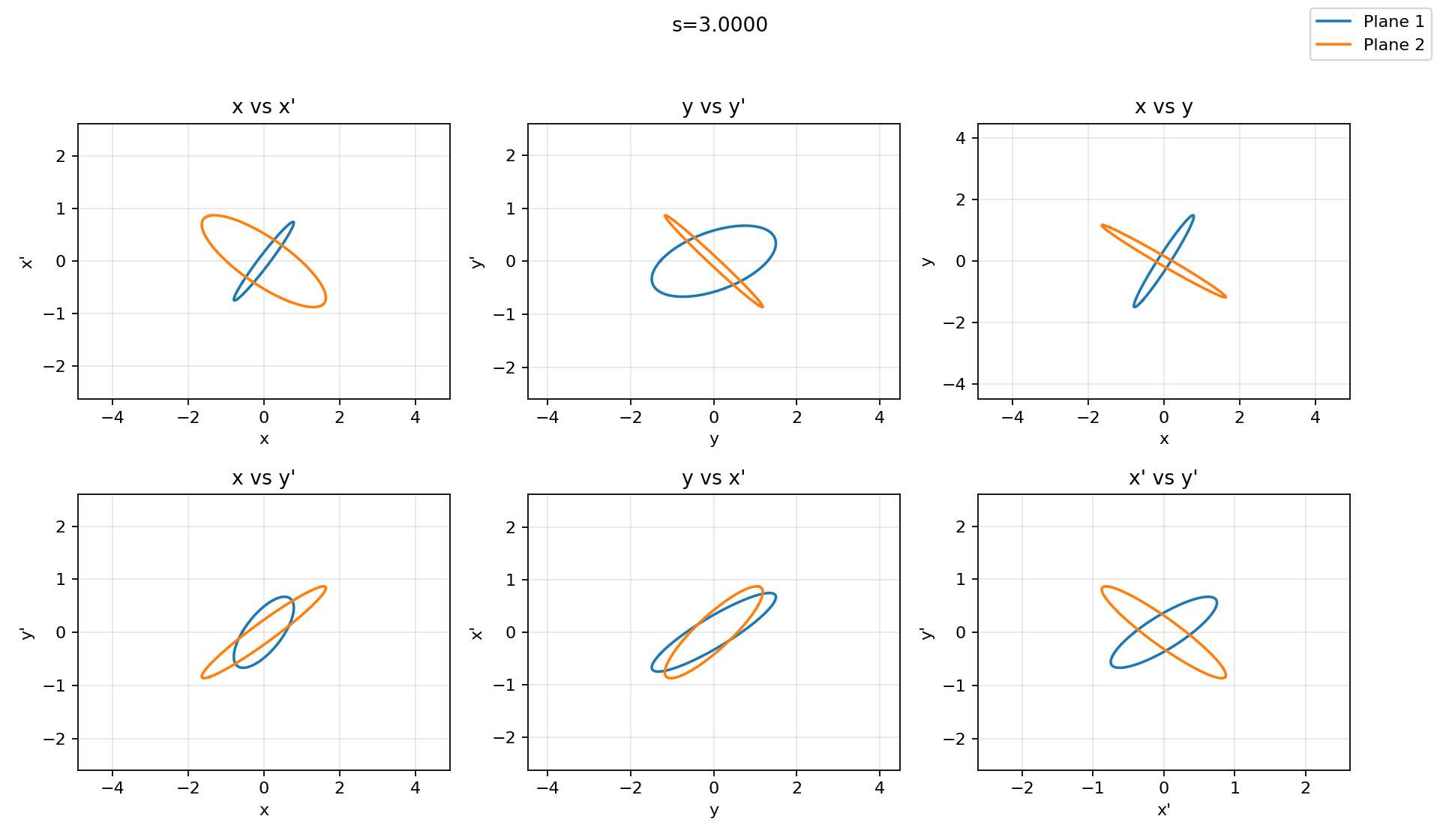}
    \caption{The ellipse projections from invariant planes $\mathcal{P}_1$ and $\mathcal{P}_2$ onto transvserse planes in the lab frame, $(x,x',y,y')$. Top plot is at the beginning of the lattice and the bottom plot is at $s=3.0$\,m.  }
    \label{fig:ellipseprojectionsskewringperi}
\end{figure}

\begin{figure}[tbp]
    \centering
    \includegraphics[width=0.49\linewidth]{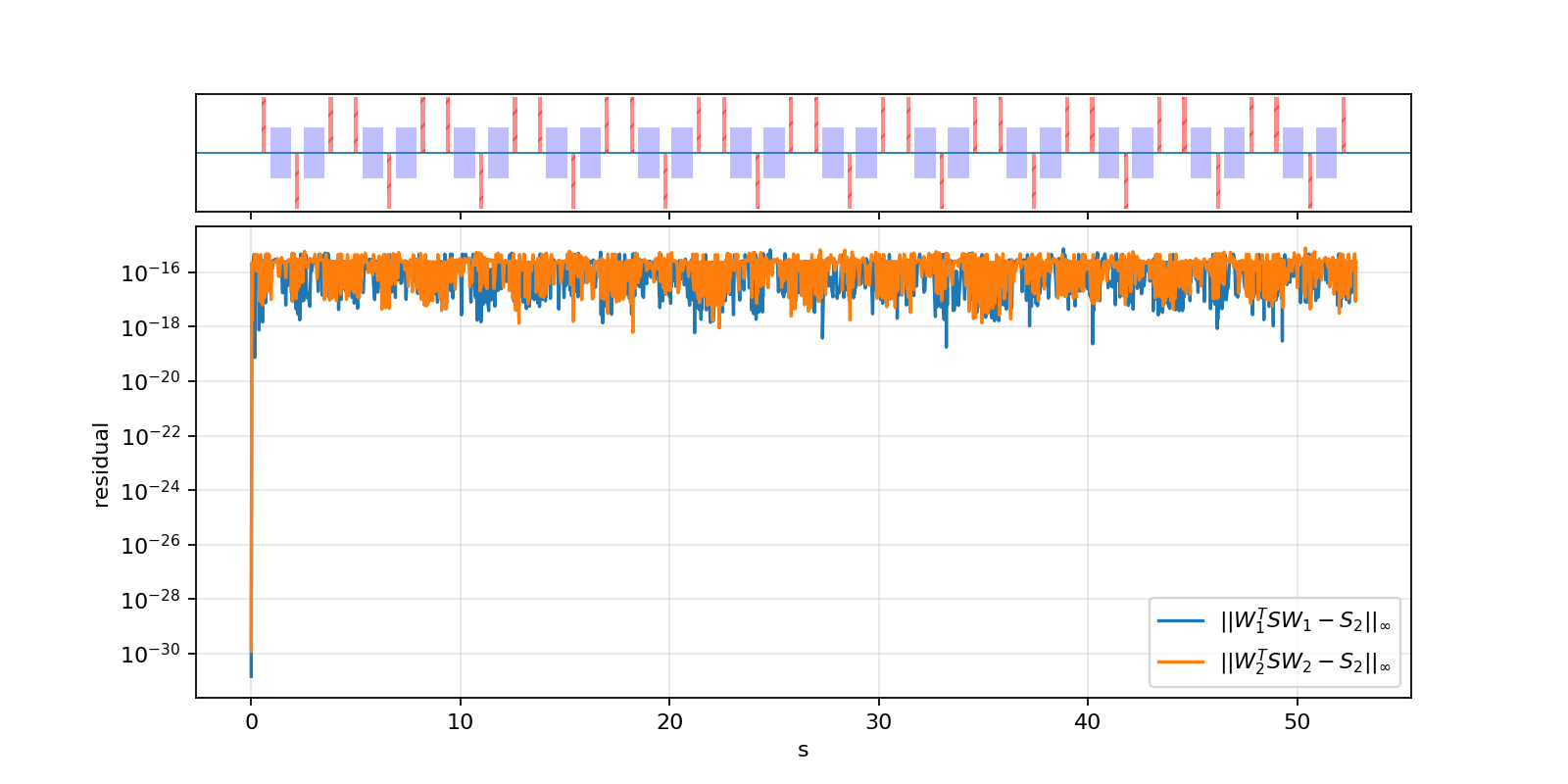}
    \includegraphics[width=0.49\linewidth]{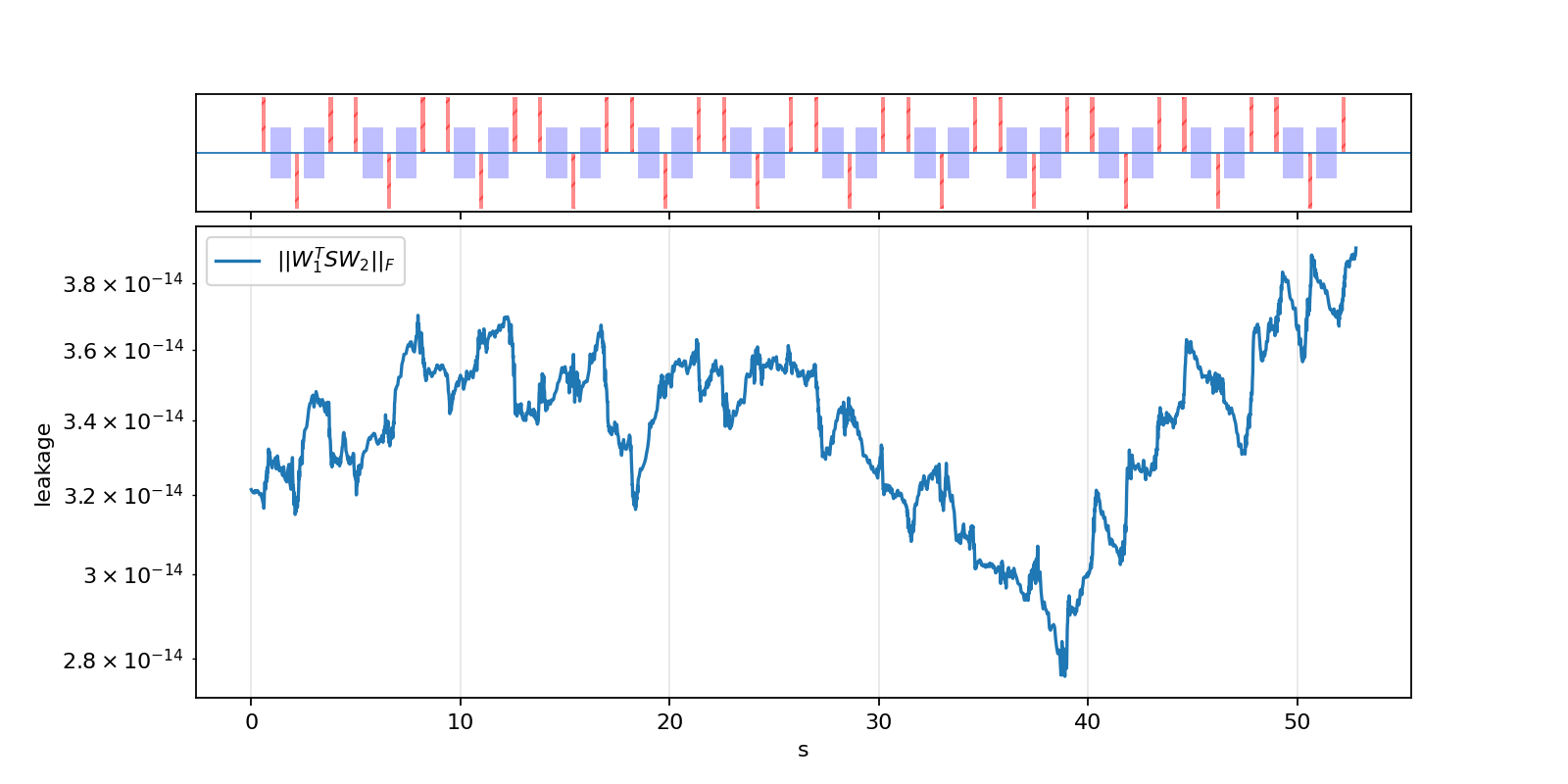}
    \caption{Invariant plane diagnostics: left plot is showing symplecticity of the basis and right plot is showing plane leakage.}
    \label{fig:planediagnosticsskewringperi}
\end{figure}

Our first application is a coupled ring constructed with skew quadrupoles and normal dipoles. The periodic cell constitutes of three skew quadrupoels and two normal dipoles. Each dipole bends with an angle of $15^\circ$ and the formed ring has a super periodicity of 12. The eigenmode tunes are calculated from the one-turn map and are they are $Q_1=0.0873$ and $Q_2=0.2369$. The optics functions and the gauge-invariant coupling strengths $u_{k,\mathrm{inv}}$ are given in Fig.~\ref{fig:skewtripletring}. The periodic solution is calculated from the eigenbasis of the one-turn map $\mathcal{M}$, which shows that the optics functions are periodic. The invariant coupling strengths indicate that mode~1 is $y$ dominant and mode~2 is $x$ dominant. The eigenmode plane projected ellipses are given in Fig.~\ref{fig:ellipseprojectionsskewringperi}, where the first figure is at the beginning of the lattice and the second figure is at $s=3.0$\,m. The second plot also shows how plane projections change within the lattice. We also provide plane diagnostics by computing plane leakage and symplecticity checks of the basis in Fig.~\ref{fig:planediagnosticsskewringperi}. The plane leakage shows whether the defined eignmodes $\mathcal{P}_{1}$ and $\mathcal{P}_2$ mix. The designed ring shows that it is straightforward to match the optics to the eigenmodes of the lattice.

\subsubsection{Fully coupled ring: solenoids and skew quadrupoles}

\begin{figure}[tbp]
    \centering
    \includegraphics[width=0.49\linewidth]{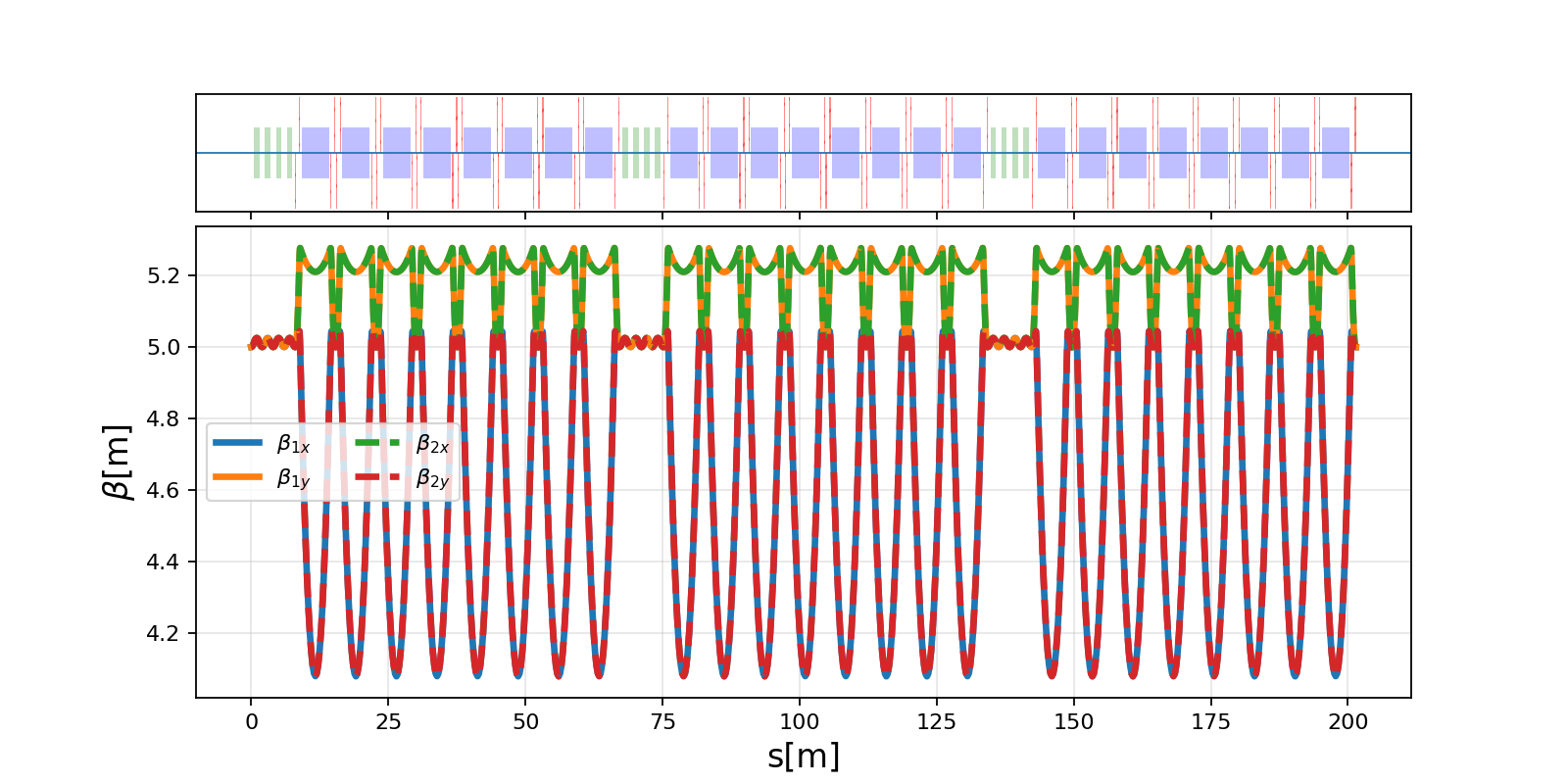}
    \includegraphics[width=0.49\linewidth]{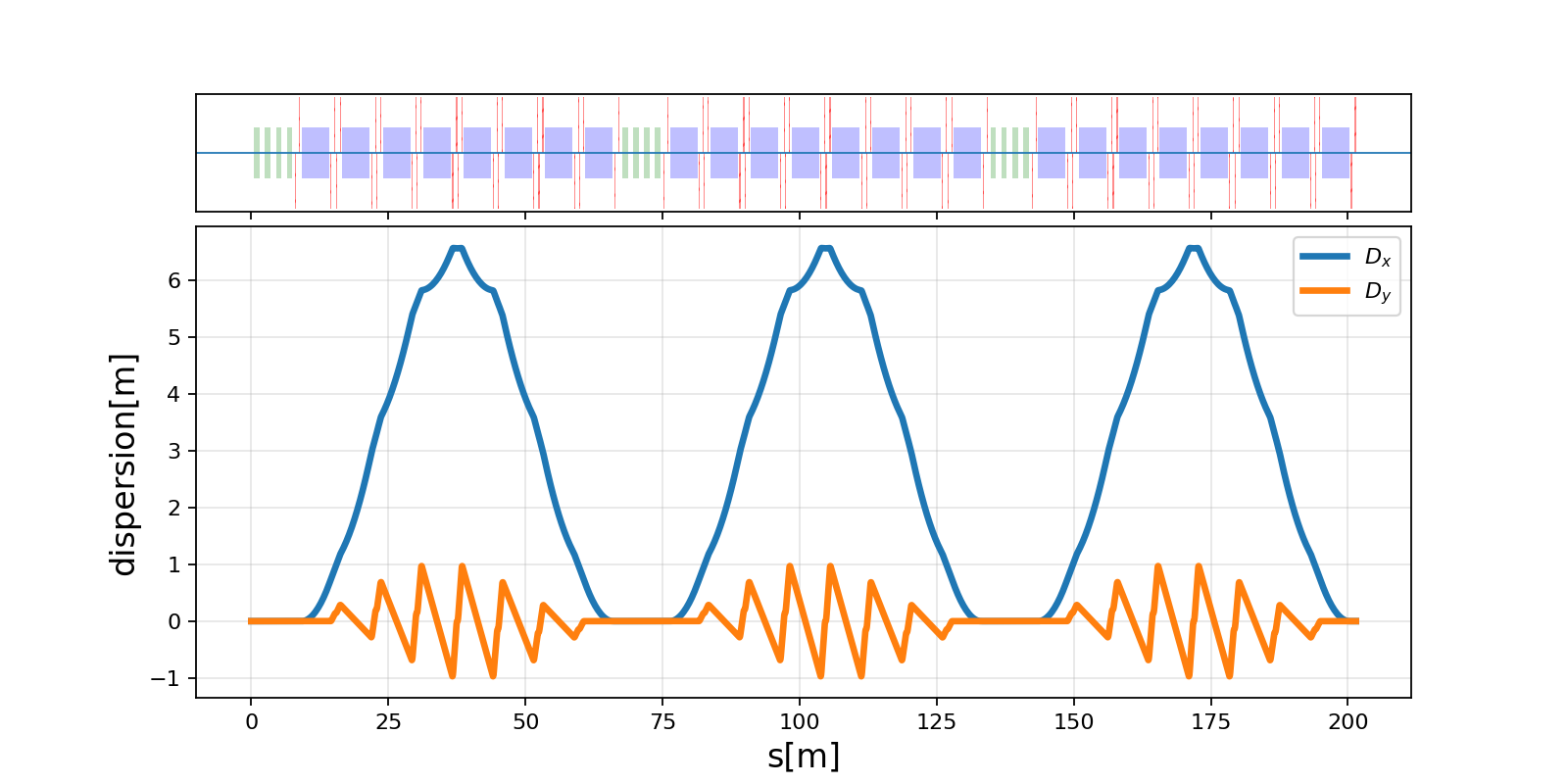}
    \includegraphics[width=0.49\linewidth]{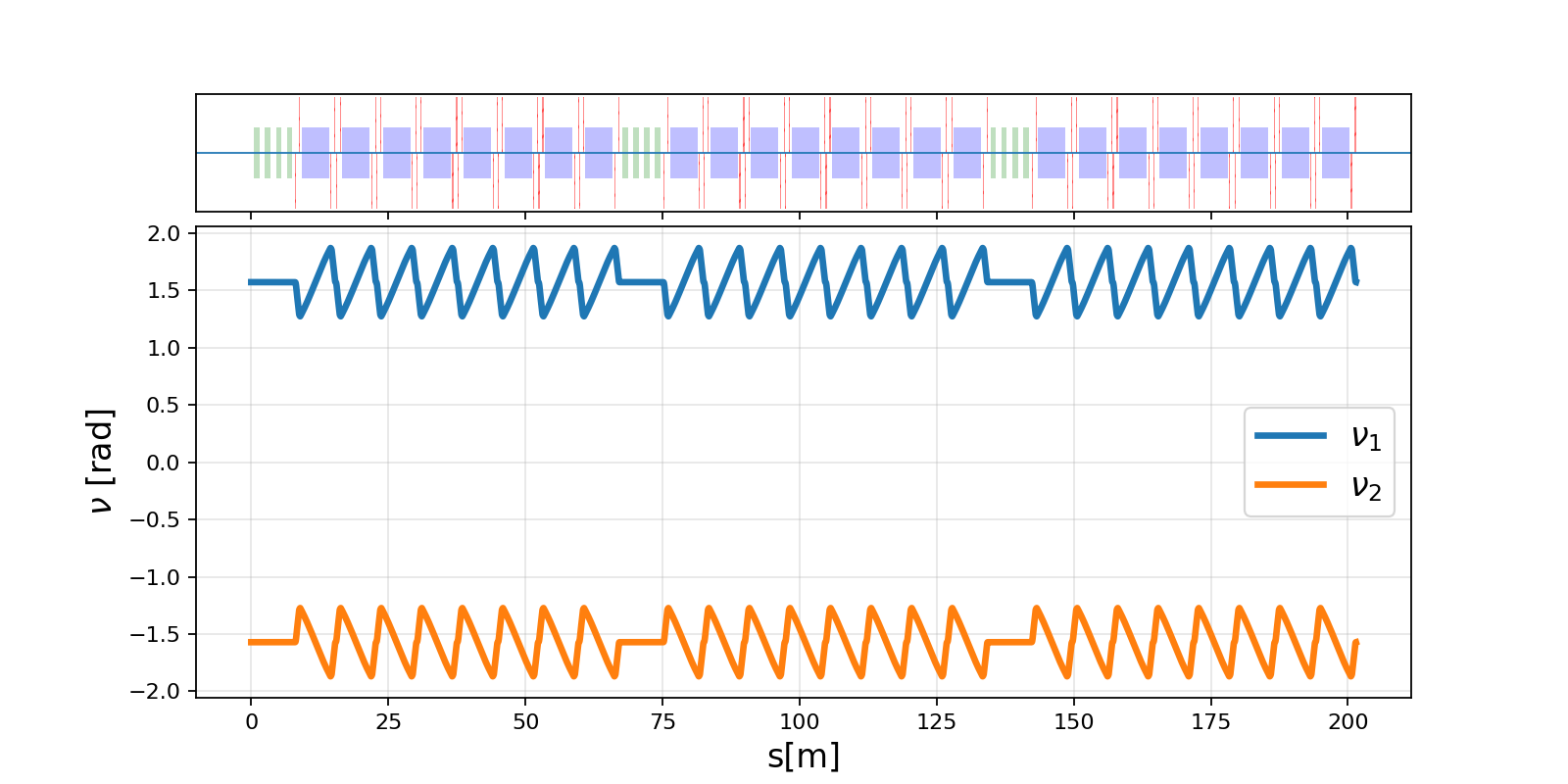}
    \includegraphics[width=0.49\linewidth]{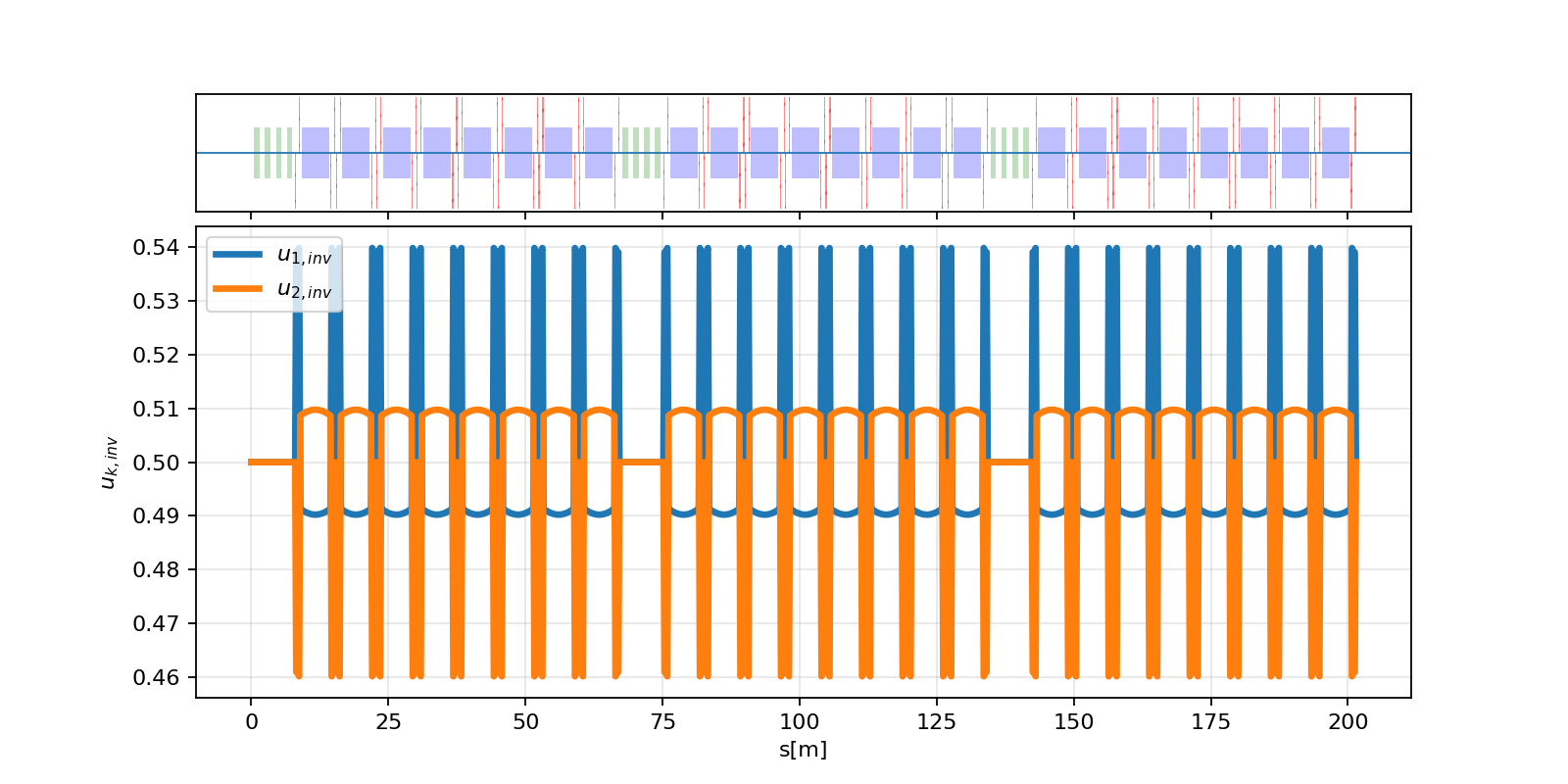}
    \caption{Coupled ring design with skew-quadrupole doublet and solenoids. Top left plot shows the coupled $\beta$ functions, top right plot shows dispersion functions, bottom left plot shows the coupling phases $\nu_{1,2}$, and bottom right plot shows the invariant coupling strengths $u_{k,\mathrm{inv}}$.}
    \label{fig:solskewringoptics}
\end{figure}

\begin{figure}[tbp]
    \centering
    \includegraphics[width=\linewidth]{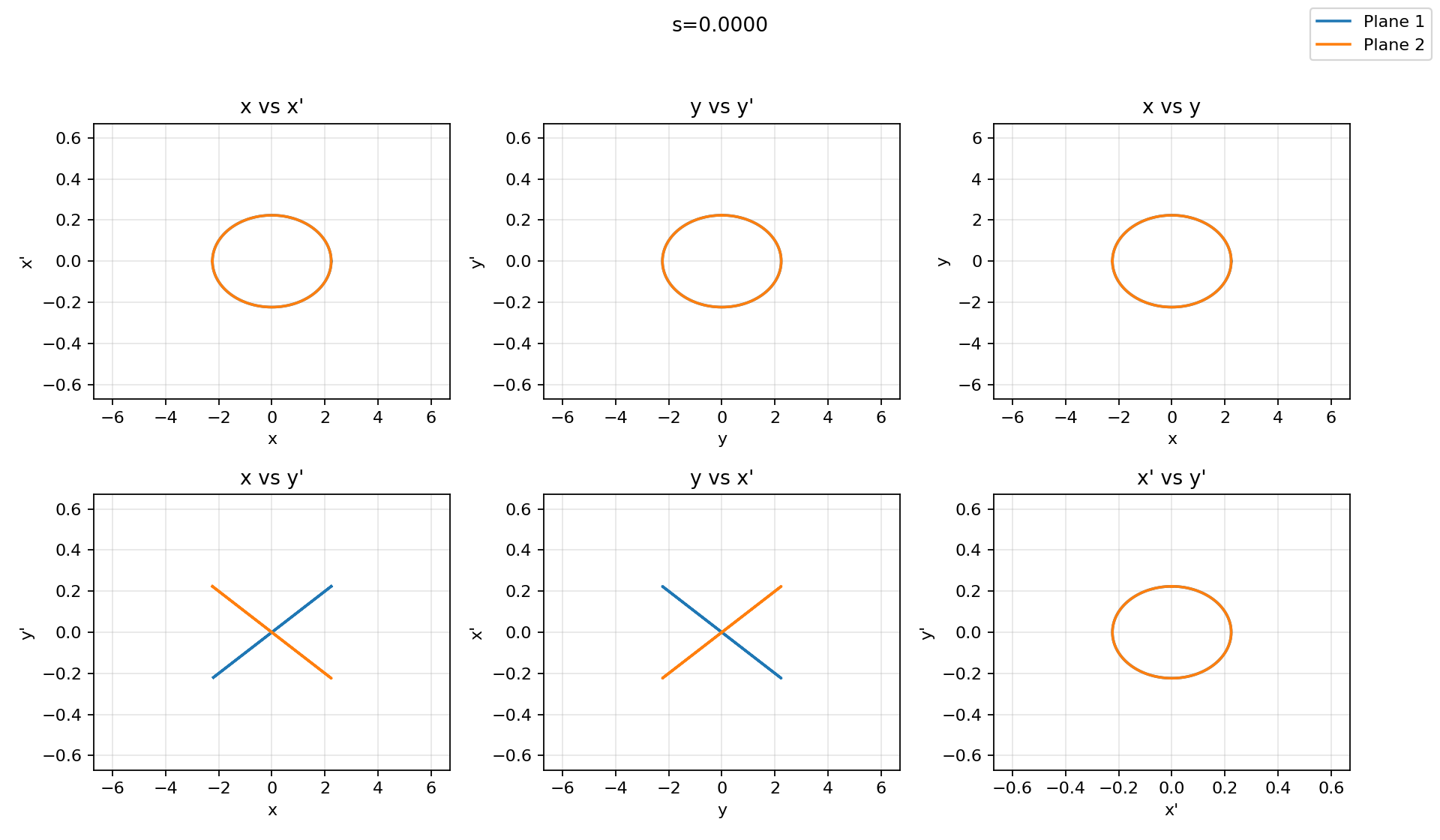}
    \includegraphics[width=\linewidth]{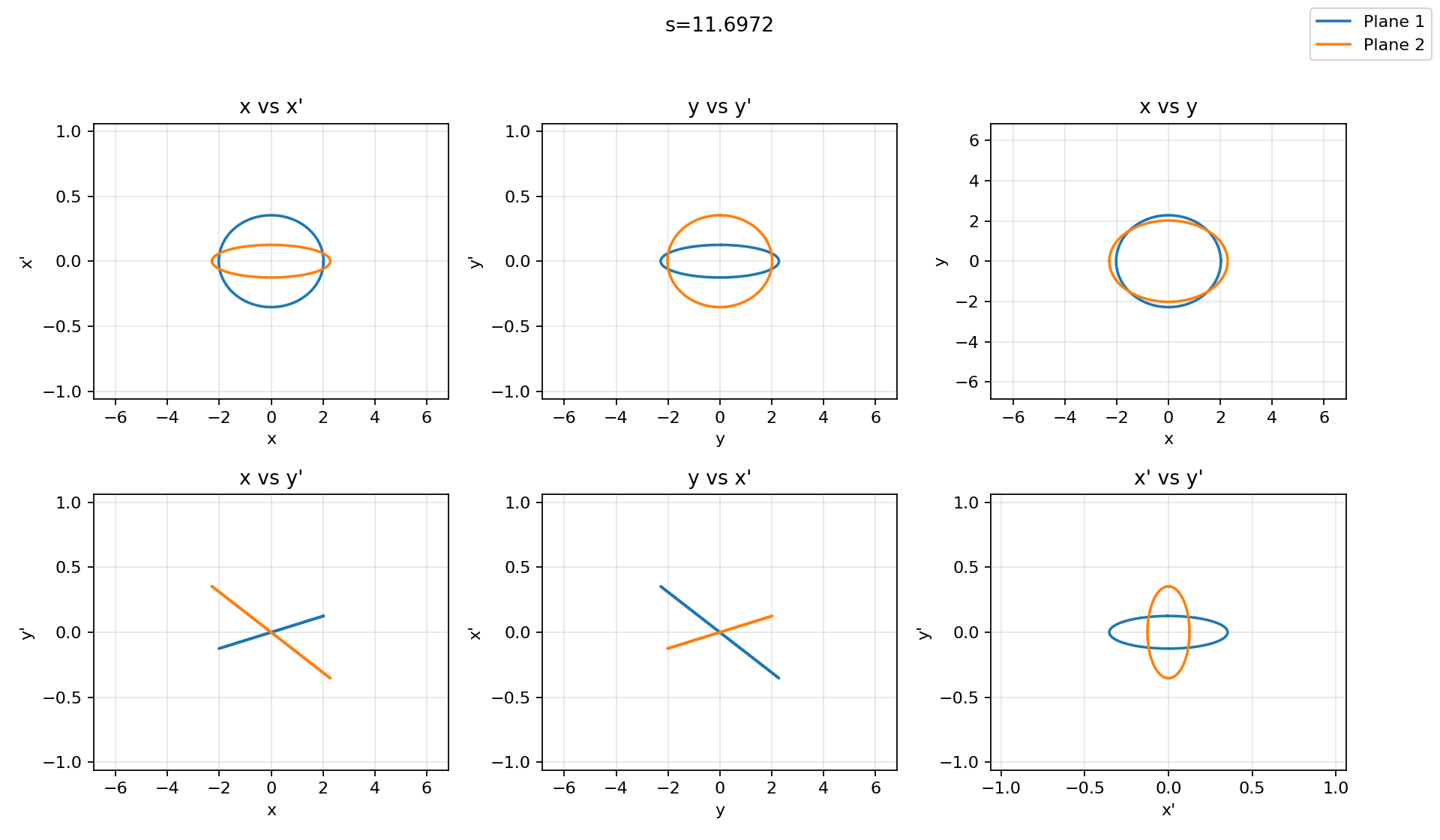}
    \caption{The ellipse projections from invariant planes $\mathcal{P}_1$ and $\mathcal{P}_2$ onto transvserse planes in the lab frame, $(x,x',y,y')$. Top plot is at the beginning of the lattice and the bottom plot is at $s=11.69$\,m.}
    \label{fig:ellipsessolskewring}
\end{figure}

\begin{figure}[tbp]
    \centering
    \includegraphics[width=0.49\linewidth]{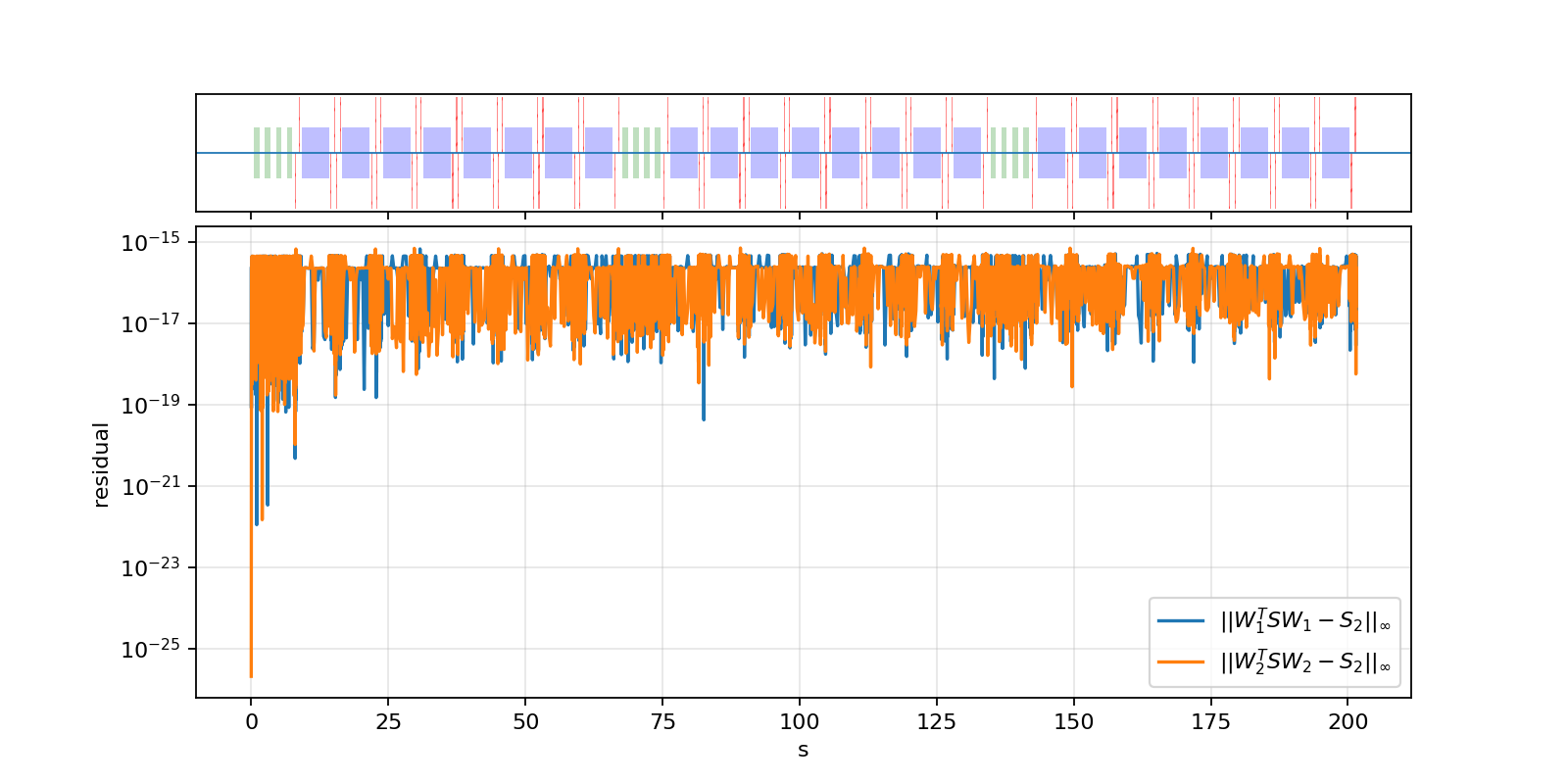}
    \includegraphics[width=0.49\linewidth]{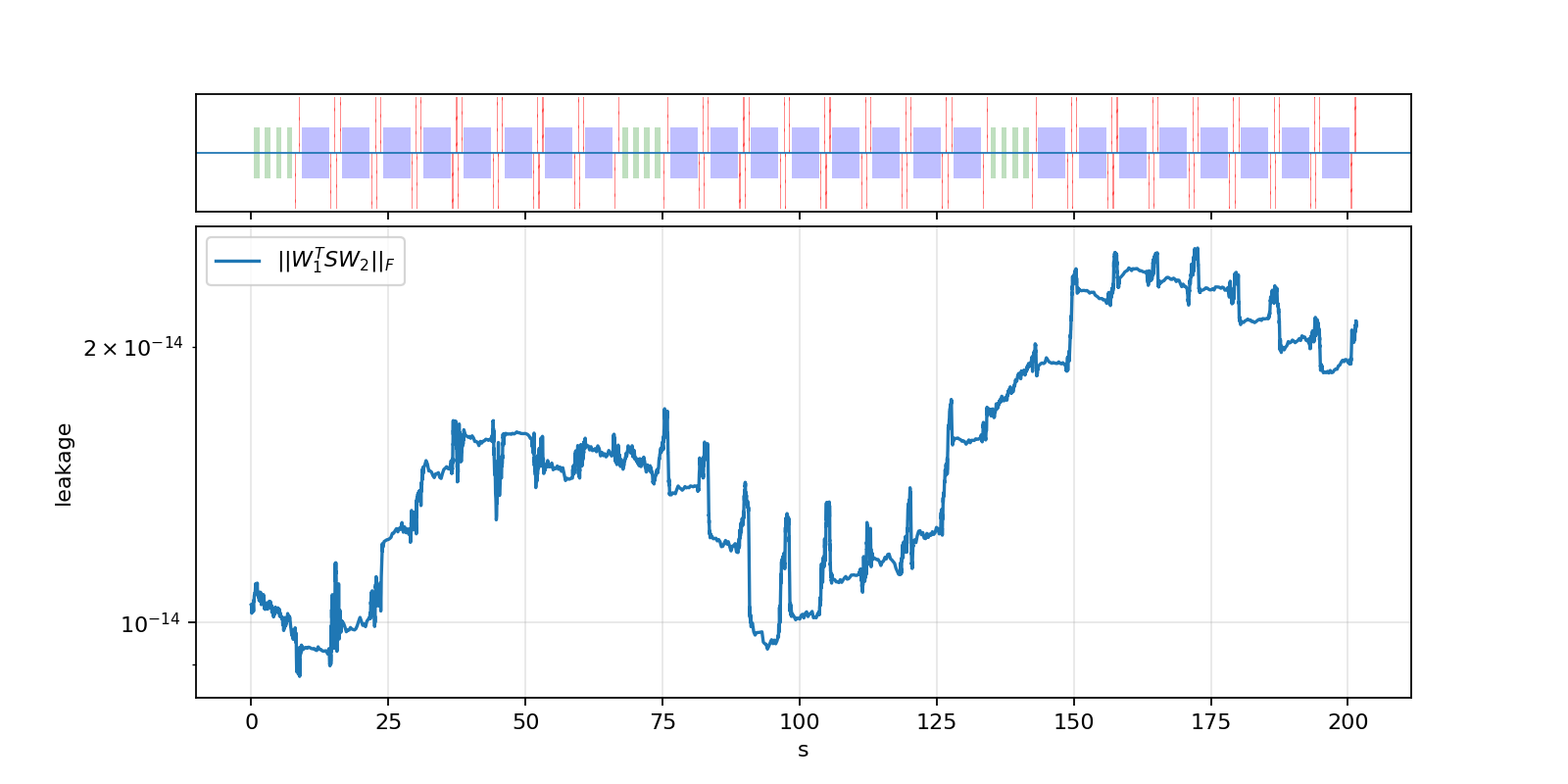}
    \caption{Invariant plane diagnostics: left plot is showing symplecticity of the basis and right plot is showing plane leakage.}
    \label{fig:planediagnosticssolskewring}
\end{figure}

For our second application, we design a fully coupled ring. The arcs are built on achromat condition with a skew-quadrupole doublet repeating cell with indexed dipoles. The indexed dipoles refer to a combined-function element with a dipole and quadrupole field with a focusing index of one half. The indexed dipoles and the skew-quadrupole doublets are degenerate cells, with equal eigenmode phase advances in the cell. Solenoids are included in the straight sections for breaking the degeneracy between the eigenmode planes. With solenoids in straight sections, the eigenmode tunes are: $Q_1=0.1116$ and $Q_2=0.3497$. The ring optics are presented in Fig.~\ref{fig:solskewringoptics}. The $\beta$ function plots show small oscillation through the lattice, which is great for maintaining a round beam. The coupling phases are periodic with values $\nu_1=\tfrac{\pi}{2}$ and $\nu_2=-\tfrac{\pi}{2}$, which results in small correlation in $(x,y)$ plane. The invariant coupling strengths are $u_{k,\mathrm{inv}}\approx 0.5$, indicating both planes have similar content in $\mathcal{X}$ and $\mathcal{Y}$ spaces. The ellipse projections are given in Fig.~\ref{fig:ellipsessolskewring}, where at the start of the lattice the $(x,y)$ plane is circular and other phase spaces are elliptic with no correlations; except, $(x,y')$ and $(y,x')$ phase planes. The projection structure of the modes are known as circular modes of the solenoids and the lattice. If a beam with intrinsic flatness ratio---i.e \ $\epsilon_2\ll\epsilon_1$---is injected into the circular modes, the beam will have non-zero angular momentum dominance~\cite{burov2002circular,gilanliogullaricircular,gilanliogullari2025circular}. The second plot in Fig.~\ref{fig:ellipsessolskewring} shows how the circular-mode projections change through the skew-quadrupole sections. The plane-leakage diagnostics are given in Fig.~\ref{fig:planediagnosticssolskewring}, and in the strong coupling region we see that the continious mode labeling fixes the bookkeeping redundancy.

\section{Conclusion}
In conclusion, we formulated coupled transverse beam optics in terms of the invariant eigenmode planes of a stable symplectic "one-turn" map. In the stable simple-spectrum case, the transverse phase space admits a unique decomposition into two invariant symplectic subspaces, while the choice of a symplectially normalized bases within each plane is not unique and is characterized by an in-plane $Sp(2)\times Sp(2)$ gauge freedom. This perspective clarifies that widely used coupled-optics parametrizations correspond to different gauge choices (bases conventions) built on the same invariant geometric principle. 

We highlighted the importance of gauge-invariance in the dynamics, and physically meaningful quantities must be built from gauge-invariant quantities. We have given an example set of gauge-invariant quantities under $Sp(2)\times Sp(2)$ group. Remarkably, the defined Twiss parameters, phase advances, and coupling measures are representation-dependent parameters and are not invariant under coordinate reparametrizations. To effectively quantify the coupling content of the eigenmode planes, we introduced a bounded, base-independent coupling measure that reflects the lab-frame horizontal/vertical content of each eigenmode plane via projector overlaps.

As mentioned in Sagan--Rubin, mode flips might occur in the regions where coupling is strong. The mode flips occur because parametrization-dependent quantities become ill-defined in strong coupling regions/near degeneracy. For instance, Lebedev--Bogacz's definition of coupling strength value, which in theory should be bounded between $[0,1]$, is also tied to the meaning of the phase advance to be positively defined. The basis-dependent parameters are meaningful when the chosen convention is enforced throughout the dynamics, and when, in certain regimes, the parameters fail to behave properly, and mode flips/swaps occur. To obtain smooth $s$-dependent optics functions and robust mode tracking, we defined and employed a continuity gauge based on Procrustes alignment, which suppresses spurious phase jumps and mode re-labeling away from true near-degeneracy. We have also provided step-by-step computations and matching of coupled optics within the framework we discussed.

Overall, the gauge-invariant plane definition provides a unified framework for comparing coupled-optics parametrizations and for defining physically interpretable coupling parameters and measurable quantities. 

\section{Acknowledgements}
This work was supported by the U.S. Department of Energy, under Contract No. DE-AC02-06CH11357. Work funded by DOE's Office of Nuclear Physics and Argonne contract No. 3J-60004-0003A to Illinois Tech, formerly Illinois Institute of Technology.

\appendix

\section{Transfer Matrices of Coupled Elements}\label{app:transfermatrices}
The transfer maps of coupling elements are computed from equations of motion from the single particle Hamiltonian that is given in~\cite{lebedev2010betatron}. A skew quadrupole can be referred as a normal quadrupole with a rotation, which in general is
\begin{equation}
\begin{split}
    \mathcal{M}_{\mathrm{skew}} &= R(\theta)\mathcal{M}_{\mathrm{quad}}R^{-1}(\theta), \\
    &= \begin{pmatrix}
        \cos\theta\,I_{2\times2} & \sin\theta\,I_{2\times2} \\
        -\sin\theta\,I_{2\times 2} & \cos\theta\,I_{2\times2}
    \end{pmatrix}\cdot\begin{pmatrix}
        M & 0 \\
        0 & N
    \end{pmatrix}\cdot\begin{pmatrix}
        \cos\theta\,I_{2\times2} & -\sin\theta\,I_{2\times2} \\
        \sin\theta\,I_{2\times 2} & \cos\theta\,I_{2\times2}
    \end{pmatrix}.
    \end{split}
\end{equation}
Here, $I_{2\times2}$ is the identity matrix, $\theta$ is rotation angle with respect to the normal quadrupole, $M$ and $N$ are $2\times2$ transfer matrices for horizontal and vertical planes. Typically, a skew quadrupole refers to a normal quadrupole rotated by $45^\circ$, which results in
\begin{equation}
    \mathcal{M}_{\mathrm{skew}} = \frac{1}{2}\begin{pmatrix}
        M+N & N-M \\
        N-M & M+N
    \end{pmatrix}. 
\end{equation}
Here, $M$ and $N$ are dependent on the quadrupole strength $k=\frac{G}{B\rho}$. For a solenoid transfer matrix, the equations are solved exactly with Larmor rotation angle. In $x,x',y,y'$ coordinate system, the solenoids transfer matrix is
\begin{equation}
    \mathcal{M}_{\mathrm{solenoid}} = \begin{pmatrix}
        \frac{1}{2}(1+\cos\chi) & \frac{1}{k}\sin\chi & \frac{1}{2}\sin\chi & \frac{1}{k}(1-\cos\chi) \\
        -\frac{k}{4}\sin\chi & \frac{1}{2}(1+\cos\chi) & -\frac{k}{4}(1-\cos\chi) & \frac{1}{2}\sin\chi \\
        -\frac{1}{2}\sin\chi & -\frac{1}{k}(1-\cos\chi) & \frac{1}{2}(1+\cos\chi) & \frac{1}{k}\sin\chi \\
        \frac{k}{4}(1-\cos\chi) & -\frac{1}{2}\sin\chi & -\frac{k}{4}\sin\chi & \frac{1}{2}(1+\cos\chi)
    \end{pmatrix},
\end{equation}
which includes the canonical transformation.

\section{Invariant Projectors of Invariant Eigenmode Planes}\label{app:InvariantProjectors}

In the simple-spectrum stable case the one-turn map $\mathcal M\in Sp(4)$ admits two real
two-dimensional invariant subspaces (eigenmode planes) $\mathcal P_1$ and $\mathcal P_2$ such that
$\mathbb R^4=\mathcal P_1\oplus\mathcal P_2$ and $\mathcal M\,\mathcal P_k\subset \mathcal P_k$.
While each plane $\mathcal P_k$ is uniquely defined by $\mathcal M$, any basis chosen inside the
plane is not unique. In computations we represent $\mathcal P_k$ by a full-rank basis matrix
$W_k\in\mathbb R^{4\times 2}$ with $\mathrm{span}(W_k)=\mathcal P_k$.
Different choices $W_k\mapsto W_kG_k$ with any invertible $2\times 2$ matrix $G_k$
span the same plane, and in particular the symplectic gauge group $G_k\in Sp(2)$ preserves the
within-plane symplectic normalization used in the main text.
To express plane-dependent quantities independently of the chosen basis, it is convenient to work
with the Euclidean (orthogonal) projector onto $\mathcal P_k$.

\subsection*{Definition and basic properties}

Let $W_k\in\mathbb R^{4\times 2}$ have full column rank. The Euclidean projector onto
$\mathrm{span}(W_k)$ is defined by
\begin{equation}
\Pi_k \;\equiv\; W_k\,(W_k^{T}W_k)^{-1}W_k^{T}.
\label{eq:Pi_def_appendix}
\end{equation}
Since $W_k^TW_k$ is symmetric positive definite for full-rank $W_k$, the inverse exists.
By direct computation, $\Pi_k$ is symmetric and idempotent,
\begin{equation}
\Pi_k^T=\Pi_k,\qquad \Pi_k^2=\Pi_k,
\label{eq:Pi_symm_idem}
\end{equation}
hence it is the orthogonal projector with respect to the standard Euclidean inner product on
$\mathbb R^4$.
Moreover, $\mathrm{range}(\Pi_k)=\mathrm{span}(W_k)=\mathcal P_k$ and $\mathrm{null}(\Pi_k)$ is the
Euclidean orthogonal complement $\mathcal P_k^{\perp}$.

\subsection*{Invariance under basis changes (gauge invariance)}

We now show that $\Pi_k$ depends only on the subspace $\mathcal P_k$ and is independent of the
chosen basis $W_k$.
Let $\widetilde W_k=W_kG_k$ for an arbitrary invertible $2\times 2$ matrix $G_k$.
Then
\begin{align}
\widetilde\Pi_k
&=\widetilde W_k(\widetilde W_k^T\widetilde W_k)^{-1}\widetilde W_k^T
= W_kG_k\Big(G_k^T(W_k^TW_k)G_k\Big)^{-1}G_k^TW_k^T \nonumber\\
&= W_kG_k\Big(G_k^{-1}(W_k^TW_k)^{-1}(G_k^T)^{-1}\Big)G_k^TW_k^T
= W_k(W_k^TW_k)^{-1}W_k^T \nonumber\\
&=\Pi_k.
\label{eq:Pi_invariance_proof}
\end{align}
Therefore $\Pi_k$ is invariant under all changes of basis $W_k\mapsto W_kG_k$,
and in particular under the within-plane symplectic gauge $G_k\in Sp(2)$.
This is precisely the sense in which $\Pi_k$ is a gauge-invariant object associated with the
eigenmode plane $\mathcal P_k$.

\subsection*{Relation to invariance under the lattice map}

Because $\mathcal P_k$ is invariant under $\mathcal M$, one has $\mathcal M W_k = W_k R_k$
for some $2\times2$ matrix $R_k$ (the reduced map). Using the definition
\eqref{eq:Pi_def_appendix},
\begin{equation}
\mathcal M\,\Pi_k
=\mathcal M W_k(W_k^TW_k)^{-1}W_k^T
= W_k R_k (W_k^TW_k)^{-1}W_k^T
=\Pi_k\,\mathcal M \quad \text{on } \mathcal P_k,
\end{equation}
and, equivalently,
\begin{equation}
\Pi_k\,\mathcal M\,\Pi_k = \mathcal M\,\Pi_k,
\label{eq:Pi_M_relation}
\end{equation}
which expresses that $\mathcal M$ maps $\mathcal P_k$ into itself.
(Outside $\mathcal P_k$ the commutator $[\Pi_k,\mathcal M]$ is not generally zero unless
$\mathcal M$ is also invariant on $\mathcal P_k^\perp$.)

\subsection*{Complementary projectors and plane decomposition}

When the spectrum is simple and stable, the planes form a direct sum
$\mathbb R^4=\mathcal P_1\oplus\mathcal P_2$.
If $W=[W_1\ W_2]$ is any full-rank $4\times4$ matrix built from bases of the two planes,
then $\Pi_1$ and $\Pi_2$ satisfy
\begin{equation}
\Pi_1\Pi_2=0,\qquad \Pi_1+\Pi_2=I_4,
\label{eq:Pi_complementary}
\end{equation}
provided $\mathcal P_2=\mathcal P_1^\perp$ with respect to the Euclidean inner product.
In general, the invariant eigenmode planes are not Euclidean-orthogonal, so
$\Pi_1\Pi_2\neq 0$ and $\Pi_1+\Pi_2\neq I_4$ need not hold.
Nevertheless, each $\Pi_k$ remains the unique Euclidean projector onto $\mathcal P_k$ and
retains the gauge-invariance property \eqref{eq:Pi_invariance_proof}.

For symplectic geometry it is often more natural to use symplectic complements rather than
Euclidean complements. If $W$ is symplectically orthonormalized such that
$W^TS_4W=\mathrm{diag}(S_2,S_2)$, then the planes are symplectically orthogonal in the sense
$W_1^TS_4W_2=0$. In that case one may also form the symplectic projector
\begin{equation}
\Pi_k^{(S)} \equiv W_k W_k^{+},\qquad
W_k^{+}\equiv -S_2 W_k^T S_4,
\label{eq:symplectic_projector}
\end{equation}
which is likewise invariant under $W_k\mapsto W_kG_k$ with $G_k\in Sp(2)$ and satisfies
$\Pi_k^{(S)}\Pi_k^{(S)}=\Pi_k^{(S)}$ and $\mathrm{range}(\Pi_k^{(S)})=\mathcal P_k$.
The Euclidean projector $\Pi_k$ and the symplectic projector $\Pi_k^{(S)}$ coincide only in
special gauges where the plane is orthonormal with respect to both structures.
In the main text we use the Euclidean projector \eqref{eq:Pi_def_appendix} to define
basis-independent coupling fractions via traces with coordinate projectors.

\section{Formalism Relations to Standard Parametrizations and Gauge Freedoms}\label{app:formalismrelations}

\subsection{Edwards--Teng Parametrization}\label{subapp:ETparam}

Edwards-Teng parametrization finds a symplectic decoupling matrix $T\in\mathrm{Sp}(4)$ that decouples the one-turn map $\mathcal{M}\in\mathrm{Sp}(4)$, by a transformation
\begin{equation}
    T^{-1}\mathcal{M}T = \begin{pmatrix}
        m_1 & 0 \\
        0 & m_2
    \end{pmatrix}, \qquad m_k\in\mathrm{Sp}(2). 
    \label{eq:ETdecouplingprocess}
\end{equation}
The decoupling matrix $T$ is parametrized as rotation and symplectic shear
\begin{equation}
    \begin{split}
        T = \begin{pmatrix}
            \cos\phi I_{2\times 2} & -\sin\phi D^{-1}_{2\times 2} \\
            \sin\phi D_{2\times 2 } & \cos\phi I_{2\times 2}
        \end{pmatrix},
    \end{split}
    \label{eq:ETdecouplingmatrix}
\end{equation}
where $T^{T}S_4T=S_4$ condition is satisfied. The decoupling matrix $T$ is not unique. If $T$ satisfies Eq.~\eqref{eq:ETdecouplingprocess}, then 
\begin{equation}
T' \;=\; T\,G,
\qquad
G \;=\;
\begin{pmatrix}
G_1 & 0\\
0 & G_2
\end{pmatrix},
\qquad G_k\in Sp(2),
\label{eq:ET_block_gauge}
\end{equation}
produces an equally valid decoupling, because
\begin{equation}
(TG)^{-1}\mathcal{M}(TG)
=
G^{-1}\left(T^{-1}\mathcal{M}T\right)G
=
\begin{pmatrix}
G_1^{-1}m_1G_1 & 0\\
0 & G_2^{-1}m_2G_2
\end{pmatrix}.
\label{eq:ET_block_similarity}
\end{equation}
Therefore each reduced map $m_k$ is defined only up to similarity within $Sp(2)$, while its spectrum
(and hence the tunes) is invariant. Geometrically, \eqref{eq:ET_block_gauge} is a change of canonical
coordinates within each eigenmode plane. It leaves the invariant planes $\mathcal{P}_k$ unchanged,
but modifies the particular basis used to represent them.

The Edwards--Teng decoupling parametrization connects to our formalism with a gauge transformation. As mentioned in Section~\ref{subsec:geometricsubsection}, the gauge transformation in this context is a coordinate transformation. From our Eq.~\eqref{eq:uniqueRmat}, the one turn map $\mathcal{M}$ produces a unique reduced map $R_k$ as:
\begin{equation}
    \mathcal{M}W_k = W_kR_k,
\end{equation}
which is rewritten as
\begin{equation}
    R_k = W_k^{+}\mathcal{M}W_k, \qquad W_k^+ = -S_2W_k^TS_4.
\end{equation}
$R_k$ represents individual invariant plane reduced dynamics, and we can construct the full reduced map $R=\mathrm{diag}(R_1,R_2)$ with
\begin{equation}
    R = W^{-1}\mathcal{M}W, \qquad W=[W_1 \, W_2].
\end{equation}
Naturally, Edwards--Teng decoupling matrix $T$ can be written as
\begin{equation}
    T = W\cdot G, \quad G=\begin{pmatrix}
        G_1 &0 \\
        0 & G_2
    \end{pmatrix}, \quad G_{1,2}\in Sp(2).
\end{equation}
The decoupling process becomes:
\begin{equation}
    G^{-1}W^{-1}\mathcal{M}WG = G^{-1}RG = \begin{pmatrix}
        G_1^{-1}R_1G_1 & 0 \\
        0 & G_2^{-1}R_2G_2
    \end{pmatrix}.
\end{equation}
The eigenvalues of the reduced map $R$ corresponds to the eigenmode phase advance and naturally to the eigenmode tunes, therefore the decoupling from Edwards--Teng formalism is related to our $W$ basis matrix by a similarity relation which leaves the eigenvalues and symplectic normalization unchanged. The Edwards--Teng decoupling matrix $T$ is therefore related to our $W$ basis matrix by a block-diagonal gauge transformation $G$:
\begin{equation}
    T = W\cdot G.
    \label{eq:ETrelationtoW}
\end{equation}
Important clarification is that the matrix $T$ exists but the choice of parametrizing $T$ as Eq.~\eqref{eq:ETdecouplingmatrix} is a gauge choice. One can find a $G$ from Eq.~\eqref{eq:ETrelationtoW} to change the basis $W$ to Edwards--Teng gauge convention. 

\subsubsection{Example computation}

Let us take a single periodic solenoid cell with a computed transfer map
\begin{equation}
    \mathcal{M} = \begin{pmatrix}
        0.97044113 & 1.96214437 &  0.13774626 & 0.2785105 \\
        -0.01961854 & 0.97044113 & -0.00278469 & 0.13774626 \\
        -0.13774626 & -0.2785105 & 0.97044113 &  1.96214437 \\
        0.00278469 & -0.13774626 &  -0.01961854 & 0.97044113
    \end{pmatrix}. 
\end{equation}
The eigenvalues are related to the tunes which are $Q_1= 0.05419$ and $Q_2=0.0093$. The basis $W=[W_1 \; W_2]$ are constructed from the eigenvectors and yield:
\begin{equation}
    W = \begin{pmatrix}
        -2.23615072 & 0.0 &  0.0 & 2.23615072 \\
        0.0 &  -2.23598523*10^{-1} &  -2.23598523*10^{-1} & 0.0 \\
        0.0 &  -2.23615072 &  2.23615072 & 0.0 \\
        2.23598523*10^{-1} & 0.0 &0.0 &  2.23598523*10^{-1}
    \end{pmatrix}.
\end{equation}
The reduced map is: $R = \mathrm{diag}(R_1,R_2)=W^{-1}\mathcal{M}W$:
\begin{equation}
    R = \begin{pmatrix}
         9.42592155*10^{-1}&  3.33946178*10^{-1}& 0.0& 0.0 \\
         -3.33946178*10^{-1} & 9.42592155*10^{-1} &0.0 &  0.0 \\
         0.0 & 0.0 & 9.98290105*10^{-1} & 5.84536580*10^{-2} \\
         0.0 & 0.0 & -5.84536580*10^{-2} & 9.98290105*10^{-1}
    \end{pmatrix}.
    \label{eq:reducedmapexample}
\end{equation}
The tunes computed from the reduced map and one-turn map are identical. The reduced map $R$ is already in the rotation matrix format, $R\in SO(2)\times SO(2)$, which yields the circular normal-mode coordinates with action-angle parametrization. Therefore, it is valid to say the $W$ matrix is a decoupler and is related to Edwards--Teng $T$ matrix. We can set $T=W$ for a decoupler matrix, however the form of $W$ is not Edwards--Teng decoupling matrix in representation. Mathematically, they have to be related for this example, with a gauge transformation: $T=W.G$ for $G=\mathrm{diag}(G_1,G_2)\in Sp(2)\times Sp(2)$. For $u_{k,inv}=0.5$, the Edwards--Teng decoupling angle is $\phi=\tfrac{\pi}{4}$, which yields
\begin{equation}
    \begin{split}
        \frac{1}{\sqrt{2}}\begin{pmatrix}
            I_2 & -D^{-1} \\
            D & I_2
        \end{pmatrix} = \begin{pmatrix}
            A & B \\
            C & F
        \end{pmatrix} \cdot \begin{pmatrix}
            G_1 & 0 \\
            0 & G_2
        \end{pmatrix}.
    \end{split}
    \label{eq:solvingforG}
\end{equation}
Solving the Eq.~\eqref{eq:solvingforG}, yields $G$ as:
\begin{equation}
    G = \begin{pmatrix}
        0.4471970 & 0.0 & 0.0 & 0.0 \\
        0.0 & 4.47230 & 0.0 & 0.0 \\
        0.0 & 0.0 & 0.4471970 & 0.0 \\
        0.0 & 0.0 &0.0 & 4.47230
    \end{pmatrix}. 
\end{equation}
Using $G$ and the calcualted basis $W$, we get Edwards--Teng representation matrix:
\begin{equation}
    \begin{split}
        T &= W\cdot G \\
        & = \frac{1}{\sqrt{2}}\begin{pmatrix}
            1 & 0 & 0 & 10.0 \\
            0 & 1 & -0.1 & 0 \\
            0 & -10.0 & 1 & 0 \\
            0.1 & 0 & 0 & 1
        \end{pmatrix}. 
    \end{split}
\end{equation}
Therefore, this shows that Edwards--Teng decoupling matrix $T$ and our basis $W$ is related with a gauge transformation. Finally, computing the diagonlization from the $T$ matrix yields:
\begin{equation}
    \begin{split}
        \mathcal{M}_{\mathrm{diag}} &= T^{-1}\cdot\mathcal{M}\cdot T \\
        &= \begin{pmatrix}
            9.98290105*10^{-1} & 5.84579842*10^{-1} & 0.0 & 0.0 \\
            -5.84493320*10^{-3} & 9.98290105*10^{-1} & 0.0  &0.0 \\
            0.0 & 0.0 & 9.42592155*10^{-1} & 3.33970894 \\ 
            0.0 & 0.0 & -3.33921464*10^{-2} &  9.42592155*10^{-1}
        \end{pmatrix}
    \end{split}
\end{equation}
Finally, the eigenmode tunes from the diagonalized map yields: $Q_1=0.0093$ and $Q_2=0.05419$. Which shows that the Edwards--Teng decoupling matrix produces the same eigenmode tunes but with a mode swap, which is a valid mode labeling convention. The diagonalized map produced by Edwards--Teng decoupling matrix $T$ is different than the reduced map in Eq.~\ref{eq:reducedmapexample}, which also shows the gauge-freedom between different parametrizations.

\bibliographystyle{JHEP}
\bibliography{biblio.bib}

\end{document}